\documentclass[manuscript]{aastex}
\usepackage{amsfonts,amsmath,graphicx,color}

\shorttitle{Investigation of the orientation of galaxies in clusters}
\shortauthors{Pajowska, God{\l}owski, Zhu, Popiela, Panko \& Flin}
\begin{document}
\title{Investigation of  the orientation of galaxies in clusters: the importance, methods and results of research}

\author{Paulina Pajowska}
\affil{Uniwersytet Opolski, Institute of Physics, ul.  Oleska  48, 45-052 Opole, Poland}
\email{paoletta@interia.pl}

\author{W{\l}odzimierz God{\l}owski}
\affil{Uniwersytet Opolski, Institute of Physics, ul.  Oleska  48, 45-052 Opole, Poland}
\email{godlowski@uni.opole.pl}

\author{Zong-Hong Zhu}
\affil{Department of Astronomy, Beijing Normal University, Beijing 100875, China}
\email{zhuzh@bnu.edu.cn}

\author{Joanna Popiela}
\affil{Uniwersytet Opolski, Institute of Physics, ul.  Oleska  48, 45-052 Opole, Poland}
\email{joa\_ols@wp.pl}

\author{Elena Panko}
\affil{I.I.Mechnikov Odessa National University, Theoretical Physics and Astronomical Department, Park Shevchenko, Odessa, 65014, Ukraine}
\email{panko.elena@gmail.com}

\author{Piotr Flin}
\affil{Jan Kochanowski University, Institute of Physics, ul. Swietokrzyska 15, 25-406 Kielce, Poland}
\email{sfflin@cyf-kr.edu.pl}

\begin{abstract}
Various models of structure formation can account for various aspects of 
the galaxy formation process on different scales, as well as for various 
observational features of structures. Thus, the investigation of galaxies
orientation constitute a standard test of galaxies formation scenarios since
observed variations in angular momentum represent fundamental constraints for 
any model of galaxy formation.  We have improved the method of analysis of 
the alignment of galaxies in clusters. Now, the method allows to analyze both 
position angles of galaxy major axes and two angles describing the spatial 
orientation of galaxies. The distributions of analyzed angles were tested 
for isotropy by applying different statistical tests. For sample of analyzed 
clusters we have computed the mean values of analyzed statistics, checking 
whether they are the same as expected ones in the case of random distribution 
of analyzed angles. The detailed discussion of this method has been performed. 
We have shown how to proceed in many particular cases in order to improve the 
statistical reasoning when analyzing the distribution of the angles in the 
observational data. Separately, we have compared these new results with those 
obtained from numerical simulations. We show how powerful is our method on 
the example of galaxy orientation analysis in 247 Abell rich galaxy clusters. 
We have found that the orientations of galaxies in analyzed clusters are not 
random. It means that we genuinely confirmed an existence of the alignment of 
galaxies in rich Abells' galaxy clusters. This result is independent from the 
clusters of Bautz-Morgan types.
\end{abstract}

\section{Introduction}

Solving the problem of the structures formation is one of the most
significant issue of modern extragalactic astronomy.  Many authors investigated the scenarios
of structures formation since \citet{Peebles69,Zeldovich70}.
New scenarios are mostly modifications and improvements of the older ones
\citep{Lee00,Lee01,Lee02,Navarro04,MO05,Bower05,Trujillio06,Brook08,Paz08,Shandarin12,Codis12,Varela12,Giah14,Blaz15}.

The final test of veracity in a given scenario is the convergence of its predictions 
with observations. 
One of the possibilities of such test is analysing the angular momenta of
galaxies. Investigating the orientation of galaxy planes in space is of
great importance since various scenarios of cosmic structures formation and
evolution predict different distributions of galaxies angular momentum,
\citep{Peebles69,Doroshkevich73,Shandarin74,Silk83,Catelan96,Li98,Lee00,Lee01,Lee02,Navarro04,Trujillio06,Zhang13},
i.e. provide distinct predictions concerning the orientation of objects at different
levels of structure -- in particular clusters and superclusters of galaxies.
Our model assumes that normals to the planes of galaxies are their rotational
axes, which seems to be quite reasonable, at least for the spiral galaxies.
Various models can account for various aspects of the galaxy formation process
on different scales, as well as for various observational features of
structures. This provides us with a method for testing scenarios of galaxy
formation. In other words, the observed variations in angular momentum
give us simple but fundamental test for different models of galaxy formation
\citep{Rom12,Joachimi15,kiess15}.

From the observational point of view it is not very difficult to investigate the
distribution of the angular momenta for the luminous matter i.e. real galaxies and 
their structures. One should note however, that in real Universe, observed luminous 
matter of galaxies are surrounded by dark matter halos that are much more extended 
and massive. Direct observation of dark mater halos and theirs angular momentum
is not so easy. Fortunately, there are the relation between luminous and dark mater 
(sub)structures. As a result we have a dependence between dark matter halos and luminous 
matter (real galaxies) orientation \citep{Trujillio06,Paz08,Pe08,Bett10,Paz11,Kim11,Varela12}.
Recently, the analysis of the Horizon-AGN simulation shows the similar dependence \citep{Okabe18,Codis18}.
It means that the analysis of angular momentum of luminous matter gives us also 
information about angular momentum of the total structure hence the analysis of 
the angular momentum of real (luminous) galaxies is still useful as a test of 
galaxy formation. The investigation of the galaxies orientation in clusters are 
also very important with regard to investigation of weak gravitational lensing 
For more detailed discussion of the  significance of this problem see  
\citet{heav00,Hey04,kiess15,sg15,Codis16}.

Since the angular momenta of galaxies and also the directions of galaxy spin
are usually unknown, the orientations of galaxies are investigated
instead. In order to acquire this, either the distributions of galaxy position
angles only \citep{h4} or the distributions of the angles giving the orientation
of galaxy planes \citep{Jaaniste78,f4} are examined. Many authors investigated
the orientation of galaxies in different scales. The review of the observational
results on the problem of galaxies orientation and structures formation was
presented both theoretically \citep{Sch09} and observationally 
\citep{g2011a,Rom12}. 

One of the most meaningful aspects of the problem of the origin of galaxies involves
the investigation of the orientation of galaxies in clusters.  During the analysis of the
angular momentum of a galaxy cluster, in principle we should take into account that total 
angular momentum of the cluster could come from both the angular momentum of each galaxy
member and from the rotation of the cluster itself. However, one should note that there
is no evidence for rotation of the groups and clusters of galaxies themselves. So, it
is commonly agreed that such structures do not rotate (for example
\citet{regos89,dia97,dia99,rines03,Hwang07,Tovm15}, see however \citet{kal05} for 
the opposite opinion). An especially important result is obtained by \citet{Hwang07}. 
They have examined the dispersions and velocity gradients in 899 Abell clusters and 
have found a possible evidence for rotation in only six of them. This allowed us to 
conclude that any non-zero angular momentum in groups and clusters of galaxies should 
arise only from possible alignment of galaxy spins. Moreover, the stronger alignment 
means the larger angular momentum of such structures.

For many years, astrophysisicts payed a lot of attention to the orientation of galaxies 
in clusters. It was investigated both theoretically (see for example \citep{Ci94,Ci98}) 
and observationally. Generally, summarizing the research results provided by various 
authors, it can be stated that we have no satisfactory evidence for the alignment
of galaxies in groups and poor clusters of galaxies, while there is an ample
evidence of this kind for rich clusters of galaxies \citep{Godlowski05}
(see also \citet{g2011a} for an improved analysis and \citet{sg15} for review).

Thus, an interesting problem arises if there are any dependence on the alignment
to the mass of the structure. \citet{Godlowski05} suggested that the alignment 
of galaxies in clusters should increase with the mass of the cluter. Thus, 
\citet{Godlowski05} hinted that the alignment should increase with the number of 
objects (richness) in a particular cluster, too. These suggestions  were later 
confirmed by \citet{Aryal07}. 
These autors analyzed a total of 32 clusters of different richness and BM types. 
They  confirmed that the alignment is changing with the richness and moreover 
that they change  with BM type of the clusters.
 However, one should note that both \citet{Godlowski05} and \citet{Aryal07} investigations
 were qualitative only.  The next step is to test this hypothesis also quantitatively.

This was the reason why \citet{g10a} examined the orientation of galaxies in
clusters both qualitatively and quantitatively. In this paper it was found that
the alignment of galaxy orientation increased with numerousness of the cluster. 
However, the problem that we may obtain is whether we found a significant alignment in analyzed 
sample of 247 rich Abell clusters, or increasing alignment with cluster richness only. 
For this reason \citet{g2011b} analyzed the distribution of position angles using 
$\chi^2$ test, Fourier test and autocorrelation test as 
well as Kolmogorov test, showing that it is not random.

\begin{figure}
\includegraphics[angle=0,scale=1.00]{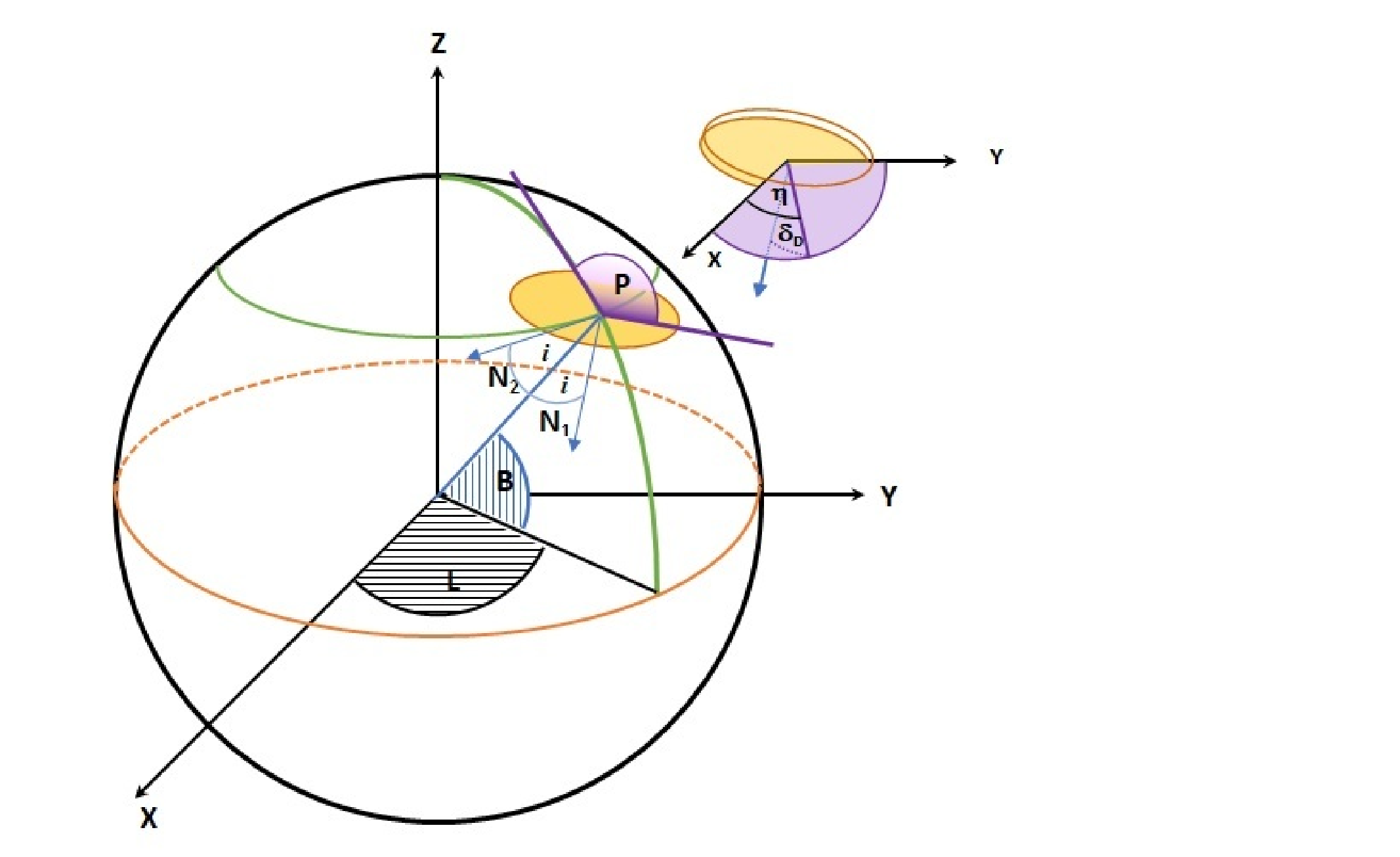}
\caption{
A schematic illustration of angles 
$\delta_D$ (the polar angle between the normal to the galaxy plane and the 
main plane of the coordinate system) and  $\eta$ (the azimuth angle between 
the projection of this normal onto the main plane and the direction towards the 
zero initial meridian). $i$ is the inclination angle with respect to the
observer's line of sight, $P$ is the position angle in the reference system, $N_1$ and $N_2$
are posible positions of the  normal to galaxy plane 
while $L$ and $B$ are the longitude and latitude of the 
reference coordinate system (for more details see \citet{f4,aks08}).
\label{fig:f0}}
\end{figure}

In the present paper, following the ideas of \citet{g2011b} and \citet{Panko13}, we improved 
method allows us to analyze the distributions both of the position angles $p$ and distribution of 
two angles giving spatial orientation of galaxies. 
We denote $\delta_D$ angle (the polar angle between the normal to the galaxy 
plane and the main plane of the coordinate system) and the $\eta$ angle (the azimuth 
angle between the projection of this normal onto the main plane and the direction
towards the zero initial meridian), see Figure \ref{fig:f0} for geometry of the angles. 
The main idea of our method is to analyze the distributions of these angles using statistical tests. 
We have analyzed in more details and improved the statistical tests used 
in \citep{g2011b} as well as introduce new statistical tests into the method.
We analyzed how the tests changes if expected values of galaxies in particlular bins varies 
(as in the case of analysis of the $\delta_D$ angle). It slightly changes for autocorelation 
test but it is very important for Fourier test and Kolmogorov-Smirnov test.
We also introduced to our improved method of investigation of galaxy alignment in clusters,
the "control tests" that neglects a possible asymmetry of the distribution according to 
main coordinate plane. 
The idea of such tests is to analyze only the difference between more "parallel" or 
more "perpendicular" orientation according  to the coordinate system main plane 
(or main direction towards the zero initial meridian in the case $\eta$ angle). 
We have checked how the Kolmogorov-Smirnov test behaves in the investigation of 
the orientation of galaxies in cluters. We have introduced alternative tests, namely Cr\'amer-von Mises 
and Watson that showed more explicitly that the allignment truly exists.

Usually the effect of aligment of galaxies in structures is not very 
strong and its analysis requires precise statistical considerations. In such a case 
it is very important to verify that no other observational systematics can affect.
To avoid a problem with the possible impact on the obtained results by data systematics,
we think it is necessary to test the method on a well-tested sample of galaxy clusters.
We have decided to use a sample of the galaxy clusters selected on the basis of the
PF catalog \citep{Panko06}. Hence, on the example of analysis of position angles in 247 rich 
Abell clusters we show how the method works in case of observational data.
For our sample of 247 clusters, we computed the mean values of the analyzed statistics. 
Our null hypothesis $H_0$ is that the mean value of the analyzed statistics is 
as expected in the case of random distribution of analyzed angles. At first, we have 
compared the theoretical prediction with the results obtained from numerical simulations.
Later, they are compared with the results obtained from the real sample of the 247 Abell  
clusters. Separately, we analyzed the sample when only galaxies brighter than $m_3+3^m$ 
were considered. Moreover, we decided to analyze if there are any differences in alignment 
of galaxies in the clusters belonging to different Bautz-Morgan (BM) types.
In order to exclude the case that the obtained results comes from errors in 
observational measurements, we have used two separate methods. We have analyzed the sample 
assuming random errors in position angles and additionaly we have used jackknife method 
especially to investigate the possible influence of background objects.

The novelty of our approach is to gather many methods of analysis of statistics of all 
angles $p$, $\delta_D$ and $\eta$ not only for some particular galaxy clusters but also 
for big samples of clusters. Unfortunately, such approach inevitably turns to the analysis 
on a case by case basis. That is why in each case, we point out possible difficulties 
and show which method has to be used. At first glance, most cases looks very similar, but 
one has to be careful not to omit the crucial differences. The advantage of the approach 
is that by analysing much more data at once, we are able to draw more general conclusions.

\section{Observational data}

In the present paper we have analyzed the sample of 247 rich Abell
clusters containing at least 100 members galaxies each
\citep{g10a,g2011b}. The sample was selected on the basis of the
PF catalogue \citep{Panko06}. The structures in the
\citet{Panko06} catalogue were extracted from the Muenster Red Sky
Survey (MRSS hereafter) \citep{MRSS03} using the 2D Voronoi
tessellation technique \citep{Ram01}, see \citet{Panko09} for
details. Note that the confidence level for cluster search was $95\%$
\citep{Ram01} and the list of clusters is reliable.

During analysis of the orientation of galaxies in structures
the curcial point is to remove the non-galactic objects - mostly
stars and artifacts.
The advantage of MRSS list of galaxies is that the author of the
catalogue very carefully analyzed the classification of all
objects in the survey. Basic data for MRSS are 217 ESO Southern
Sky Atlas R Schmidt plates covering an area of about 5000 square
degrees in the southern hemisphere, with $b < -45^{\circ}$. All
plates were digitized with the two PDS 2020GM$^{Plus}$
microdensitometers of the Astronomisches Institut M\"{u}nster 
with a step width of 15 microns (1600 dpi), corresponding to $1.01$ 
arcseconds per pixel. Objects search in digitized plates was made 
using the program SEARCH based on FOCAS algorithm \citep{Jarvis81}, 
see also \citet{Ungruhe97,MRSS03} for detailed aplication to MRSS 
catalogue. The analysis of
variations of background densities, in particular vignetting of
the telescope, and the influence of threshold of the background
for the objects detection were made carefully. The influence of
nearest objects was minimized due to final analysis inside the
small frame including the object. Final galaxy search based on 6
parameters allows to select only galaxies. More than 2700000
uncertain objects were checked visually; they were faint objects
mainly \citep{MRSS03}. So, all selected objects are galaxies.

Resulting list of MRSS galaxies contains more than 5 millions ones
till to $r_{F}=21^{m}$ detected on the best plates. However, the
limit of completeness of MRSS is $r_{F}=18.3^{m}$ \citep{MRSS03}.
This short list contains 1200000 galaxies with reliable
definitions parameters. The ellipticity and position angle for
each galaxy were calculated using the covariance ellipse method
(\citep{Carter 80}). The ellipticities and positional
angles of galaxy images were calculated using both intensities and
coordinates, so inside intensity distribution was accounted
\citep{MRSS03}. The problem of possible systematic effects was 
analyzed by \citet{Ungruhe97,MRSS03} while   the detailed study of 
uncertainties on the position
angles, including vary with galactocentric distance, was executed
in \citet{Biernacka 11}. They confirmed the results of \citet{Nilson74}. 
Following the results, we supposed the uncertainties on
the position angles of galaxies on the level $2^{\circ}$ for galaxies
elongated  images of galaxies, in the worst case the uncertainties
were on the level $5^{\circ}$. Obviously the uncertainties quickly
increase for rounded images.

The PF Catalogue was created using only MRSS galaxies inside the
completeness limit $r_{F}=18.3^{m}$. The PF Catalogue defines a
cluster as a structure which contains at least ten galaxies in the
magnitude range between $m_3$ and $m_3+3^{m}$, where $m_3$ is the
magnitude of the third brightest galaxy located in the considered
structure region. The criterion of $m_3+3^{m}$ is a well known
criterion to galaxy membership for the cluster if, as in the case
\citet{MRSS03} we have no information about radial velocities of
particular galaxies. \citet{Panko09} checked the correctness of 
this limit using statistical completed sample contained 547 PF 
structures. 

The full PF \citep{Panko06} catalogue includes
6188 galaxy clusters and groups and  contains positions of the
clusters, their radii, areas, the number of all galaxies in the 
field of structure, number of galaxies within the
magnitude range $m_3$ and $m_3+3^{m}$, as well as an estimated
number of background galaxies, ellipticity and position angles for
each structure, magnitudes of the first, the third and the tenth
galaxy in a structure (taken from the MRSS). The full PF catalogue 
contains not only the list of the clusters
but also the lists of galaxies belonging to each structures, where
the data for each galaxy member were taken from the MRSS. This
data includes: the equatorial coordinates of galaxies ($\alpha$,
$\delta$), the diameters of major and minor axes of the galaxy
image ($a$ and $b$ respectively) and the position angle of the
major axis $p$ (see also \citet{g2011b}). Because the position
angles in MRSS serve in clockwise system, we recomputed original
position angles from MRSS clockwise system to standard
counterclockwise system. We performed our computation in
Equatorial and Supergalactic Coordinate System defined in
\citet{f4}. In the case of Supergalactic Coordinate System
position angles $p$ were recomputed to supergalactic position
angles $P$. The photometrical redshifts were calculated for each
cluster using the relation $z(m_{10})$ (\citet{Bier09}) and rich PF
clusters have redshifts $z<0.12$ while median z=0.08.  The positions 
of PF and APM \citep{Dalton97} galaxy clusters are in good agreement
\citep{Bier09}.

In the present paper, as in \citet{g10a} and \citet{g2011b} we
have analyzed the sample of rich clusters that have at least 100
members and belong to ACO clusters \citep{ACO}. The advantage of
such sample is that they have the Bautz--Morgan morphological
types (BM types). There are 239 such objects in the PF catalogue.
Moreover, 9 objects can be identified with two ACO clusters. We
decide to include them in our investigation and increased our
sample to 248 objects. We excluded the structure A3822, which
potentially has substructures \citep{Biviano97,Biviano02}.
Therefore, finally our analyzed sample contains 247 clusters.
Because all analyzed clusters have ACO identification, the
distances can be found from literatures or extrapolated from 10th
brightest galaxy in clusters \citep{Panko06, Bier09}. The numbers
of clusers with particular BM types are from $35$ (BM I) till $59$
(BM II).

In our investigation we have decided to analyze two subsamples of
 data. The first one contains all galaxies lying in the region
regarded as cluster. In the second one, for avoiding possible role
of background object, only galaxies brighter than $m_3+3^{m}$ were
taken into account.

\section{The method of investigation}

The analysis of the
orientation of galaxies has usually been examined by two main methods. In the first 
one \citep{h4} the distribution of the position angles of the major axis of galaxies 
is performed. During the analysis of position angles, we exclude from examination 
all galaxies with axial ratio $q=b/a>0.75$, because  for the face--on galaxies position 
angles give only marginal information connected with the orientation of galaxy.
In the second method based on the de--projection of the galaxy images, we have 
analyzed spatial orientation of galaxies. This idea was introduced by
\citet{Opik70}, applied by \citet{Jaaniste78} and significantly
modified by \citet{f4,g2,g3}. In this method, we take into account both galactic
position angles $p$ and another important parameter -- the galaxy inclination
with respect to the observers' line of sight $i$. Using these two angles we
have determined two possible orientations of the galaxy plane in space, which gave
two possible directions perpendicular to the galaxy plane. As was discussed in the
introduction, it is expected that one of these normal corresponds to the direction
of galactic rotation axis. One should note however, that de--projection of galaxy 
images on the celestial sphere gives four solutions for the angular momentum vector.
Usually we consider only two distinguishable solutions since we do not know the direction
of galaxy rotation.

The inclination angle has been computed from the galaxy image according to the
formula: $cos^2 i=(q^2 -q^2_0 )/(1-q^2_0)$, where the observed axial ratio $q=b/a$
and $q_0$ is "true" axial ratio. This formula is valid for oblate spheroids
\citep{Holmberg46}. The value $q_0=0.2$ is used in the case when we have
no information about morphological types of galaxies (as in MRSS catalog).
For each galaxy we determined two angles $\delta_D$ and $\eta$.  
Following \citet{g2011b} we performed our computation both in Equatorial and 
Supergalactic  coordinate systems (\citet{f4} based on \citet{TS 76}). The relations 
between angles ($\delta_D$, $\eta$) and  ($i$, $P$) in the Supergalactic coordinate 
system ($L$, $B$) (Figure \ref{fig:f0}) are the following ones (similar formulae may 
be obtained for Equatorial coordinate system) 
\begin{equation}
\label{eq:a1}
\sin\delta_D  =  -\cos{i}\sin{B} \pm \sin{i}\cos{r}\cos{B},
\end{equation}
\begin{equation}
\label{eq:a2}
\sin\eta  =  (\cos\delta_D)^{-1}[-\cos{i}\cos{B}\sin{L} + \sin{i}
(\mp \cos{r}\sin{B}\sin{L} \pm \sin{r}\cos{L})],
\end{equation}
\begin{equation}
\label{eq:a3}
\cos\eta  =  (\cos\delta_D)^{-1}[-\cos{i}\cos{B}\cos{L} + \sin{i}
(\mp \cos{r}\sin{B}\cos{L} \mp \sin{r}\sin{L})],
\end{equation}
where $r=P-\pi/2$. As a result of the reduction of our analysis into two
solutions only, it is necessary to consider the sign of the expression:
$S=-\cos{i}\cos{B} \mp \sin{i}\cos{r}\sin{B}$
and for $S\ge 0$ we should reverse the sign of $\delta_D$ respectively (see \citet{g10a}).
Please note the usualy the researchers use the simplified version of the method,
taking into account only Equations \ref{eq:a1} and  \ref{eq:a2}, which however caused 
serious problems in the interpertation of the results of the analysis of the spatial 
orientation of galaxies.

The essential progress of the investigation of galaxy alignment was made by \citet{h4}.
Their method of investigation of galaxies orientation is based on statistical
analysis of the distribution of galaxies position angles. The essence of the method
was to use three type of statistical tests: $\chi^2$-test,  Fourier and First
Autocorrelation. It was shown later that this methodology can also be used to study
the spatial orientation of galaxies planes \citep{f4,Kindl87,Ar00,g10a,Panko13,Aryal16}.

The main idea of the paper is to show how to make the statistical methods more 
reliable and to interpret the obtained results. We show how the methods work in 
particular cases and apply it to the analysis of the distribution of position 
angles of observational sample of 247 rich Abell clusters. In particular we 
determine if the orientations of galaxies in clusters are isotropic or not. 

The essence of the method \citet{g2011b} is to compute the mean values of
analyzed statistics for the whole sample of analyzed cluster and compare 
them with that obtained from numerical simulations. Our null hypothesis
$H_0$ is that the mean value of the analyzed statistics is as expected 
in the case of random distribution of analyzed angles.
In all tests, the range of the $\theta$ angle
(where for $\theta$ one can put $\delta_D+\pi/2$, $\eta$, $p$ (or $P$)
respectively) is divided into $n$ bins of equal width. We have used 
$n=36$ bins. As a check, we repeated 
the division for other values of $n$, but generally we have not found 
any significant difference. There is one exception, namely is 
Kolmogorov-Smirnov test and we discuss this in detail in the section 
"Numerical Simulations and Results".

In the whole paper we denote as $N$ the total number of galaxies in analyzed clusters
while $N_k$  is the number of galaxies with orientations within the $k$-th  angular
bin  and  $N_{0,k}$ is the expected number of galaxies in the k-th bin.
In the case of the analysis of the position angles $p$ or $P$ and $\eta$ 
angles all $N_{0,k}$ are equal to $N_0$, which is also the mean number of galaxies 
per bin. In the case of the analysis of the angles $\delta_D$ of course $N_{0,k}$ 
are not equal and are obtained from the cosine distribution. The case when not all 
$N_{0,k}$ are equal was not analyzed in the paper of \citet{g2011b}, so adding such 
analysis significantly improves the method of analyzing the alignment of galaxies in 
clusters. This  improvement means that the method is now valid also in the case when 
some (or even all) $n$ bins have not equal width.

The first group of tests is based on the $\chi^2$ test:
\begin{equation}
\label{eq:c1}
\chi^2 = \sum_{k = 1}^n {(N_k -N\,p_k)^2 \over N\,p_k}= \sum_{k = 1}^n {(N_k -N_{0,k})^2 \over N_{0,k}}.
\end{equation}
where $p_k$ is the probability that a chosen galaxy falls into $k$-th bin.
Because we have $n$ bins, the number 
of degrees of freedom of the $\chi^2$ test  is $(n-1)$. This causes  that the 
mean value $E(\chi^2)=n-1$ and the variance $\sigma^2(\chi^2)=2(n-1)$.
For $n=36$ this leads to the values $E(\chi^2)=35$ and $\sigma^2(\chi^2)=70$.
When we analyzed the sample of $m$ clusters and computed the mean value of
statistic, then $E(\bar{\chi^2})=n-1$, but $\sigma^2(\bar{\chi^2})$
decreased by the factor $m$ and equaled ${\sigma^2(\chi^2) \over m}$.
For $n=36$ and $m=247$ this gives $\sigma^2(\bar{\chi^2})=0.2834$ and
$\sigma(\bar{\chi^2})=0.5324$.

In the basic investigation the range of $\delta_D$ angle is from $-\pi/2$ to $\pi/2$.
The idea of the control test is to restrict our analysis only to the case of the
absolute value of $\delta_D$ angle \citep{f4,Aryal04,Aryal05a,Aryal06,Aryal07}.
Then, we neglected a possible asymmetry of the distribution according to main 
coordinate plane and analyzed only the differences between more "parallel" or more 
"perpendicular" orientation according to the coordinate system main plane. 
So the range of $\delta_D$ angle
is from $0$ to $\pi/2$ and we divided the entire range of a $\theta$ angle into 
$18$ instead of $36$ bins. During the analysis of $247$ clusters the mean value 
of statistic is of course $E(\bar{\chi^2_c})=17$ while $\sigma^2(\bar{\chi^2_c})=0.1377$ 
and $\sigma(\bar{\chi^2_c})=0.3710$. Analogically, during analysis of the position 
and azimuthal $p$ and $\eta$ angles, we also reduced ranges of analyzed angles 
from $0$ to $\pi/2$, so we divided the entire range of a $\theta$ angle into $18$ 
instead of $36$ bins. Of course in the case of $\eta$ angle, the idea of the control 
test is to analyze the asymmetry between more "parallel" or more "perpendicular" 
projection to the normal to the galaxy plane  according to the main direction towards 
the zero initial meridian of the coordinate system.

The second group of our tests is based on the first auto-correlation test \citep{h4}.
 Probably the best test for autocorrelation is the 
von-Neumann-Durbin-Watson test. However, in our paper we do not analyse full 
autocorrelation. Our idea is, as noted above, to obtain the average value for analysed 
statistic (and later to check if the mean value of the analysed statistics is as 
expected in the case of random distribution of analysed angles).  Although, the \citet{h4} 
first autocorrelation test may not work as well as the von-Neumann-Durbin-Watson test, 
the idea presented there is also widely used (see for example \citet {Percival93}- 
especially  Chapter 6 of the book). So, it seems that the use of this test 
works well enough.  However, we analyzed its properties in more detail than \citet{h4} 
before we used it. 

The first auto-correlation test quantifies the correlations between galaxy numbers in 
neighboring angle bins. This correlation is measured by the statistic $C$: 
\begin{equation}
\label{eq:c2}
C\, = \, \sum_{k = 1}^n { (N_k -N_{0,k})(N_{k+1} -N_{0,k+1} )
\ \over \left[ N_{0,k} N_{0,k+1}\right]^{1/2} }
\end{equation}
where $N_{n+1}=N_1$. According to the original paper \citet{h4}, in the case 
of an isotropic distribution, the expected value of $C$ is $E(C) = 0$ with 
the standard deviation 
$\sigma(C) = n^{1/2}$.

In the paper \citet{g2011b}, it was shown that original \citet{h4} result was
an approximation only that is not correct in our case, since they assumed
that $N_k$ are independent from each other. Therefore, in the formula for $E(C)$ 
is present an additional term connected with the covariance between $N_k$ and $N_{k+1}$: 
\begin{equation}
\label{eq:c5a}
E(C)\, =-\sum_{k = 1}^n{{\,N\,p_k\,p_{k+1}}\over \sqrt{N_{0,k} N_{0,k+1}}}. 
\end{equation}
When all $p_k$ and hence $N_{k,0}$ are equal ($p_k=1/n$), as it was in the case of 
position angles, then $E(C)\, =-1$. Moreover, $\sigma^2(C)$ contains a term which is 
the variance of the products of $N_k$ and $N_{k+1}$ which are not independent. 
As a result, the correct value of $D^2(C)$  is only approximately equal to $n$ and the 
correct value must be computed using numerical simulations. Moreover, for one cluster 
the difference between the results in expected values of $C$ (0 or -1) is relatively 
small compared to $\sigma(C) \approx \sqrt{n}$. However, when we analyse the sample 
of $m$ clusters, the situation is different because the variance is decreased by the 
factor $m$. As a result, in the case of sample $247$ clusters, 
$\sigma(\bar{C}) \approx \sqrt{n/247} = 0.3818$ is significantly smaller than $1$ 
(which is the difference between expected values) \citep{g2011b}.

In the case of analysis of the $\delta_D$ angles the situation is more
complicated because of course $N_{0,k}$ and as results all $p_k$ are
not equal and they are obtained from cosine distribution. As a result in our case
$n=36$, $E(C)\, =-\sum_{k = 1}^n{\,N\,p_k\,p_{k+1}}=-0.9973$.
Similarly as in the case of $\chi^2$ test we introduced the control  first
auto-correlation test. Also in this case, we restricted our analysis only to 
the case of the absolute value of $\delta_D$ angle,  so the range of 
$\delta_D$ angle is from $0$ to $\pi/2$.  We divided entire range of a $\theta$ 
angle into $18$ instead of $36$ bins. 
$E(C_c)\, =-\sum_{k = 1}^n{\,N\,p_k\,p_{k+1}}=-.9701$. As in the case of the 
basic test we could approximate standard deviation of $\bar{C_c}$, even if
correct value must be obtained from numerical simulations. For the case of 
the position angles and $247$ clusters $\sigma(\bar{C_c}) \approx \sqrt{n/247} = 0.26995$ 
and is again significantly smaller than the difference in expected values  
(which is again equal to $1$). Analogically as in the case of control $\chi^2$ 
test, during analysis the $p$ and $\eta$ angles, we also reduced ranges of 
analyzed angles from $0$ to $\pi/2$, so we divided entire range of a $\theta$ 
angle into $18$ instead of $36$ bins. 

The most popular test used for analysis of galaxy alignment is the Fourier
Test \citep{h4} and its modifications, 
even if doubts are sometimes raised (we will discuss them separately below), 
as to the adequacy and the scope of applicability of this type of tests.
The idea of this test is, that if the deviation from isotropy is a slow 
varying function of the angle $\theta$ then the expected number of galaxies 
with orientations within the $k$-th  angular bin $N_k$ 
is in the most general  form
 given by formulae \citet{g3}:
\begin{equation}
\label{eq:f9}
N_k = N_{0,k} (1+\Delta_{11} \cos{2 \theta_k} +\Delta_{21} \sin{2
\theta_k}+\Delta_{12} \cos{4 \theta_k}+\Delta_{22} \sin{4\theta_k}+.....).
\end{equation}
In Fourier test the crucial is the amplitude $\Delta= (\sum_i \sum_j  \Delta_{ij})^{1/2}$
and probability  that  the amplitude $\Delta$ is greater than a fixed value.
Using maximum-likelihood method  we obtain the  expressions for the
$\Delta_{ij}$ coefficients. Usually only first or maximum two first modes are used in the investigation.

For that, Equation \ref{eq:f9} could be rewritten in the form:
\begin{equation}
\label{eq:f18}
{N_k - N_{0,k}\over N_{0,k}}  =
\Delta_{11} \cos{2 \theta_k} +\Delta_{21} \sin{2\theta_k}+\Delta_{12} \cos{4 \theta_k}+\Delta_{22} \sin{4\theta_k}.
\end{equation}
If we define $I$ vector as:
\begin{equation}
\label{eq:f19}
I=
\left(
\begin{array}{c}
\Delta_{11} \\
\Delta_{21} \\
\Delta_{12} \\
\Delta_{22}
\end{array}
\right)
\end{equation}
then the solution for ${\bf x} \equiv I$ is given by \citet{Brandt97} equation (9.2.26):
\begin{equation}
\label{eq:f20}
{\bf x}=-\left({\bf A}^TG_y{\bf A}\right)^{-1}{\bf A}^TG_y{\bf c}
\end{equation}
where:
${\bf c}$ is the vector of particular $y_i={N_k - N_{0,k}\over N_{0,k}}$:
\begin{equation}
\label{eq:f21}
{\bf c}=
\left(
\begin{array}{c}
{N_1 - N_{0,1}\over N_{0,1}}\\
{N_2 - N_{0,2}\over N_{0,2}}\\
.\\
.\\
{N_k - N_{0,k}\over N_{0,k}},
\end{array}
\right)
\end{equation}
$G_y$ is the inverse matrix  to the covariance matrix of
particular $y_i$ i.e weight matrix:
\begin{equation}
\label{eq:f22}
G_y=
\left(
\begin{array}{cccc}
g_1&.&.&.\\
.&g_2&.&.\\
.&.&.&.\\
.&.&.&g_n
\end{array}
\right)=
\left(
\begin{array}{cccc}
N_{0,1}&.&.&.\\
.&N_{0,2}&.&.\\
.&.&.&.\\
.&.&.&N_{0,n}
\end{array}
\right)
\end{equation}
and matrix  ${\bf A}$ of coefficients with particular $\Delta_{ij}$ has form:
\begin{equation}
\label{eq:f23}
{\bf A}=
-\left(
\begin{array}{cccc}
\cos{2 \theta_1}&\sin{2 \theta_1}&\cos{4 \theta_1}&\sin{4 \theta_1}\\
\cos{2 \theta_2}&\sin{2 \theta_2}&\cos{4 \theta_2}&\sin{4 \theta_2}\\
.&.&.&.\\
\cos{2 \theta_n}&\sin{2 \theta_n}&\cos{4 \theta_n}&\sin{4 \theta_n}
\end{array}
\right)
\end{equation}
while matrix $G({\bf x})={\bf A}^TG_y{\bf A}$ (see \citep{Brandt97} equation (9.2.27)) 
where covariance matrix $C_{cov}({\bf x})=G({\bf x})^{-1}$. 
For detailed form of solutions for $\Delta_{ij}$ coefficients and their covariance matrix as well 
as the formulae for probability  that  the amplitude $\Delta$ is greater than a fixed value
 in the particular cases of the analysis see  Appendix.

During analysis of alignment of galaxies, it is important not only the power
of the deviation from isotropy, but also its direction. The sign of the
coefficient $\Delta_{11}$ gives us the information about direction of such
departure from isotropy. If $\Delta_{11}<0$, then the excess of the galaxies
with $\theta$ angle near $90^o$ is observed, while $\Delta_{11}>0$ means the
excess for $\theta$ angle near $0^o$ (for detailed discussion see Appendix)

In the paper \citet{g2011b} it was discussed the properties of statistics
$\Delta_{ij}/\sigma(\Delta_{ij})$, $\Delta_{1}/\sigma(\Delta_{1})$ and the
$\Delta/\sigma(\Delta)$, in the case of the distributions of the position
angles. It was showed, that Equation \ref{eq:f8b} (\citet{h4} Equation 26)
is obtained as a result of the theorem of propagation of errors what is not good 
approximation because the theorem of propagation errors is obtained in the linear 
model whereas $\Delta_j = \left( \Delta_{1j}^2 + \Delta_{2j}^2 \right)^{1/2}$
is strictly nonlinear (see Equations \ref{eq:f6} and \ref{eq:f6b}). 
Hence, the notation $\Delta^2_j \over \sigma^2(\Delta_j)$ means
only that elements of $\Delta^2$ should be divided by elements of covariance
matrix $\Delta_{ij}$. Consequently, the notation $\Delta_j/\sigma(\Delta_j)$ does
not mean that coefficient $\Delta_j$ is divided by its  error. Such an
interpretation is only a rough approximation based on linear model \citep{g2011b}.
As a results, the correct values are the folowing:  $E(\Delta^2_{1}/\sigma^2(\Delta_{1}))=2$,
$E(\Delta^2/\sigma^2(\Delta))=4$, $D^2(\Delta^2_{1}/\sigma^2(\Delta_{1}))=4$ and
$D^2(\Delta^2/\sigma^2(\Delta))=8$. Moreover, $D^2(\Delta_{1}/\sigma(\Delta_{1}))=1/2$, and 
$D^2(\Delta/\sigma(\Delta))=1/2$. The last results are obtained again from theorem of 
propagation errors, however in the paper \citet{g2011b} it was showed that in
presently analyzed case it works quite well, even if the correct
values must be obtained from  numerical simulations.

Because $E(X)=\sqrt{E(X^2)-D^2(X)}$, we get
\begin{equation}
\label{eq:ff5}
E\left({\Delta_1 \over \sigma(\Delta_1)}\right)=
\sqrt{E\left(\Delta^2_{1} \over \sigma^2(\Delta_{1})\right)-D^2\left(\Delta_1 \over \sigma(\Delta_1)\right)}=
\sqrt{2-0.5}=1.2247
\end{equation}
and
\begin{equation}
\label{eq:ff6}
E\left({\Delta \over \sigma(\Delta)}\right)=
\sqrt{E\left(\Delta^2 \over \sigma^2(\Delta )\right)-D^2\left(\Delta \over \sigma(\Delta)\right)}=
\sqrt{4-0.5}=1.8708
\end{equation}
One should note that our sample contain 247 clusters. So
$\sigma^2(\overline{\Delta_{1}/\sigma(\Delta_{1})})$ and
$\sigma^2(\overline{\Delta/\sigma(\Delta)})$ are equal
${1/2 \over 247}=0.002024$ while standard deviations of
$\sigma(\overline{\Delta_{1}/\sigma(\Delta_{1})})$ and
$\sigma(\overline{\Delta/\sigma(\Delta)})$ are equal
$\sqrt{1/2 \over 247}=0.04499$. 

The case of the analysis of the distribution of the $\eta$ angles
is similar to the case of position angles. One should note however that
 the analysis of the distribution of the $\delta_D$ angles is more complicated.
At first, it is because not all $N_{0,k}$ are equal. It is the reason  that now
$\sigma^2(\Delta_{11})$ and $\sigma^2(\Delta_{21})$ are not exactly but only 
approximately equal $N/2$ (see Equations \ref{eq:f11a} and \ref{eq:f12a}) and what 
is crucial, when we analyze both $2\theta$ and $4 \theta$ Fourier modes 
together,  not all $\Delta_{ij}$ coefficients are independent of each other. 
 Also the case of the control test for Fourier test is more
complicated than in the case of $\chi^2$ and autocorrelation test.

The simplest situation is the statistics of $\Delta_{ij}/\sigma(\Delta_{ij})$.
There are the cases of one dimensional ($1D$) Gaussian distribution. In these
cases the situation is not changing with comparison of \citet{g2011b} and
it is very clear. Variables $\Delta_{11}/\sigma(\Delta_{11})$ and
$\Delta_{21}/\sigma(\Delta_{21})$ are still normalized gausian variables with
expected value equal $0$ and
variance equal $1$. Of course
$\sigma^2(\overline{\Delta_{ij}/\sigma(\Delta_{ij})})=1/247=0.00405$ and
$\sigma (\overline{\Delta_{ij}/\sigma(\Delta_{ij})})=0.06363$. 

Unfortunately, when we consider $\Delta_1/\sigma(\Delta_1)$ and 
$\Delta/\sigma(\Delta)$ variablesthe situation is much more complex. 
In the general case, the notation $\Delta^2_j \over \sigma^2(\Delta_j)$
or $\Delta^2 \over \sigma^2(\Delta)$ should be substituted by the use of
auxiliary value $J=\sum_i \sum_j  I_{i}^T {G_{ij} I_{j}}$ (see Equations
 \ref{eq:f36} and \ref{eq:f44}). The advantage of the extended notation is that 
it is valid  also in the situation when the covariance matrix is not the diagonal 
matrix (i.e. not all $\Delta_{ij}$ are independent of each other).
In such a case, even if we take into account only first Fourier mode i.e coefficients
$\Delta_{11}$ and $\Delta_{21}$ which are not independent to each other,
then (see Equation \ref{eq:f37}) the auxiliary value J has the form:
\begin{equation}
\label{eq:ff7}
J=A\Delta^2_{11}+2Y\Delta_{11}\Delta_{21}+B\Delta^2_{21}= {\Delta^2_{11} \over 1/A} +{\Delta_{11}\Delta_{21} \over 1/2Y} +{\Delta^2_{21} \over 1/B}
\end{equation}
In the present case $\Delta_{11}$ and $\Delta_{21}$ are not independent but it
could be very easy transformed to the form where they are independent \citep{joh92}.
In the present case the transformation:
$\Delta'_{11}= {\Delta_{11} \over \sigma(\Delta_{11})\sqrt{1-\rho^2}}- {\rho\Delta_{21} \over \sigma(\Delta_{21})\sqrt{1-\rho^2}}$
and  $\Delta'_{21}={\Delta_{21} \over \sigma(\Delta_{21})}$
(where $\sigma(\Delta_{11})$, $\sigma(\Delta_{21})$ and $cov(\Delta_{11}, \Delta_{21})$
are given by Equation \ref{eq:f38} and as a result the correlation ratio
$\rho=cov(\Delta_{11}, \Delta_{21})/\sigma(\Delta_{11})\sigma(\Delta_{21})={-Y \over \sqrt{AB}}$
gives the variables $\Delta'_{11}$ and $\Delta'_{21}$ independent to each other.
It leads to situation when $(\Delta'_{11}, \Delta'_{21})$ has the standard bivariate
normal distribution and consequently
\begin{equation}
\label{eq:ff8}
J'= {\Delta'^2_{11} \over 1} + {\Delta'^2_{12} \over 1}.
\end{equation}
where of course $J' \equiv J$.
Because $J'$ is the sum of $\Delta'^2_{ij}/\sigma'^2(\Delta_{ij})$ where
$\Delta'^2_{ij}$ are independent to each other,  then it is $\chi^2$ distributed.
As a result the value $J'$ (i.e. $\Delta'^2_{1}/\sigma^2(\Delta'_{1})$ in our "old"
notation, where $\Delta'_1 = \left(\Delta'^2_{11} + \Delta'^2_{21} \right)^{1/2}$)
given by formulae:
\begin{equation}
\label{eq:ff9}
J'=\Delta'^2_{11}/\sigma^2(\Delta'_{11})+\Delta'^2_{21}/\sigma^2(\Delta'_{21})
\end{equation}
has  $\chi^2$ distribution with $2$ degrees of freedom. So again
$E(\Delta'^2_{1}/\sigma^2(\Delta'_{1}))=2$,
$D^2(\Delta'^2_{1}/\sigma^2(\Delta'_{1}))=4$ and
$\sigma^2(\Delta'_{1}/\sigma(\Delta'_{1}))=1/2$. Consequently, the result
obtained in Equation  \ref{eq:ff5} is still valid and
$E\left({\Delta'_1 \over \sigma(\Delta'_1)}\right)=1.2247$, where
using our "new" notation, ${\Delta'_1 \over \sigma(\Delta'_1)}$
should be noted as $\sqrt{J'}$, which means $E(\sqrt{J'})=1.2247$.
Because $J'\equiv J$ then above results are valid for original
$J=\sum_i \sum_j I_{i}^T {G_{ij} I_{j}}$ and $\sqrt{J}$ so in analyzed
case $E(\sqrt{J})=1.2247$. Of course the  above approximation
(for the sample 247 clusters)
$\sigma^2(\overline{\Delta_{1}/\sigma(\Delta_{1})})$, written now as
$\sigma^2(\overline{\sqrt{J}}) \approx {1/2 \over 247}=0.002024$
and $\sigma(\overline{\Delta_{1}/\sigma(\Delta_{1})})$, now
$\sigma(\overline{\sqrt{J}}) \approx\sqrt{1/2 \over 247}=0.04499$
are still valid.

If we take into account the $2\theta$ and $4\theta$  Fourier modes together, the
situation are complicating further. When not all $N_{0,k}$ are equal as it is in
the case of the analysis of the distribution of the $\delta_D$ angles, then
even in the situation when theoretical distribution of $N_{0,k}$ are  symmetric
with respect to value $\delta_D=0$ (i.e $N_{0,k} = N_{0,n-k}$) not all
$\Delta_{ij}$ coefficients are independent (see Equation \ref{eq:f27}).
Now, $I$ is given by Equation \ref{eq:f35} and $J=\sum_i \sum_j I_{i}^T {G_{ij} I_{j}}$
has the form:
\begin{equation}
\label{eq:ff10}
J=A\Delta^2_{11}+B\Delta^2_{21}+C\Delta^2_{12}+D\Delta^2_{22}+2U\Delta_{11}\Delta_{12}+2W\Delta_{21}\Delta_{22}
\end{equation}
Amplitude $\Delta$ (see equation \ref{eq:f36}) is described by $4D$ Gaussian distribution.
Fortunately, also in this case we could transform the vector of variables $\Delta_{ij}$
(i.e $I$ vector described by formulae \ref{eq:f35}) to the form in which
variables $\Delta'_{ij}$ give a vector of independent random variables ech with
standard normal distribution. Let $I'$ denote vector constructed from the $\Delta'_{ij}$
\begin{equation}
\label{eq:ff11}
I' =
\left(
\begin{array}{c}
\Delta'_{11} \\
\Delta'_{21} \\
\Delta'_{12} \\
\Delta'_{22}
\end{array}
\right)
\end{equation}
the transformation between $I'$ and $I$ has a form \citep{joh92}:
\begin{equation}
\label{eq:ff12}
I'={\bf L^{-1}}(I-\mu)
\end{equation}
where lower triangular matrix ${\bf L}$ is obtained from Choleski decomposition of the
covariance matrix $C={\bf L}{\bf L^T}$ and $\mu$ is a vector expected value of
 $\Delta_{ij}$. In our case the covariance matrix $C$ is given by equation \ref{eq:f27}
while $\mu \equiv 0$ because all expected values of $\Delta_{ij}$ are equal $0$. 
It means that Equation \ref{eq:ff12} has in fact
simple form $I'={\bf L^{-1}}I$.

The above result means that again $J'$ is the sum of standard normalized independent
variables over theirs errors
\begin{equation}
\label{eq:ff13}
J'=\Delta'^2_{11}/\sigma^2(\Delta'_{11})+\Delta'^2_{21}/\sigma^2(\Delta'_{21})+\Delta'^2_{21}/\sigma^2(\Delta'_{21})+\Delta'^2_{22}/\sigma^2(\Delta'_{22}).
\end{equation}
(where now all $\sigma(\Delta'_{ij}=1$) and has $\chi^2$ distribution
with $4$ degrees of freedom. As a result $E(\Delta'^2/\sigma^2(\Delta'))=4$,
$D^2(\Delta'^2/\sigma^2(\Delta'))=8$ and $\sigma^2(\Delta'/\sigma(\Delta'))=1/2$.
Consequently, the result obtained in Equation \ref{eq:ff6} is still valid and
$E\left({\Delta'_1 \over \sigma(\Delta'_1)}\right)=1.8708$, where using
our "new" notation, ${\Delta'_1 \over \sigma(\Delta'_1)}$ should be noted
as $\sqrt{J'}$, hence $E(\sqrt{J'})=1.8708$. Because $J'\equiv J$ it
leads to the conclusion that above results are valid for  original
$J=\sum_i \sum_j  I_{i}^T {G_{ij} I_{j}}$ and $\sqrt{J}$ so in the present case
$E(\sqrt{J})=1.8708$. Analogically as in the case of approximation
$\sigma^2(\overline{\Delta_{1}/\sigma(\Delta_{1})})$ the approximation for
 $\sigma^2(\overline{\Delta/\sigma(\Delta)})\approx 0.002024$ and
$\sigma(\overline{\Delta/\sigma(\Delta)}) \approx 0.04499$ are also still
valid. Of course all above  results will be valid also when the theoretical
distribution is not symmetric according to the value of $\delta_D=0$ presented 
by formulae (\ref{eq:f33} - \ref{eq:f35}).

Similarly, as in the case of $\chi^2$ and auto-correlation tests we introduce
a control Fourier test. The Fourier test requires the range of the $\theta$ angle
$(0;\pi)$. Because the original idea of control test is to restrict our analysis
only to the case of the absolute value of $\delta_D$ angle then it is natural
to do it in the following way. In the control test, the bins are equidistant
located oppositely to the zero value of $\delta_D$ angle.
So $N'_{k'}=(N_k+N_{37-k})/2$ (for $k'=k \le.36$). Analogically, we repeat this
procedure during analysis of the position $p$ and azimuthal $\eta$ angles.
One should note, that in the case of the Fourier test the number of bins in the
basic and the control test is the same  and it is equal to $n = 36$.

However, if we restrict our analysis only to the case of the absolute value of
$\delta_D$ angle, it is clear that we are able to neglect $\Delta_{21}$ and
$\Delta_{22}$ coefficients, because they are equal to zero (see also
\citep{f4,Aryal04,Aryal05a,Aryal06,Aryal07,g10a}). In that case $\Delta_1$ is
reduced to $|\Delta_{11}|$, while $\Delta$, now denoted as $\Delta_c$, is the
function of coefficients $\Delta_{11}$ and $\Delta_{12}$ only \citep{g10a}.
The above-mentioned observation is still correct in the case of the analysis of the
position $p$ and azimuthal $\eta$ angles.

Now, we compute the expected value of $|\Delta_{11}/\sigma(\Delta_{11})|$.
Because the auxiliary variable $z={\Delta_{11} \over \sigma(\Delta_{11})}$ has the
standard normal distribution then the expected value of
$|z|=|{\Delta_{11} \over \sigma(\Delta_{11})}|$ can be obtained from the following formulae:
\begin{equation}
\label{eq:ff14}
E(|z|)= \int_{-\infty}^{+\infty}|z|f(|z|)d|z|=
{2 \over \sqrt{2\pi}}\int_{0}^{+\infty}|z| \exp{(-{z^2 \over 2})}dz= \sqrt{2/\pi}\approx\sqrt{0.635742}\approx0.7973.
\end{equation}
Now, the variance of $|z|$ is given by formulae: $D^2(|z|)=E(|z|^2)-(E(|z|))^2$.
Because $z$ has the standard normal distribution (i.e. $E(z)=0$; $D^2(z)=1$) 
then $E(z^2)= D^2(z)+(E(z))^2=1$ what means that
\begin{equation}
\label{eq:ff16}
E(z^2)={1 \over \sqrt{2\pi}}\int_{-\infty}^{+\infty}z^2\exp{(-{z^2 \over 2})}dz=1.
\end{equation}
From the above formulae it is easy to see that the expected value of $|z|^2$ 
must by equal to the expected value of $z^2$ i.e. $E(|z|^2)=E(z^2)=1$.
So the variance of $|z|=|\Delta_{11}/\sigma(\Delta_{11})|$ is equal
\begin{equation}
\label{eq:ff17}
D^2(|\Delta_{11}/\sigma(\Delta_{11})|)=1-(E(|\Delta_{11}/\sigma(\Delta_{11})|)^2=1-0.635742=0.364258
\end{equation}
So, the error of $\sigma(|\Delta_{11}/\sigma(\Delta_{11})|)=0.6035$. 
Of course because our sample has $247$  clusters then
$\sigma^2(\overline{|\Delta_{11}/\sigma(\Delta_{11})|})=\sigma^2(|\Delta_{11}/\sigma(\Delta_{11})|)/247=0.001474$
and $\sigma (\overline{|\Delta_{11}/\sigma(\Delta_{11})|})=0.0384$

Fortunately, the case of $\Delta_c$ is much easier to solve. It is because
the coefficients $\Delta_{21}$ and $\Delta_{22}$ are equal to $0$ so the
Equation \ref{eq:ff10} is reduced to the form
\begin{equation}
\label{eq:ff18}
J=A\Delta^2_{11}+C\Delta^2_{12}+2U\Delta_{11}\Delta_{12}
\end{equation}
It is very easy to see that this equation  is analogical to Equation \ref{eq:ff7}
 with only differences that $\Delta_{21}$ is substituted by $\Delta_{12}$
and consequently (see  equations \ref{eq:f25} and \ref{eq:f32}) in Equation
\ref{eq:ff18} instead of coefficients $Y$ we have $U$ and instead of $B$ we have $C$.
It means that the reasoning carried for the cases when we take into account only
first Fourier mode with coefficients $\Delta_{11}$ and $\Delta_{21}$ which
are not independent to each other, (see equations from \ref{eq:ff7} to
\ref{eq:ff9}) are still valid. As a result we obtain that
\begin{equation}
\label{eq:ff19}
J'=\Delta'^2_{11}/\sigma^2(\Delta'_{11})+\Delta'^2_{12}/\sigma^2(\Delta'_{12})
\end{equation}
has  $\chi^2$ distribution with $2$ degrees of freedom and
$E(\Delta'^2_{c}/\sigma^2(\Delta'_{c}))=2$,
$D^2(\Delta'^2_{c}/\sigma^2(\Delta'_{c})=4$ while
$\sigma^2(\Delta'_{c}/\sigma(\Delta'_{c}))=1/2$. Consequently
$E\left({\Delta'_c \over \sigma(\Delta'_c)}\right)=1.2247$,
(using our "new" notation, $E(\sqrt{J'})=E(\sqrt{J})=1.2247$) and
(for the sample 247 clusters) again the approximation
$\sigma^2(\overline{\Delta_{c}/\sigma(\Delta_{c})})$, now
$\sigma^2(\overline{\sqrt{J}}) \approx 0.002024$ 
and $\sigma(\overline{\Delta_{c}/\sigma(\Delta_{c})})$, now
$\sigma(\overline{\sqrt{J}}) \approx0.04499$ are still valid.

As in  \citet{g2011b}, we investigate the isotropy of the resultant 
distributions of $\theta$ angles with the help of Kolmogorov- Smirnov test
(K-S test). We assume that the theoretical random distribution contains the
same number of objects as the observed one. In such studies the key is 
statistics $\lambda$:
\begin{equation}
\label{eq:k1}
\lambda=\sqrt{n_p}\,D_n
\end{equation}
which is given as the limit of the Kolmogorov distribution, where 
\begin{equation}
\label{eq:k2}
D_n= sup|F(x)-S(x)|
\end{equation}
$n_p$ is number of investigated points and F(x) and S(x) are theoretical and 
observational distributions of $\theta$.
Now, our interest is to compute the expected value and the standard deviation 
of the statistic for the real sample (247 rich Abell clusters). As in the previous 
case (especially $\chi^2$ test)  we introduce the K-S control test. Again, the range 
of $\delta_D$, $p$ and $\eta$ angles is from 0 to $\pi/2$ and we divided the entire 
range of a $\delta_D$,  $p$ and $\eta$ angles into 18 instead of 36 bins. Because 
of the reason discussed in the next section, in all cases of the basic and control
 K-S tests, the expected values of $\lambda$, as well as  their  standard deviations 
 are obtained from numerical simulations.

\section{Numerical Simulations and Results}

In the beginning, we would like to check whether the statistical 
methods used in our investigation lead to reliable statistical tests that are 
suitable to solve investigated problems. Since the statistics described by 
Equation \ref{eq:c1} has only limit chi-squared distribution \citep{cher54,SC67,Dom79,Krys98}, 
the first question is whether the approximation of the resulting 
distribution by chi-squared distribution is acceptable. Secondly, does the $C$ 
statistic described by Equation \ref{eq:c2} have normal distribution with 
standard deviation approximated by equation $\sigma(C) = n^{1/2}$? 
Thirdly, we want to check if the Fourier transform \citep{h4} works well,
what  in practice means to check if exponential formulae (\ref{eq:f7}) are valid 
in the investigated case. This problem  has been tested  by 1000 simulations 
of the sample of 2227 galaxies in \citet{g92} using build-in Fortran Lahey 
generator (the quality ot this generator was tested in \citet{g2011b}),
but the answers to above questions were never discussed in referred journals. 

In \citet{g92} it was  found that the answers for all above questions are yes, 
although this thesis is available only in Polish and the answers are only quantitatively.
In the present paper, in more detail, this is checked with the help of Kolmogorov- Smirnov test.
We have checked if we could reject the hypothesis $H_0$ that the distributions obtained 
from the simulations are the same as those approximated. They should be, in the case 
of statistics presented by equation \ref{eq:c1}, the $\chi^2$ distribution 
(with 35 degrees of freedom), in the case $C$ statistic described by equation 
\ref{eq:c2},  normal distribution with mean value $E(C)=-1$ and standard 
deviation $\sigma(c)=6$, while in the case of $\Delta_{11}$ (equation \ref{eq:f10a})
the normal distribution with mean value equal 0 and standard deviation given 
by formulae \ref{eq:f12a}. The results of these tests for analysis  $\delta_D$ and 
$\eta$ as well as  for position angles $P$ is presented in  Table \ref{tab:t0}.
We present in the Table \ref{tab:t0} value of statistics $\lambda$ (see equation 
\ref{eq:k1}). At the significance level $\alpha = 0.01$ the value 
$\lambda_{cr} = 1.628$. In any case, the obtained value of statistic $\lambda$ has not 
exceded  the critical value. It means, that the  result of Kolmogorov- Smirnov test has 
not excluded, on the assumed level of significance, the hypothesis, that analyzed 
distribution are as expected. Moreover, only in one case ($C$ statistic for 
$\delta_D$ angle) obtaining value of $\lambda$ is grater than critical value 
$\lambda_{cr}= 1.358$ at the  significance level $\alpha = 0.05$. Above results 
mean that our approximations work well  in the case of analysis of the distribution 
of the $\delta_D$ and $\eta$ angles while  in the case of analysis of position 
angles the approximations work perfectly well.

In our analysis we have $11$ tests. We have analyzed $\chi^2$, $\chi^2_{c}$,
$\Delta_1/\sigma(\Delta_1)$, $\Delta/\sigma(\Delta)$, $\Delta_c/\sigma(\Delta_c)$,
 $C$, $C_{c}$, $\lambda$, $\lambda_c$, $\Delta_{11}/\sigma(\Delta_{11})$ and
$|\Delta_{11}/\sigma(\Delta_{11})|$ statistics. For most of them we have
theoretical prediction given in the previous section. The exception  are variances
of $C$ and $C_c$ statistics, where we only  have approximation statistics 
 and statistics $\lambda$ and $\lambda_c$ describing Kolmogorov- Smirnov test.
 Moreover, the standard 
deviation of the $\Delta_{1}/\sigma(\Delta_{1})$, $\Delta/\sigma(\Delta)$ 
and $\Delta_c/\sigma(\Delta_c)$ statistics are obtained from theorem of propagation
errors. As a result,  (see equations \ref{eq:ff5} and \ref{eq:ff6})
theoretically we have good prediction for means of the $\Delta_{1}/\sigma(\Delta_{1})$, 
$\Delta/\sigma(\Delta)$ and $\Delta_c/\sigma(\Delta_c)$ statistics, hence in reality we
should obtain them also from numerical simulations. 
However, in all cases it is possible to perform the simulations and obtain
Cumulative Distribution Function (CDF) and Probability Density Function (PDF).

The basic problem in numerical simulations is the choice of a random number
generator. Unfortunately, many of the popular generators fail to give correct
results in multidimensional simulations \citep{Luescher94}. This problem,
with respect to analysis of alignment of galaxies, was analyzed in detail in \citet{g2011b}. 
In the paper it was shown that most suitable is RANLUX (level 4) generator 
\citep{Luescher94,James94,Luescher10} and this generator has been chosen as 
our base generator. The detailed discussion of different types of Random Generator 
showing the superiority of RANLUX was carried also for example by \citet{Shchur98}.

At first, using Monte-Carlo simulations, we simulated 247 fictitious clusters, each with 
2360 random oriented members of galaxies.
The details of our procedure are the following. For each galaxy we 
simulated the position angle (assuming uniform distribution) and inclination angle 
(cosine distribution). We have performed this procedure twice, first with galaxies 
in the clusters with coordinates distributed as in the real clusters and second 
independently for galaxies randomly distributed around the whole celestial sphere. 
Now, we obtain from Equations \ref{eq:a1}, \ref{eq:a2} and  \ref{eq:a3} the 
value of $\delta_D$ and $\eta$ angles. We performed 1000 simulations and on that 
basis we obtained PDF and CDF of analyzed angles.

Please note that instead of uniform and cosine distribution we could take 
any theoretically motivated  distribution as for example von Mises circular distribution,
which  is useful during testing the effects of theoretical model of galactic formations 
\citep{g92,g93,g94}. Instead of simulating distribution of position $p$ and inclination 
angles $i$ and then computing the value of $\delta_D$ and $\eta$ angles, it is possible, if 
necessary,  simulate directly $\delta_D$ and $\eta$ angles according distribution
motivated theoreticaly, for example from  Horizon-AGN simulation \citep{Okabe18,Codis18}.

On the basis of obtained PDF, now it is possible to compute the mean values of
analyzed statistics and its standard deviations. We compute the standard 
deviation of the mean (denoted as $\sigma(\bar{x})$) and estimate $S^{*}$ denoted 
in the tables as $\sigma(x)$, that is the estimator of the standard deviation 
in the sample as well as the standard deviation of 
$S^{*}$ which is equal to $\sigma(S^{*})=S^{*}/\sqrt{2(l-1)}$ \citep{Brandt97}.
We repeated these simulations again with 247 fictitious clusters each with number
of member galaxies the same as in the real cluster. The reason of this is that 
the number of galaxies in our real clusters is small in some cases what could 
influence the results of statistical tests. 
In Tables \ref{tab:t1a} - \ref{tab:t1c} we present mean values of the analyzed 
statistics, its standard deviation, standard deviation in the sample as 
well as its standard deviation for the sample of 247 clusters each with 2360 galaxies 
in the case of the analysis of the angles $p$, $\delta_D$ and $\eta$ respectively.
First of all, we have analyzed how results of the simulation for $p$ angles in 
Table \ref{tab:t1a} are in agreement with theoretical predictions. Our first test is 
the $\chi^2$ test. The theoretical value is $\chi^2=n-1=35$ while we have obtained the 
value 34.9978. It means that the difference is less than $1 \sigma(\bar{\chi^2})=0.0172$. 
Because we analyzed $247$ clusters then for $n=36$ the theoretical variance is equal 
${\sigma^2(\chi^2) \over m}=0.2834$ and standard deviation  equals $0.5324$. 
The simulations have given $0.5442$, so the differences are again less than the value 
of $\sigma(S)$. The similar situation is in the case of the $\chi^2$ control 
test, however differences between theoretical and simulated values of standard 
deviation is on the $2\,\sigma$ level. In the case of autocorrelation test, in both 
 basic and control tests, the simulated and theoretical mean values of $C$  
agree. The obtained value of standard deviation of $C$ also is not significantly 
deviated from value going from approximation $\sqrt{n/247} = 0.3818$. One should note 
however that in the case of control test the difference is bigger (more than $3\,\sigma$). 
When we analyzed the statistics $\Delta_{11}/\sigma(\Delta_{11})$ we have again obtained 
that simulations and theoretical mean ($0.0000$) and standard deviation ($0.06363$) 
values agree. Also in the case of analyzed $|\Delta_{11}/\sigma(\Delta_{11})|$ statistics 
we obtained a perfect agreement between values obtained from simulations and theoretical 
computations predicted by formulae \ref{eq:ff14}, \ref{eq:ff17}. 

In the case of $\Delta_{1}/\sigma(\Delta_{1})$, $\Delta/\sigma(\Delta)$ and 
$\Delta_c/\sigma(\Delta_c)$ statistics the situation is a little bit more complicated. 
It is because for obtaining the theoretical mean values we need to know their variance 
(see formulae  \ref{eq:ff5}, \ref{eq:ff6} and \ref{eq:ff19}). Unfortunately, we have the 
theoretical prediction in relation to the squares of statistics 
only, and variances of analyzed statistics are obtained by theorem of propagation of errors. 
Moreover, the function $y=\sqrt{x^2}$ is non linear so the results obtained 
by theorem of propagation of errors is only an approximation. According to this approximation 
$\sigma^2(\Delta_{1}/\sigma(\Delta_{1}))=\sigma^2(\Delta/\sigma(\Delta))=\sigma^2(\Delta_{c}/\sigma(\Delta_{c}))=1/2$.
In our cases for 247 clusters it leads to the value $\sigma(x)=0.04499$. From the 
inspection of Table \ref{tab:t1a} it is easy to see that only for statistics 
$\Delta/\sigma(\Delta)$ the differences between theoretical and observed values of $\sigma(x)$ 
is on the $1\,\sigma$ level, while for remaining two statistics the differences are a 
little bit more than $3\,\sigma$. Consequently the mean values of all three statistics 
varies from theoretical predictions and only for $\Delta/\sigma(\Delta)$ statistic 
the difference is not very high ($1.8794$ instead of $1.8708$ with $\sigma(\bar{x})=0.0014$). 
It clearly indicates that the correct values  of $\Delta_{1}/\sigma(\Delta_{1})$, 
$\Delta/\sigma(\Delta)$ and $\Delta_c/\sigma(\Delta_c)$ statistics must be obtained 
from numerical simulations.

The more difficult is the problem when we analyzed the distribution 
of $p$  angle (the easiest one), using Kolmogorov- Smirnov test. The investigated 
statistic are $\lambda$ and $\lambda_c$. The original Kolmogorov- Smirnow test is 
a nonparametric test of the equality of continuous, one-dimensional probability 
distributions.  The  test could also be adapted for discrete variables and also for 
the case when theoretical distribution depends on the estimated parameters. One 
should note that distribution of statistics $D_n$ and consequently $\lambda$ 
and $\lambda_c$ (see equation \ref{eq:k1}) depend both on bining process as well as on 
theoretical distributions $F(x)$ and on true (but unknown) value of estimated parameters.  
The limiting form for the distribution function of Kolmogorov's $D_n$ was analyzed by
\citet{Wang03}. They showed that the mean and variance of $\lambda=\sqrt{n}D_n$ are 
$\mu=0.868731$ and $\sigma^2=0.067773$ what led to the value  $\sigma=0.260333$. However,
because the distribution depends on binning process the Monte Carlo or other methods of 
the simulations  are required. In our cases we simulated the sample of 247 cluster each 
with 2360 galaxies and performed 1000 simulations. In the case of analysis of the position 
angle $p$, the range of analyzed angle is divided into $n$ bins of equal width. In the  
basic test the number of bins $n=36$ while in the case of control test it is reduced to $n=18$. 
In analyzed case we have obtained the expected value $\lambda=0.7708$ and $\lambda_c=0.7314$. 
Of course for different binning the value of galaxies in clusters, the number of clusters  
and number of simulations we will obtain different values. For example for 1000 simulations 
of the sample of 500 clusters each with 10000 galaxies having axial ratio $q=b/a>0.75$,
binned on $n=100$, the expected value $\lambda=0.8106$. 
For the case of simulated sample of 247 clusters each  with 2360 galaxies and 
$n=36$ bins the standard deviation of $\lambda$ and $\lambda_c$  equals $0.0166$, 
what is in perfect agreement with the result of \citet{Wang03} i.e. 
$\sigma=0.260333/\sqrt{247}=0.01656$. Unfortunately, during analysis of $\delta_D$ 
and $\eta$ angles the agreement is not so good (see Tables \ref{tab:t1a} - \ref{tab:t1c}).
The above analysis clearly shows that expected value of $\lambda$ as well as their 
standard deviation, although valid in particular cases, must be obtained from numerical 
simulations.

For investigation of the uniformity on a circle, the alternative to the Kolmogorov--Smirnov 
test are Cram\'er--von Mises test \citep{Cramer28,vonMises28} and Watson test \citep{Watson62,Dom79} 
based on the staistics:
\begin{equation}
\label{eq:k3}
\omega^2=\int_{-\infty}^{+\infty}(F(x)-S(x))^2d(F(x))
\end{equation}
where again $F(x)$ is the theoretical distribution and  $S(x)$ is the empirically observed distribution.
The asymptotic form of above  distribution was analyzed by \citet{Watson61}.

In Cram\'er--von Mises test one uses statistics:
\begin{equation}
\label{eq:k4}
W^2=\sum_{i = 1}^n\left(F(x_i)-{2i-1 \over 2n}\right)^2 +{1 \over 12n}
\end{equation}
while in the advanced modification, called Watson test \citep{Watson62,Dom79},
one uses statistics:
\begin{equation}
\label{eq:k5}
U^2=W^2-n(\bar{F}(x)-{1 \over 2})^2
\end{equation}
where the average value $\bar{F}(x)={1 \over n} \sum_{i = 1}^nF(x_i)$.

However, because such tests are based on diferences between observed and theoretical 
distribution like Kolmogorov--Smirnov test in power two, it should be checked the dependence 
of the distribution  on binning process, as in the case of the Kolmogorov--Smirnov test.
Again, as in the case Kolmogorov--Smirnov test, we simulated the sample of 247 cluster each 
with 2360 galaxies and performed 1000 simulations and compute the average values $\bar{W^2}$ 
and $\bar{U^2}$. In the  case of  test with the number of bins $n=36$  of equal width   we have 
obtained the expected value $\bar{W^2}=0.01278$ and $\bar{U^2}=0.00401$ with standard deviation 
$\sigma(W^2)=0.00049$ and $\sigma(U^2)=0.000070$ respectively. Unfortunately, again  for 
different binning of the value of galaxies in clusters, the number of clusters  and number of 
simulations we have obtained different values. For example for 1000 simulations of the sample 
of 500 clusters each with 10000 galaxies with axial ratio $q=b/a>0.75$, binned on $n=100$, 
the expected value $\bar{W^2}=0.00501$ and $\bar{U^2}=0.00167$ with standard deviation 
$\sigma(W^2)=0.00015$ and $\sigma(U^2)=0.000024$. The  above analysis clearly shows that the 
expected value of statistics $\bar{W^2}$ and $\bar{U^2}$ as well as their standard deviations 
must be obtained from numerical simulations. Unfortunately, this means that the use of 
Cram\'er--von Mises and Watson tests instead of the Kolmogorov--Smirnov test do not give 
significant progress in our method.

\begin{figure}
\includegraphics[angle=270,scale=0.30]{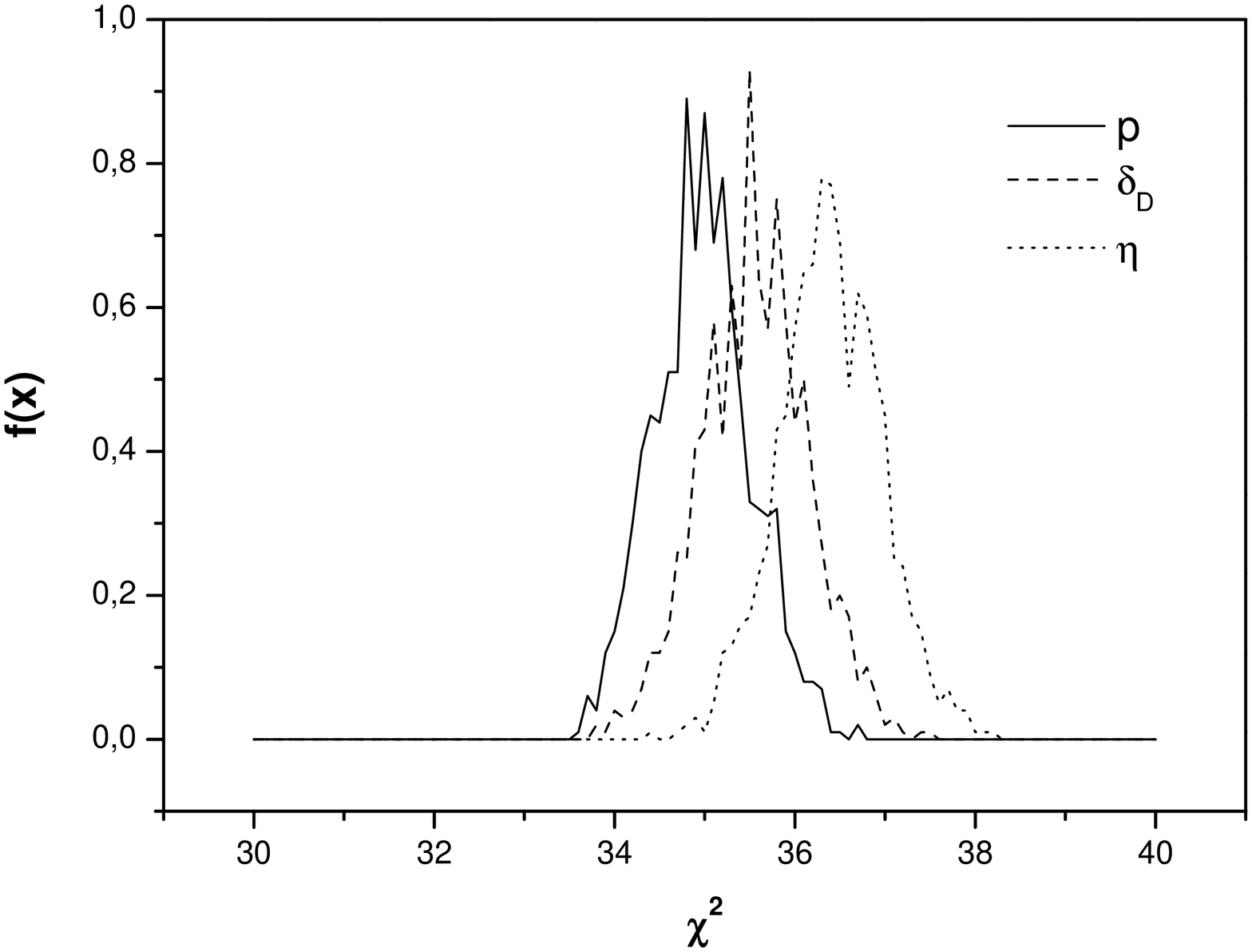}
\includegraphics[angle=270,scale=0.30]{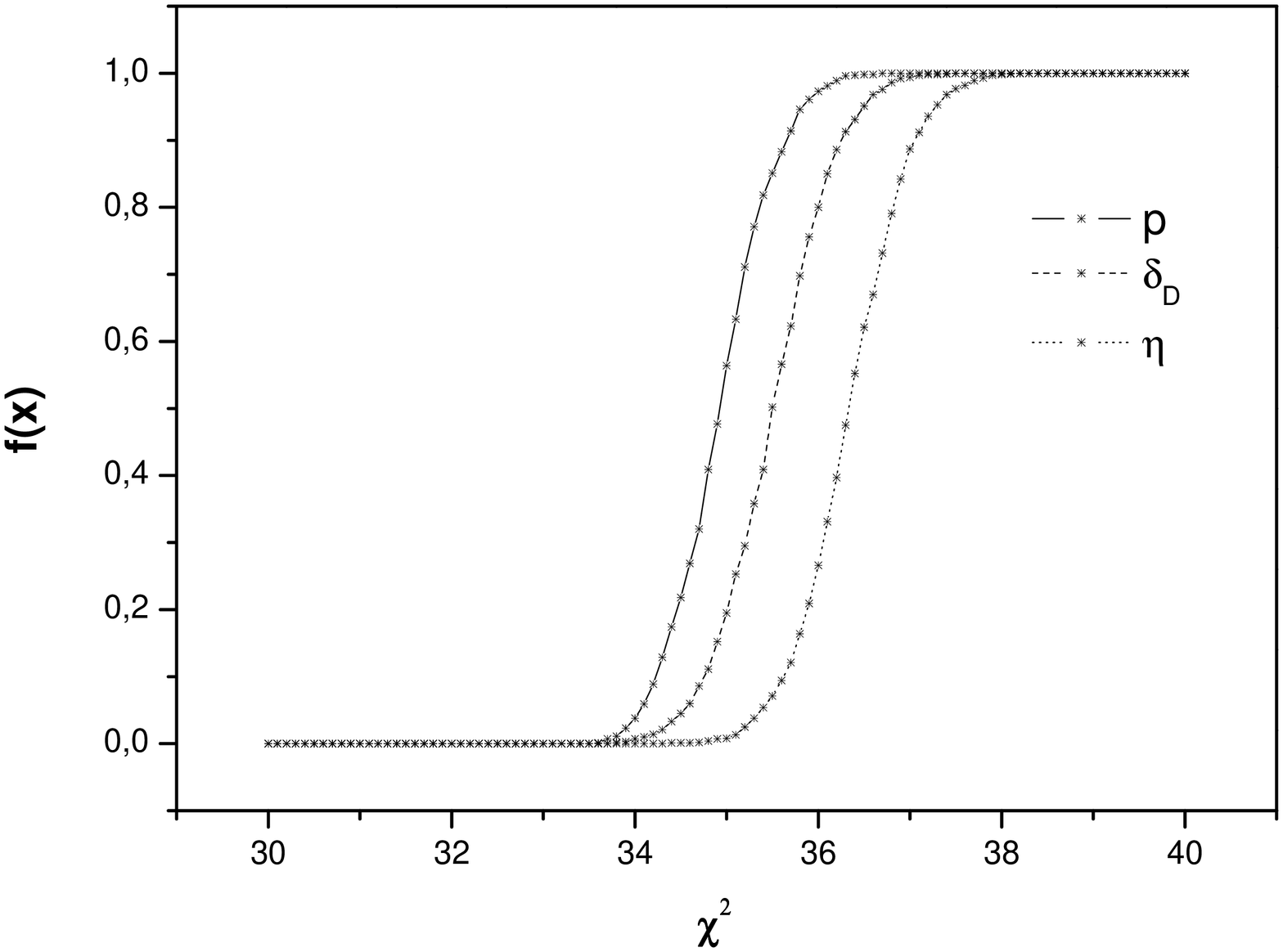}\\
\includegraphics[angle=270,scale=0.30]{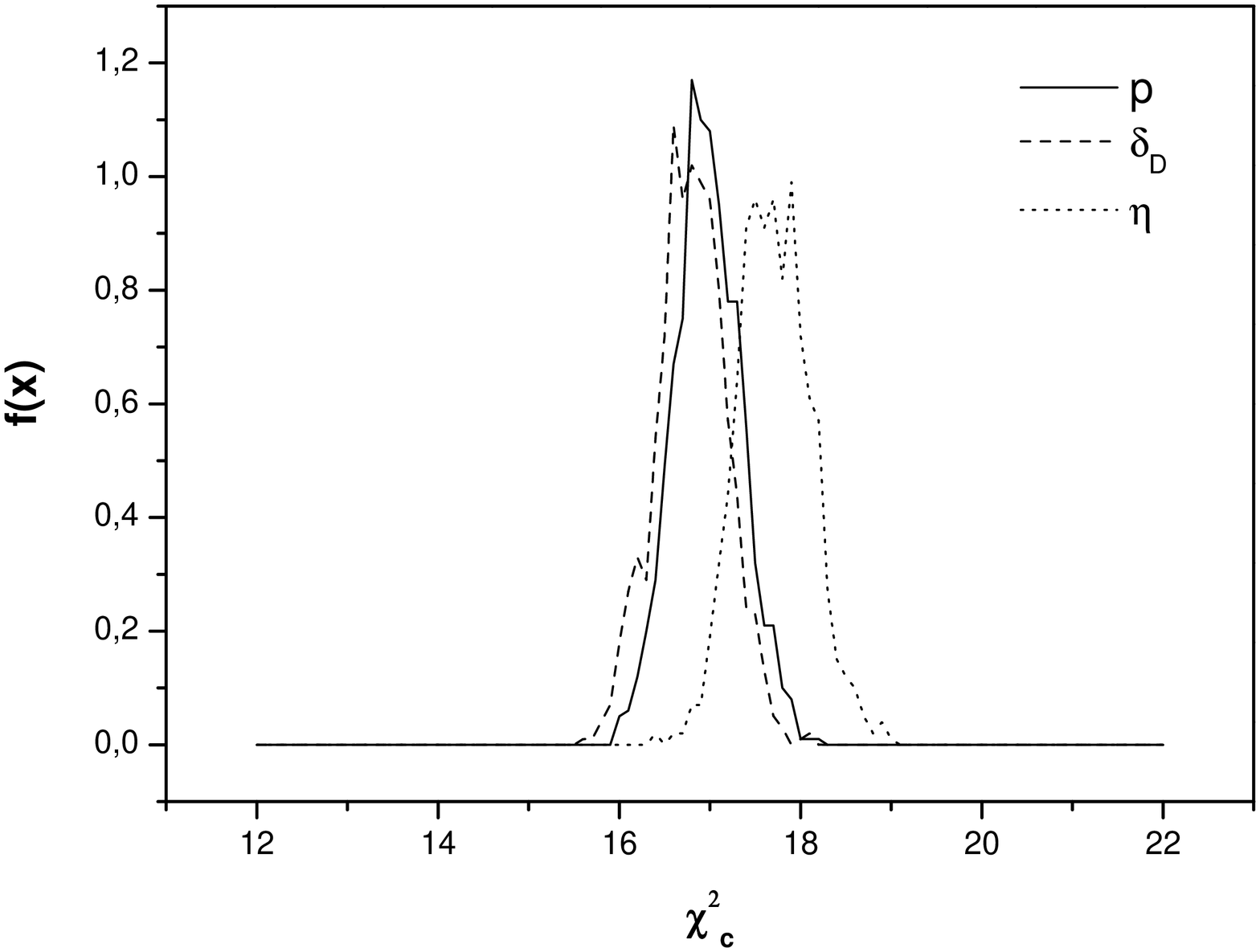}
\includegraphics[angle=270,scale=0.30]{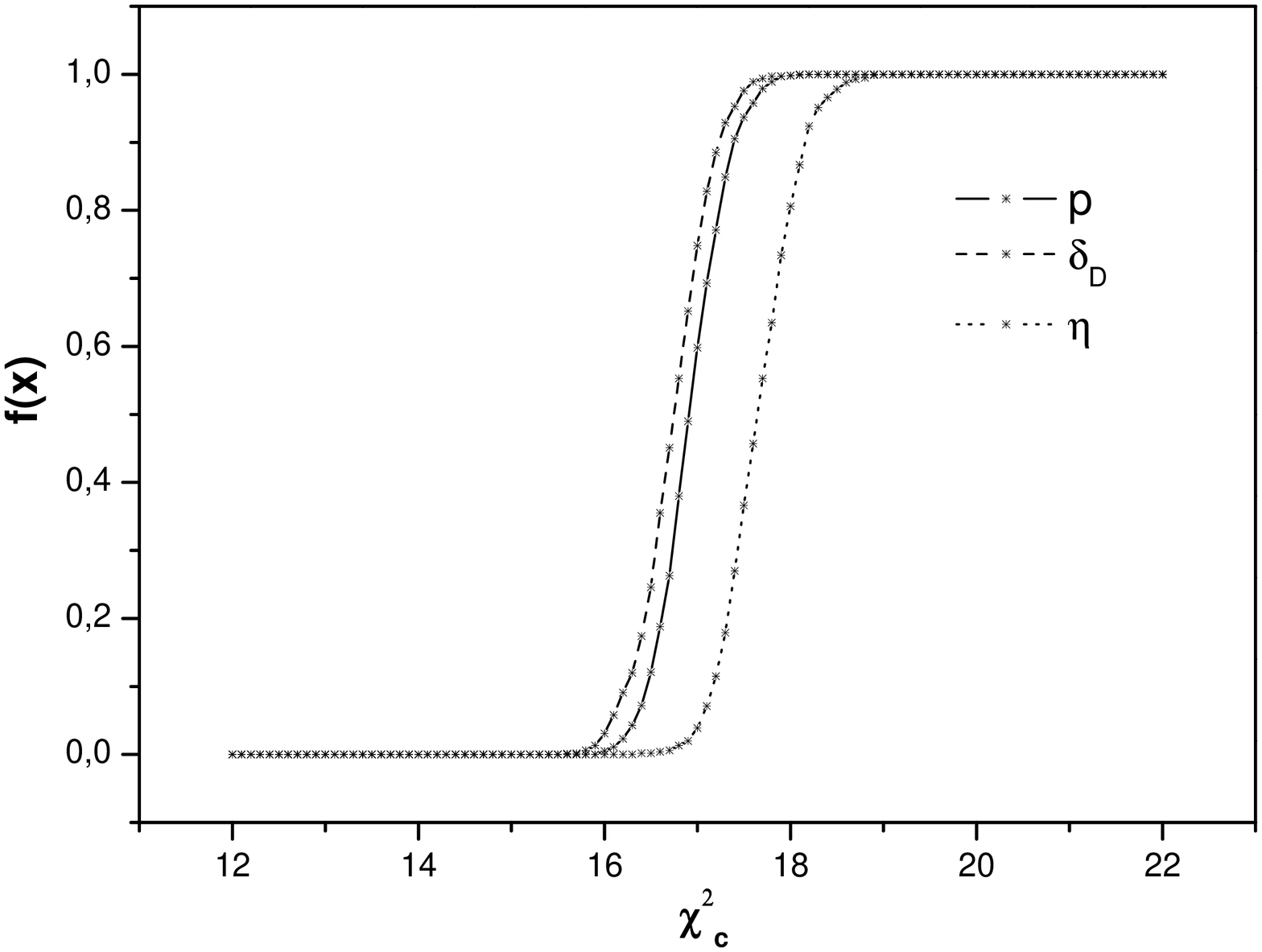}\\
\includegraphics[angle=270,scale=0.30]{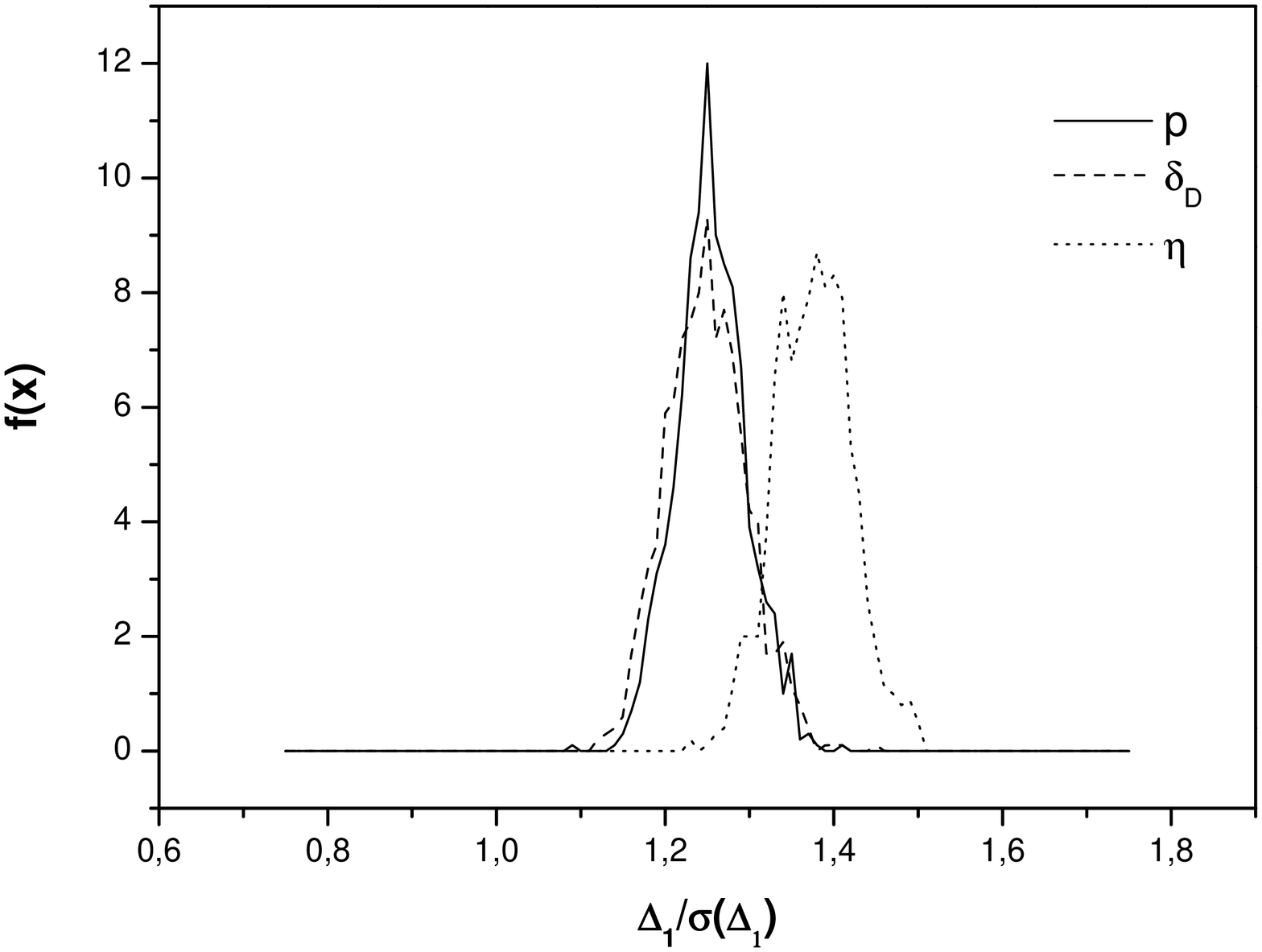}
\includegraphics[angle=270,scale=0.30]{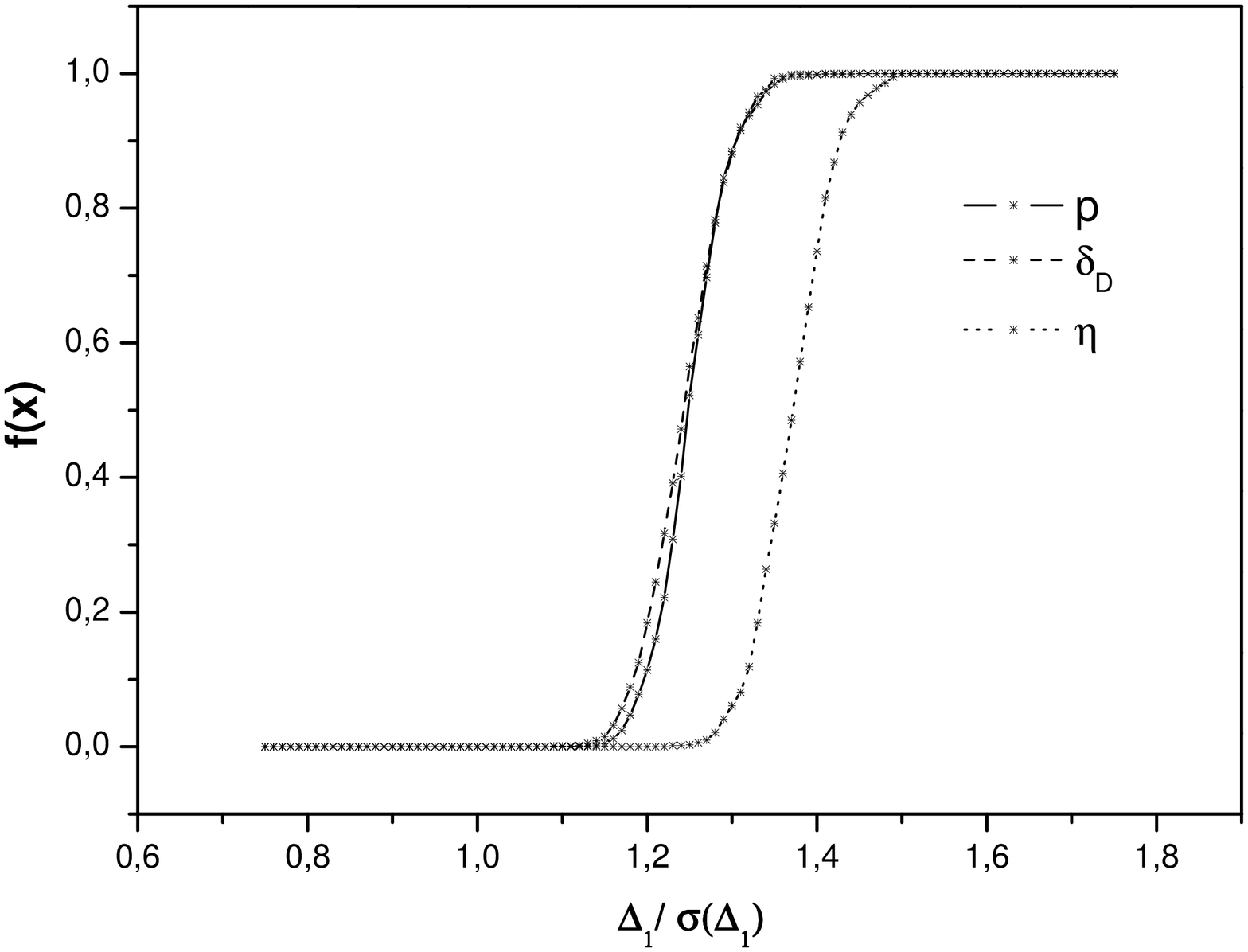}\\
\caption{The Probability Density Function (PDF) (left panel) and
Cumulative Distribution Function (CDF) (right panel) for analyzed statistics.
The figure was obtained from 1000 simulations of samples of 247 clusters
each with number of members galaxies the same as in the real clusters.
From up to down we present statistics:
$\chi^2$,$\chi_c^2$, $\Delta_{1}/\sigma(\Delta_{1})$.
\label{fig:f1}}
\end{figure}

\begin{figure}
\includegraphics[angle=270,scale=0.30]{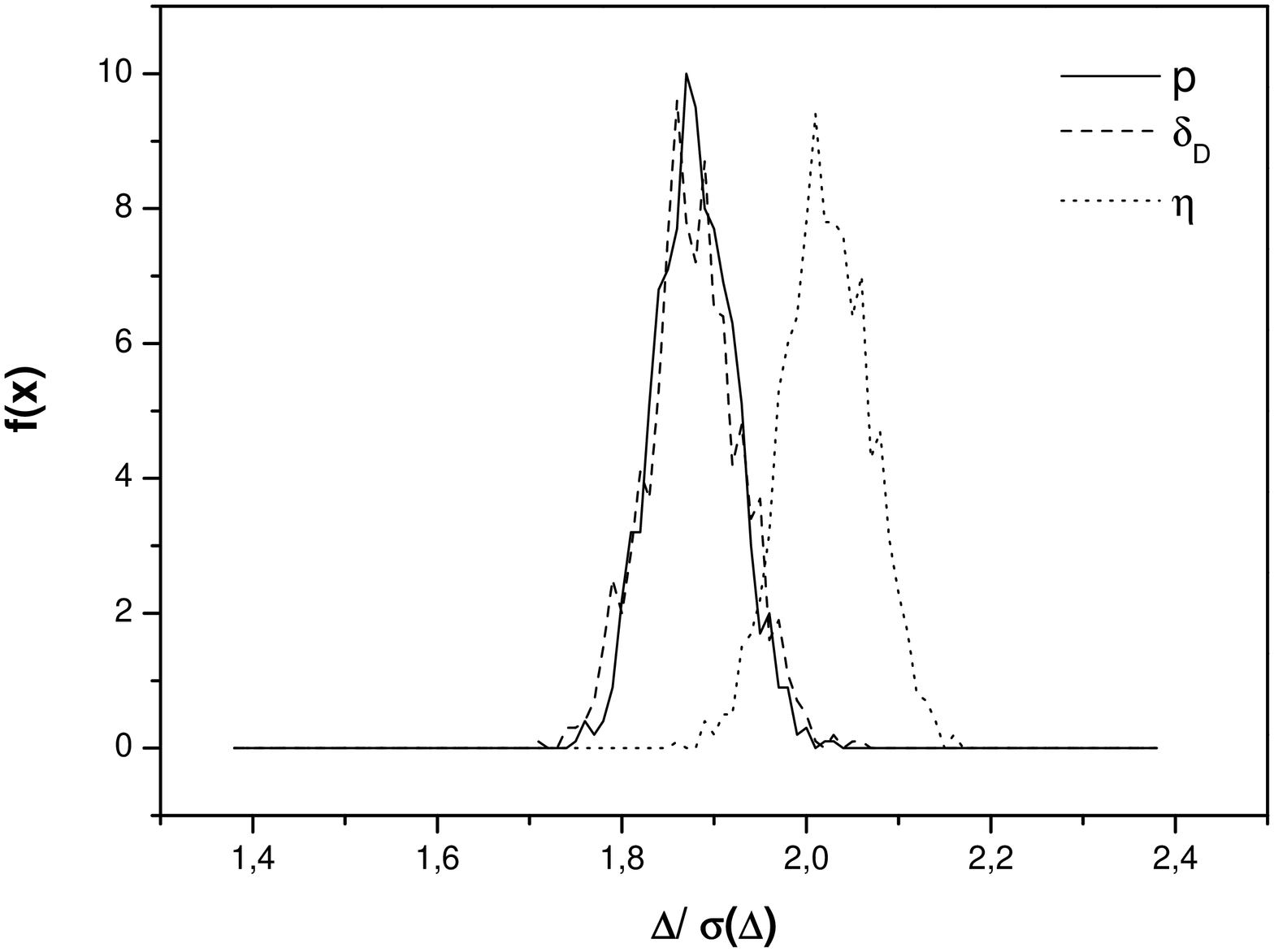}
\includegraphics[angle=270,scale=0.30]{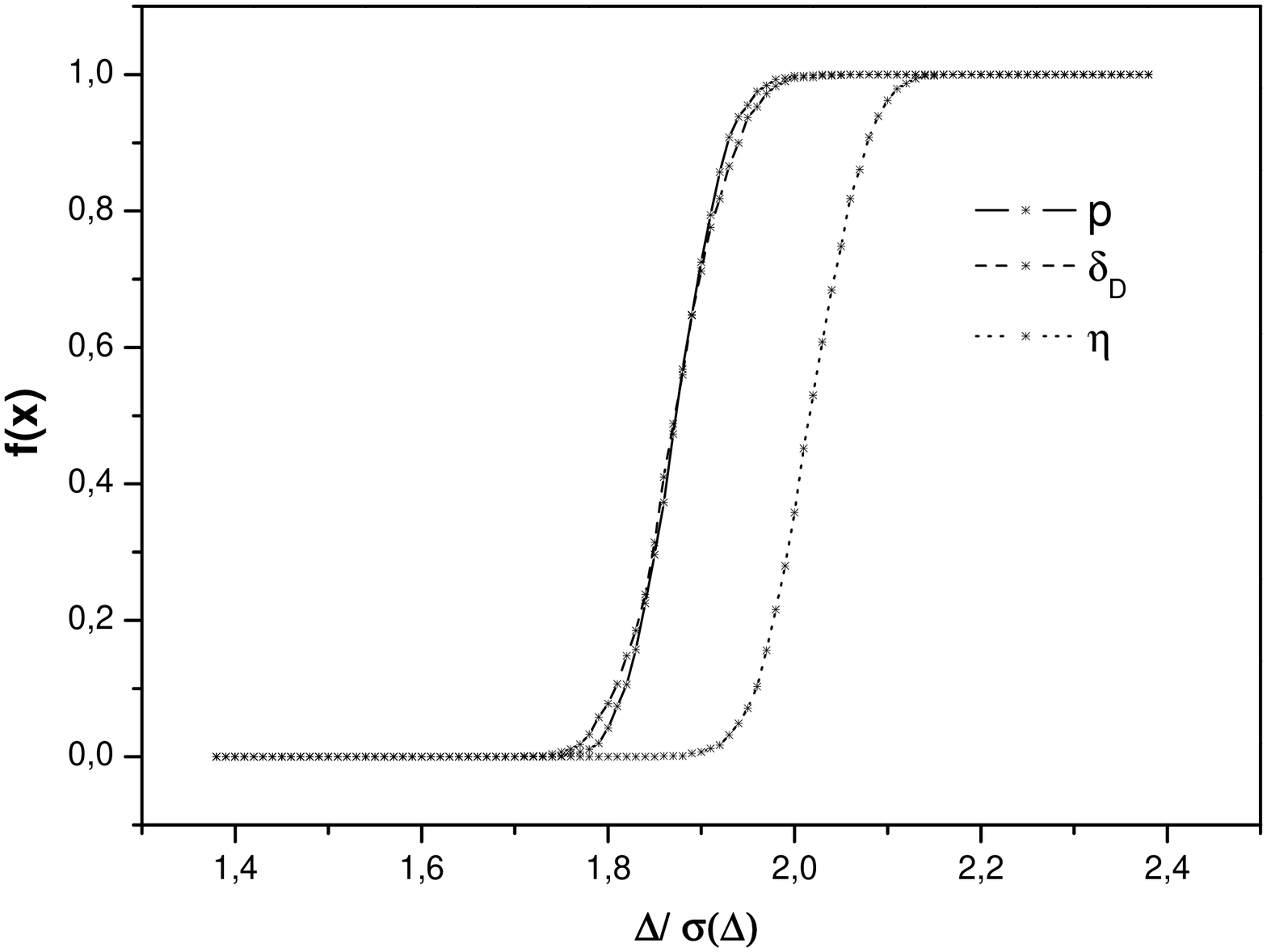}\\
\includegraphics[angle=270,scale=0.30]{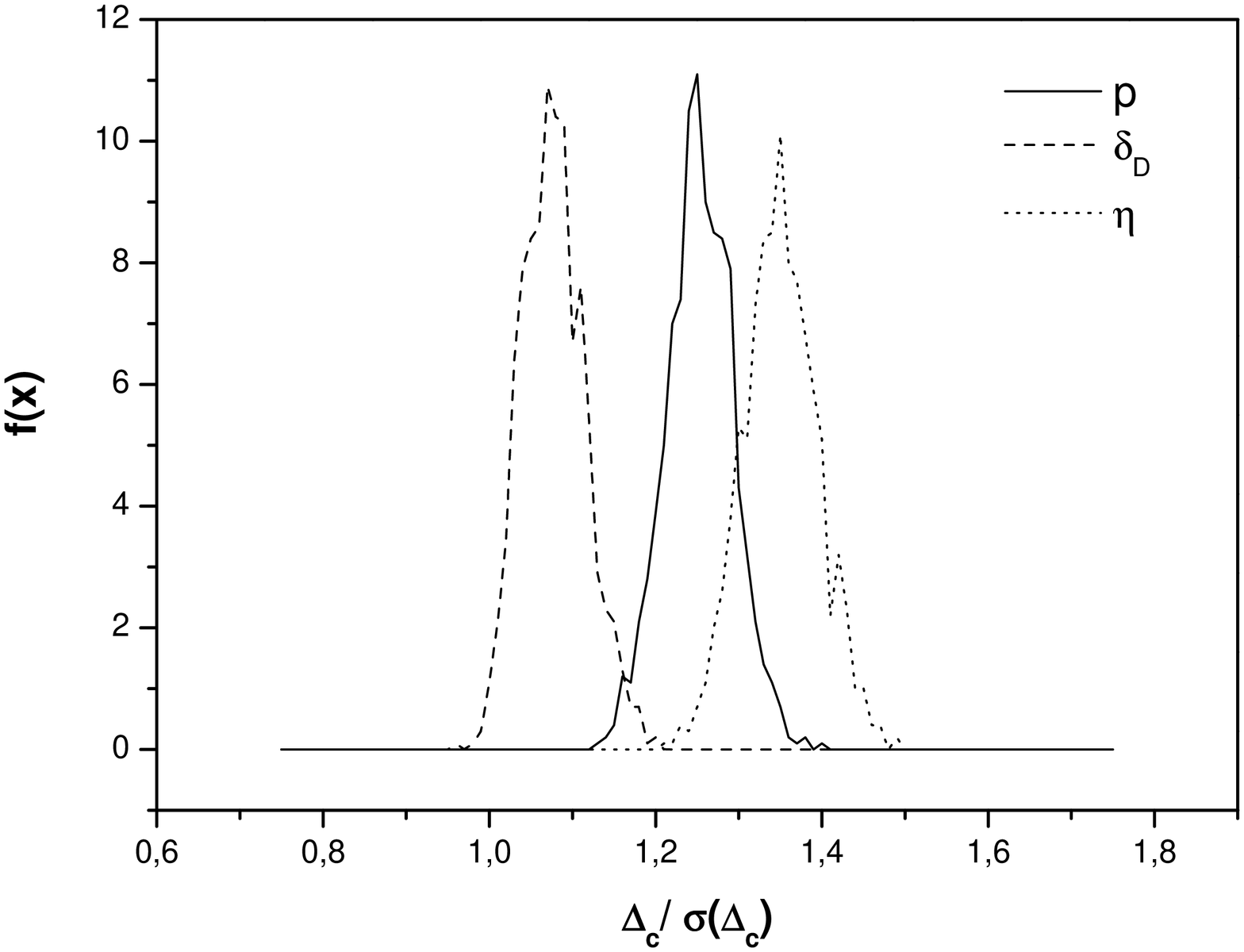}
\includegraphics[angle=270,scale=0.30]{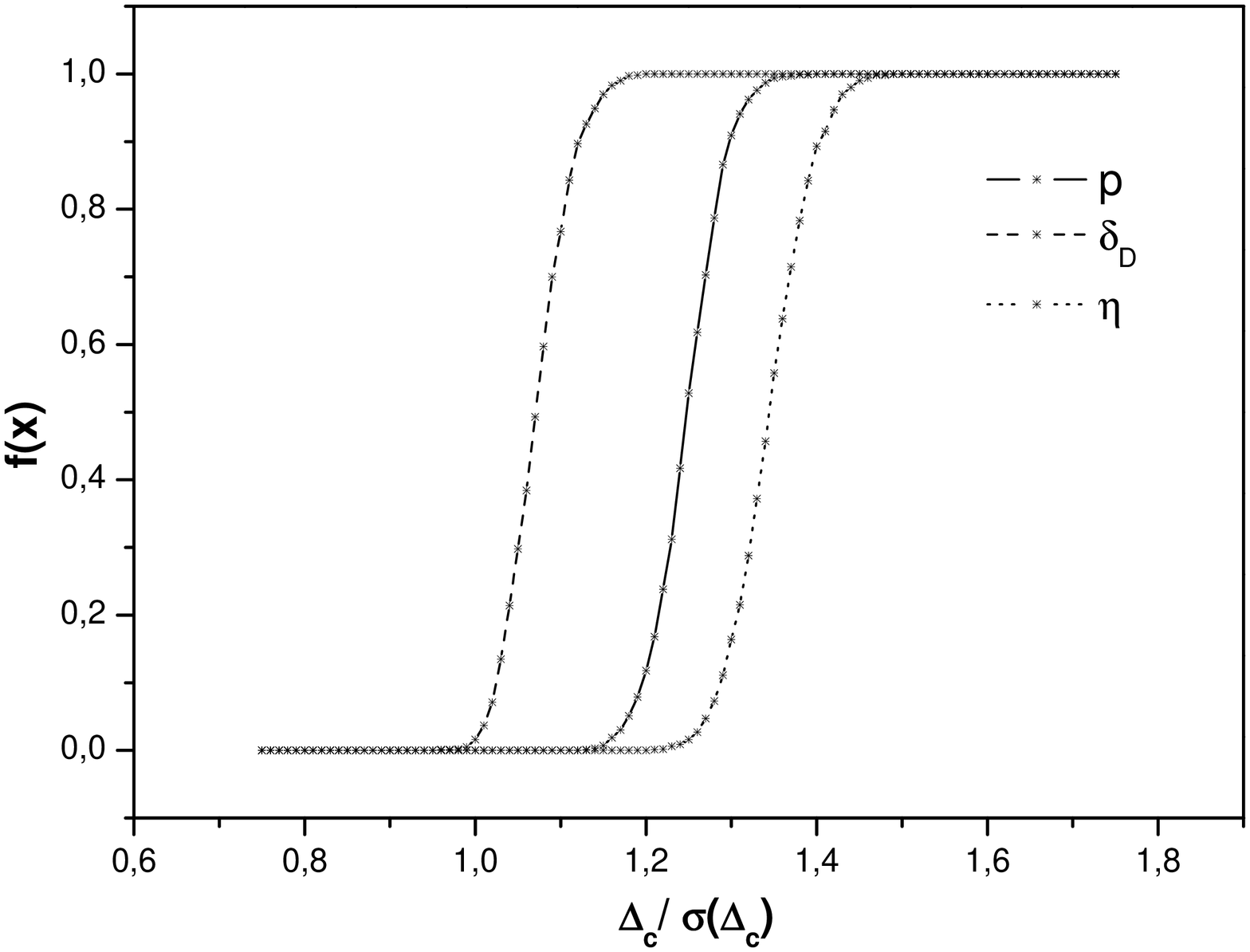}\\
\includegraphics[angle=270,scale=0.30]{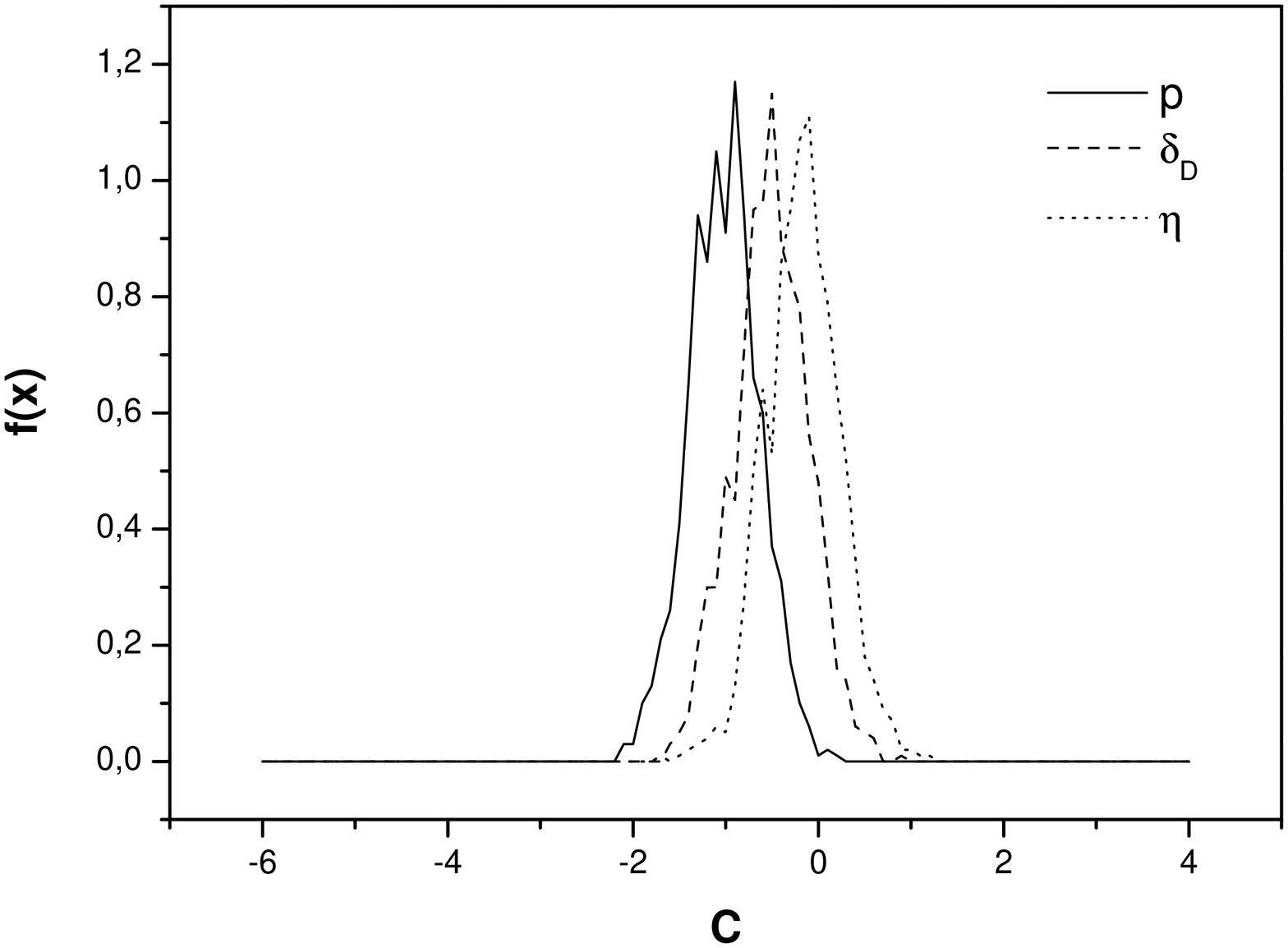}
\includegraphics[angle=270,scale=0.30]{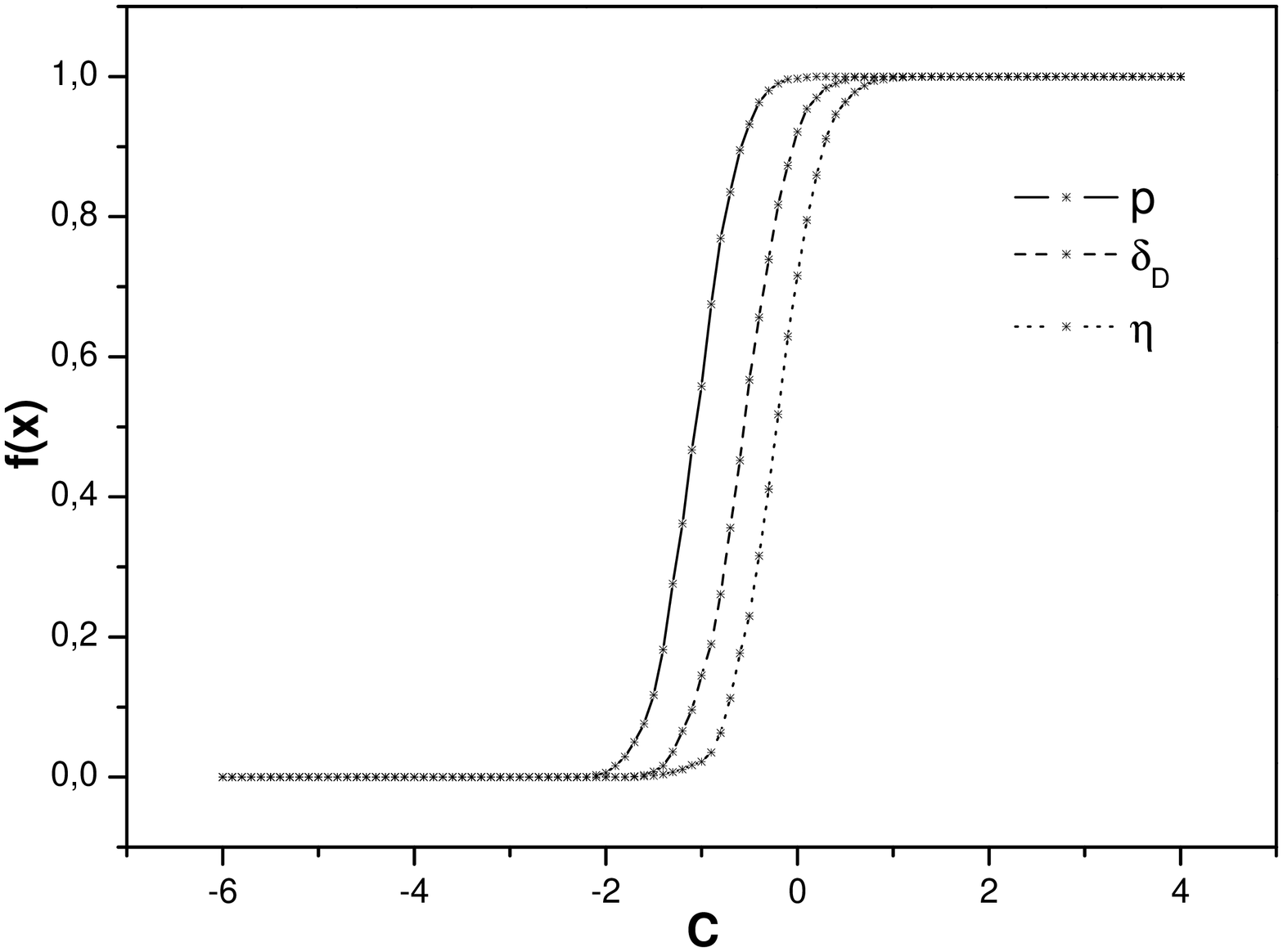}\\
\caption{The Probability Density Function (PDF) (left panel) and
Cumulative Distribution Function (CDF) (right panel) for analyzed statistics.
The figure was obtained from 1000 simulations of samples of 247 clusters
each with number of members galaxies the same as in the real clusters.
From up to down we present statistics:
$\Delta/\sigma(\Delta)$,$\Delta_c/\sigma(\Delta_c)$, $C$.
\label{fig:f2}}
\end{figure}

\begin{figure}
\includegraphics[angle=270,scale=0.30]{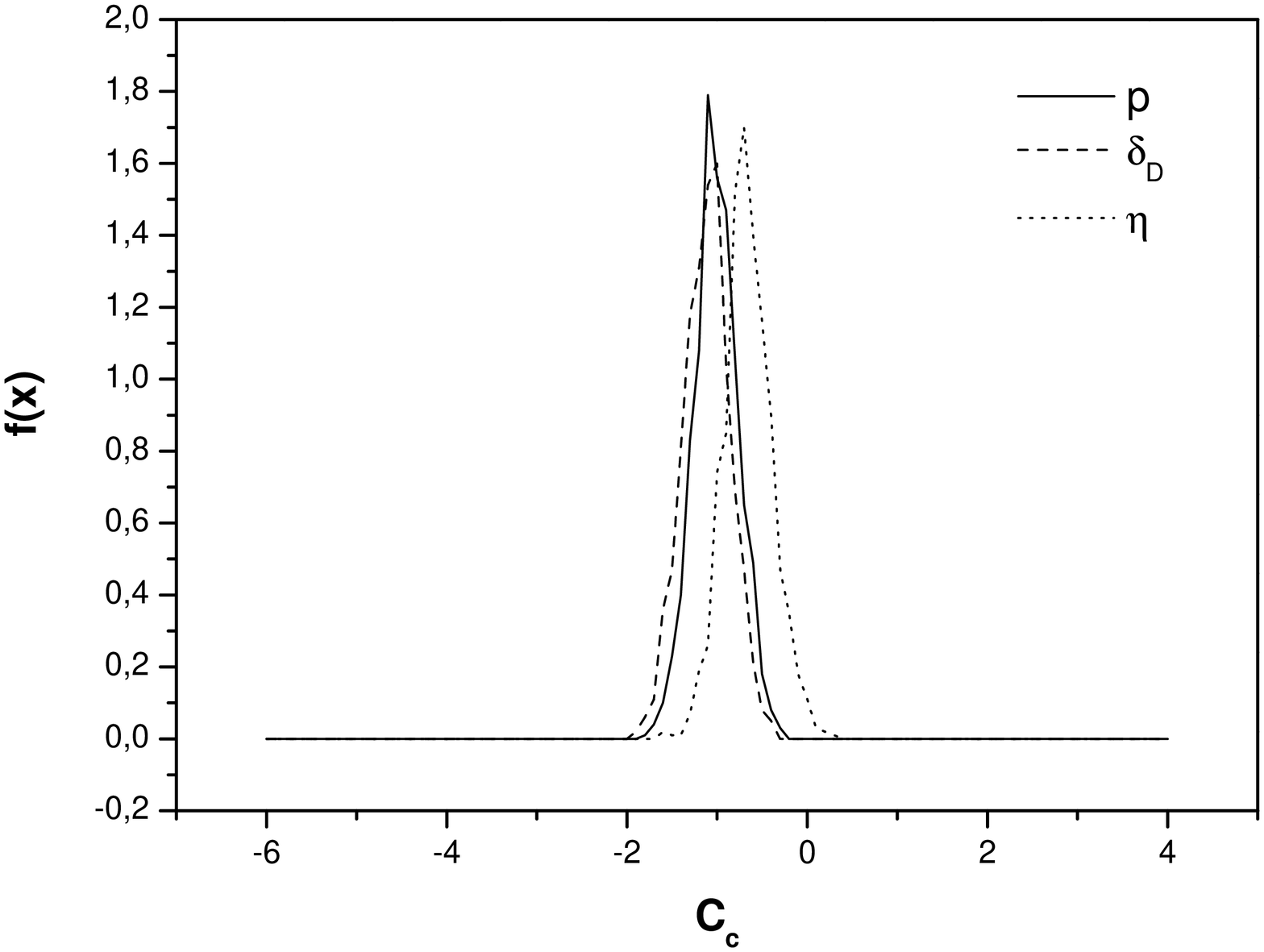}
\includegraphics[angle=270,scale=0.30]{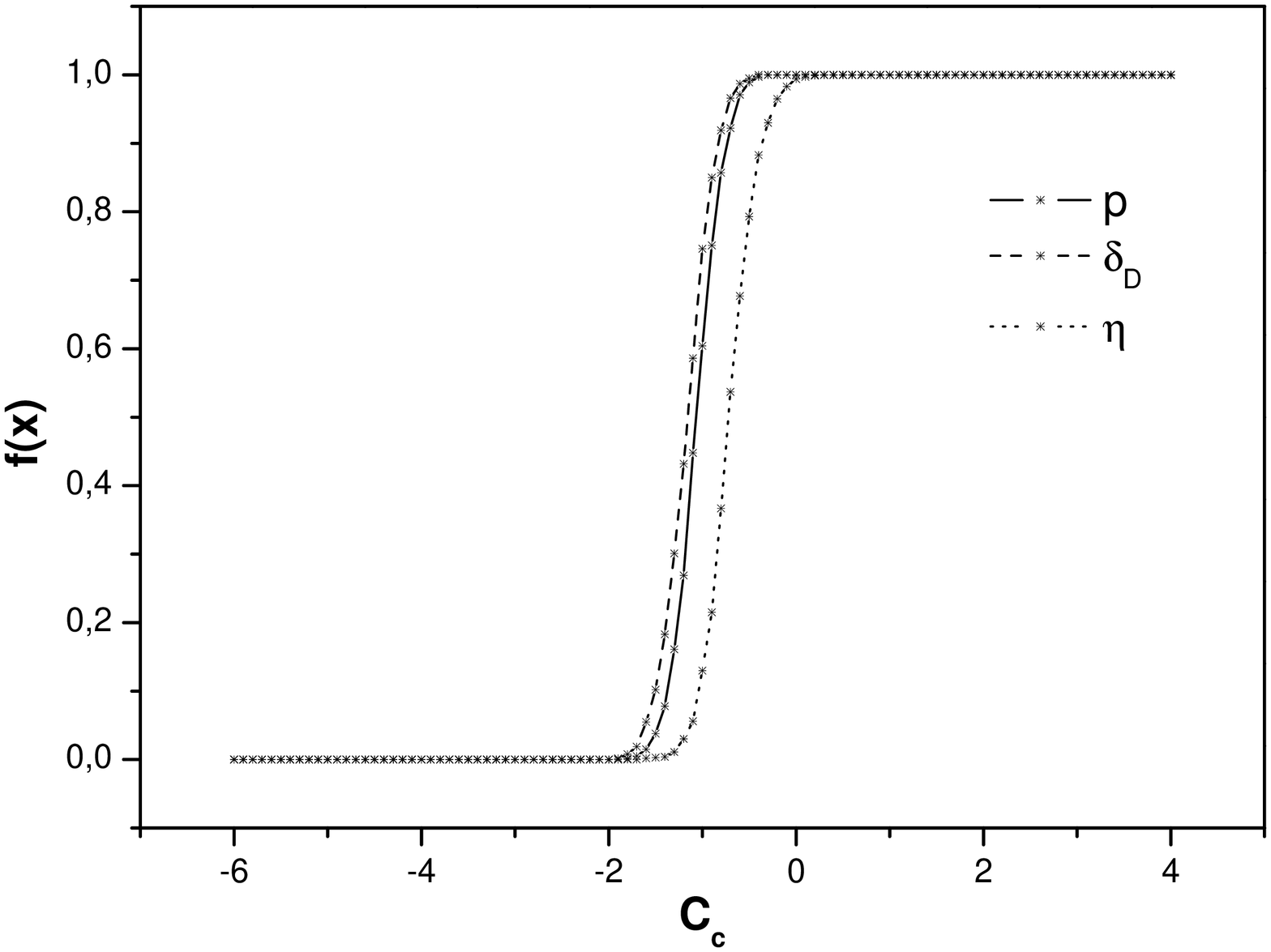}\\
\includegraphics[angle=270,scale=0.30]{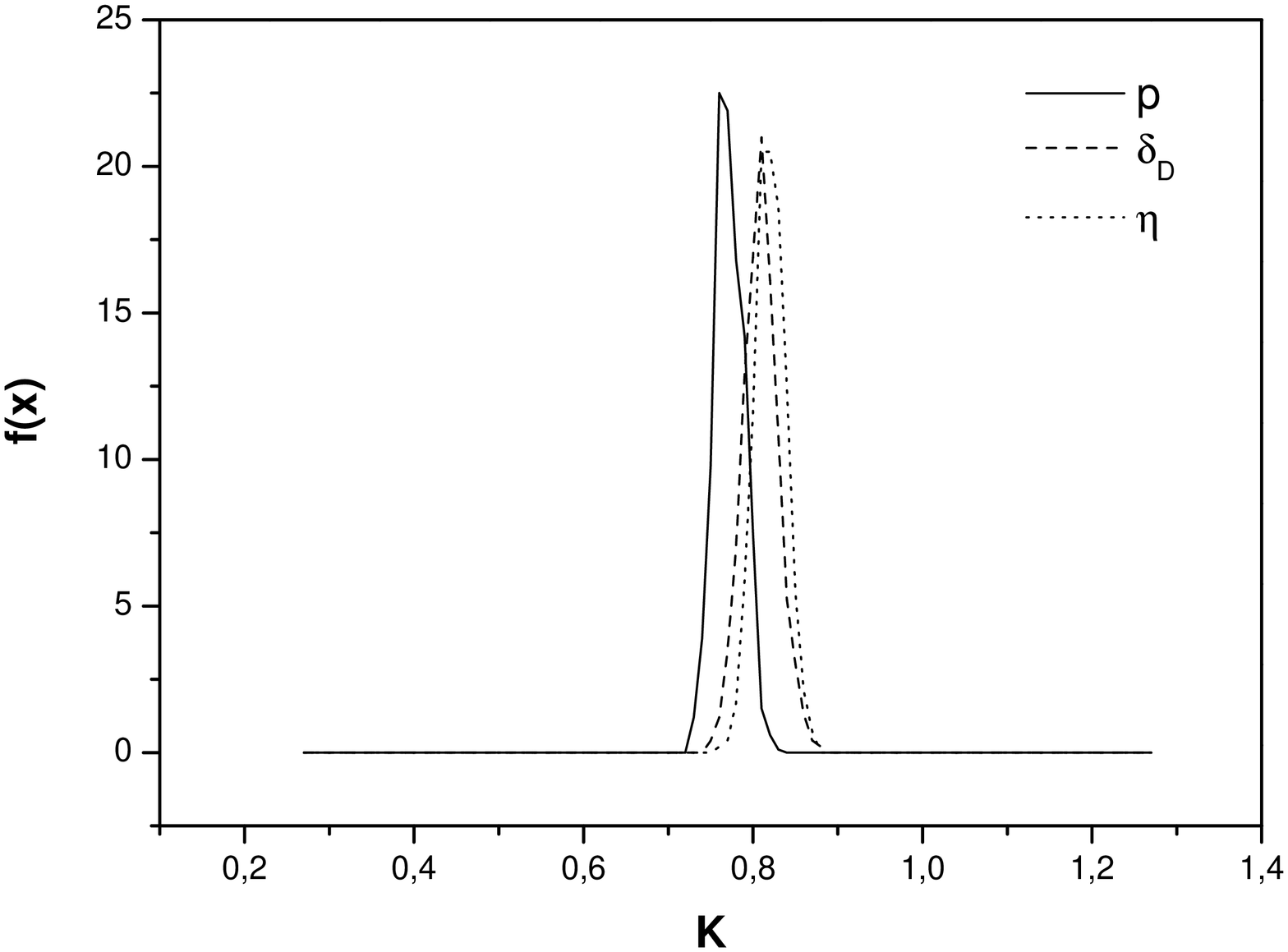}
\includegraphics[angle=270,scale=0.30]{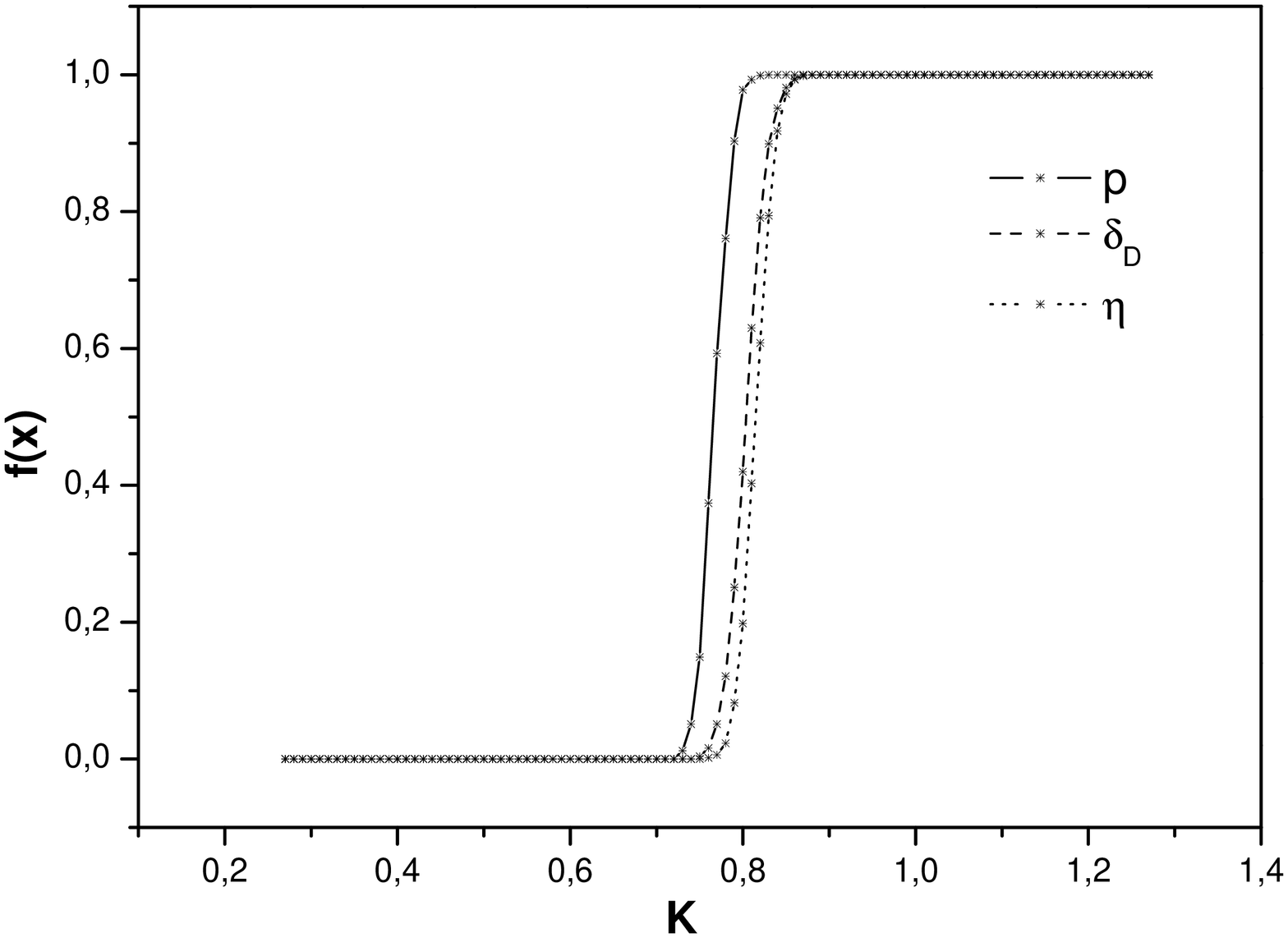}\\
\includegraphics[angle=270,scale=0.30]{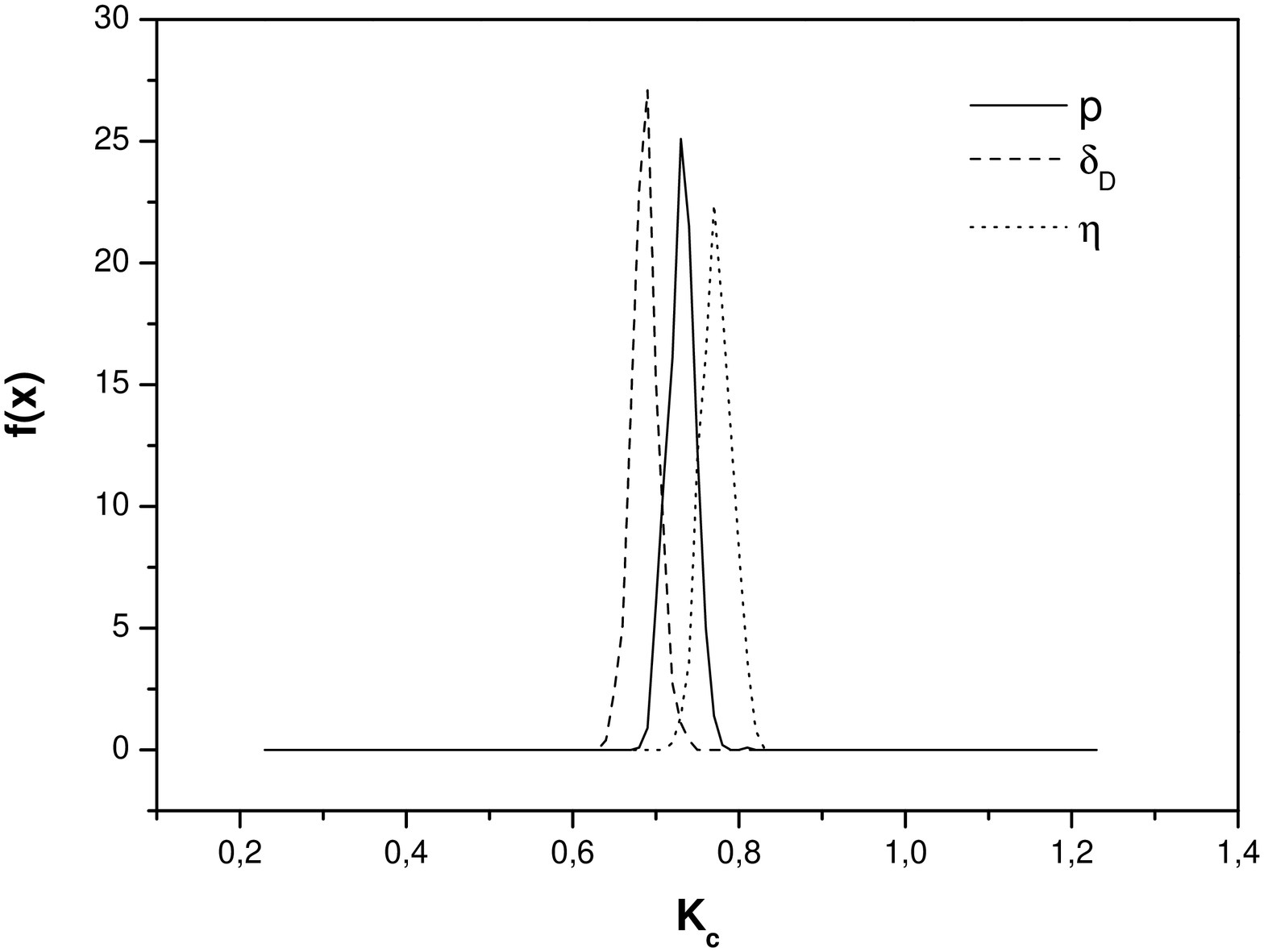}
\includegraphics[angle=270,scale=0.30]{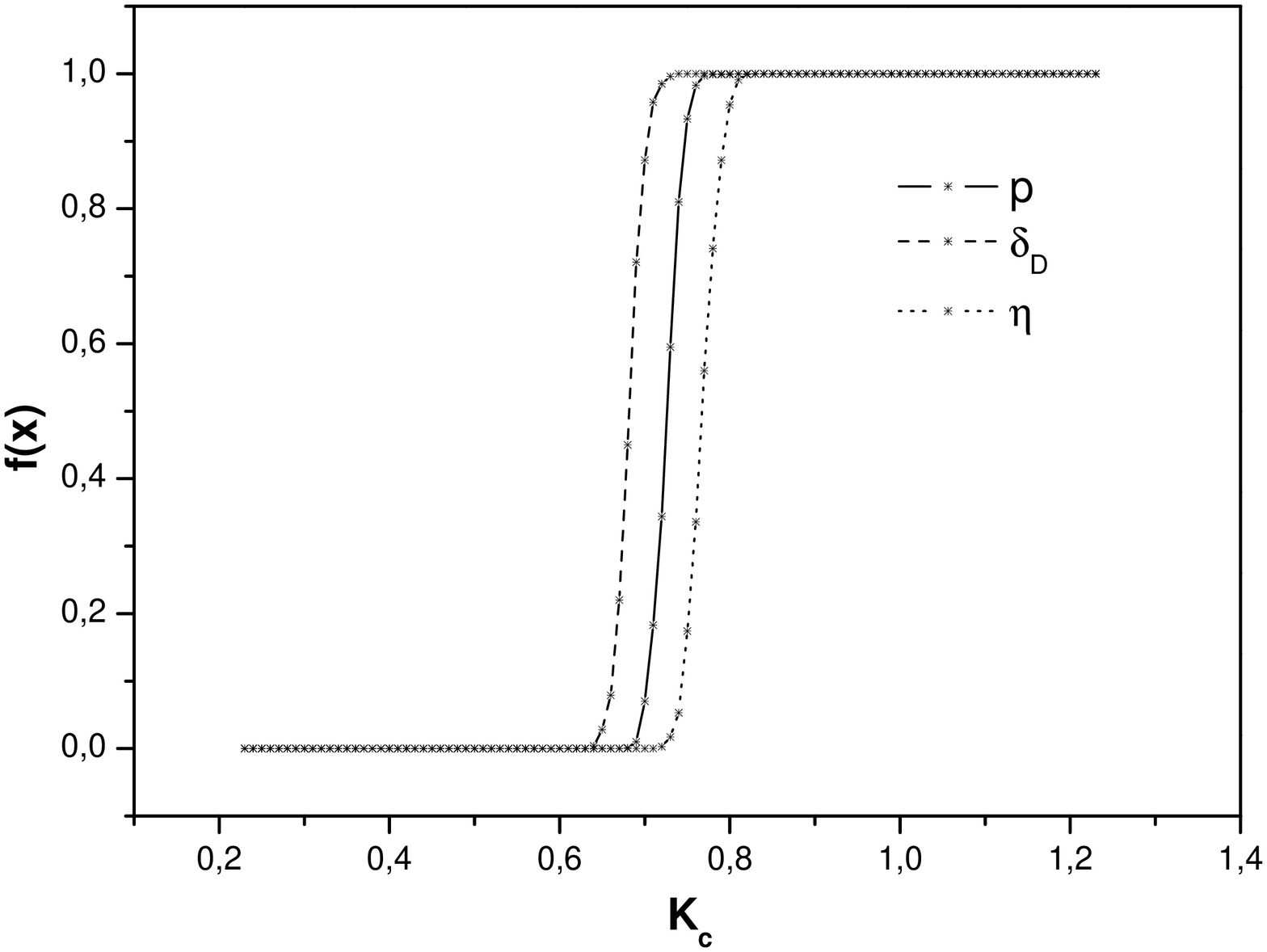}\\
\caption{The Probability Density Function (PDF) (left panel) and
Cumulative Distribution Function (CDF) (right panel) for analyzed statistics.
The figure was obtained from 1000 simulations of samples of 247 clusters
each with number of members galaxies the same as in the real clusters.
From up to down we present statistics:
$C_c$, $\lambda$, $\lambda_c$.
\label{fig:f3}}
\end{figure}

\begin{figure}
\includegraphics[angle=270,scale=0.30]{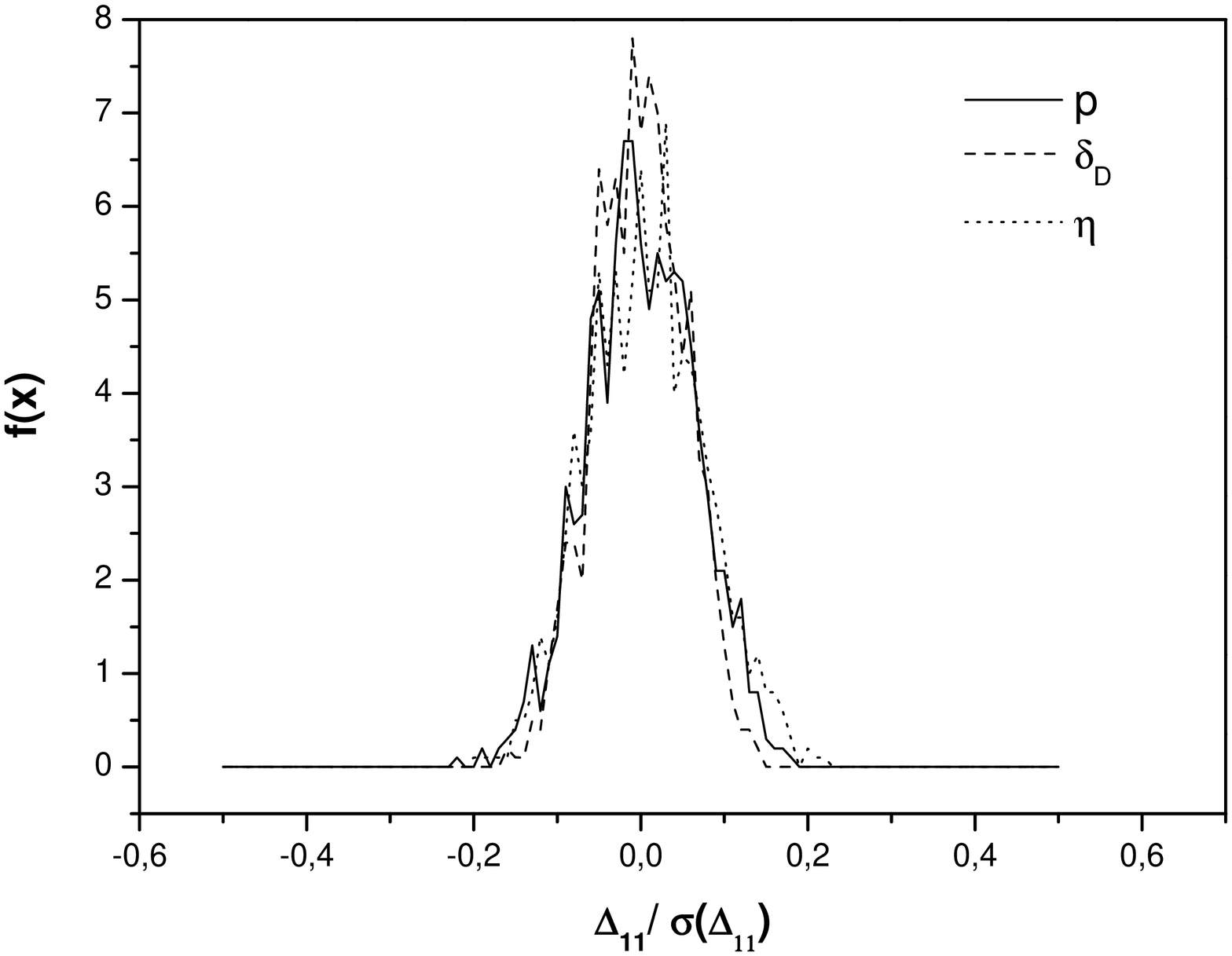}
\includegraphics[angle=270,scale=0.30]{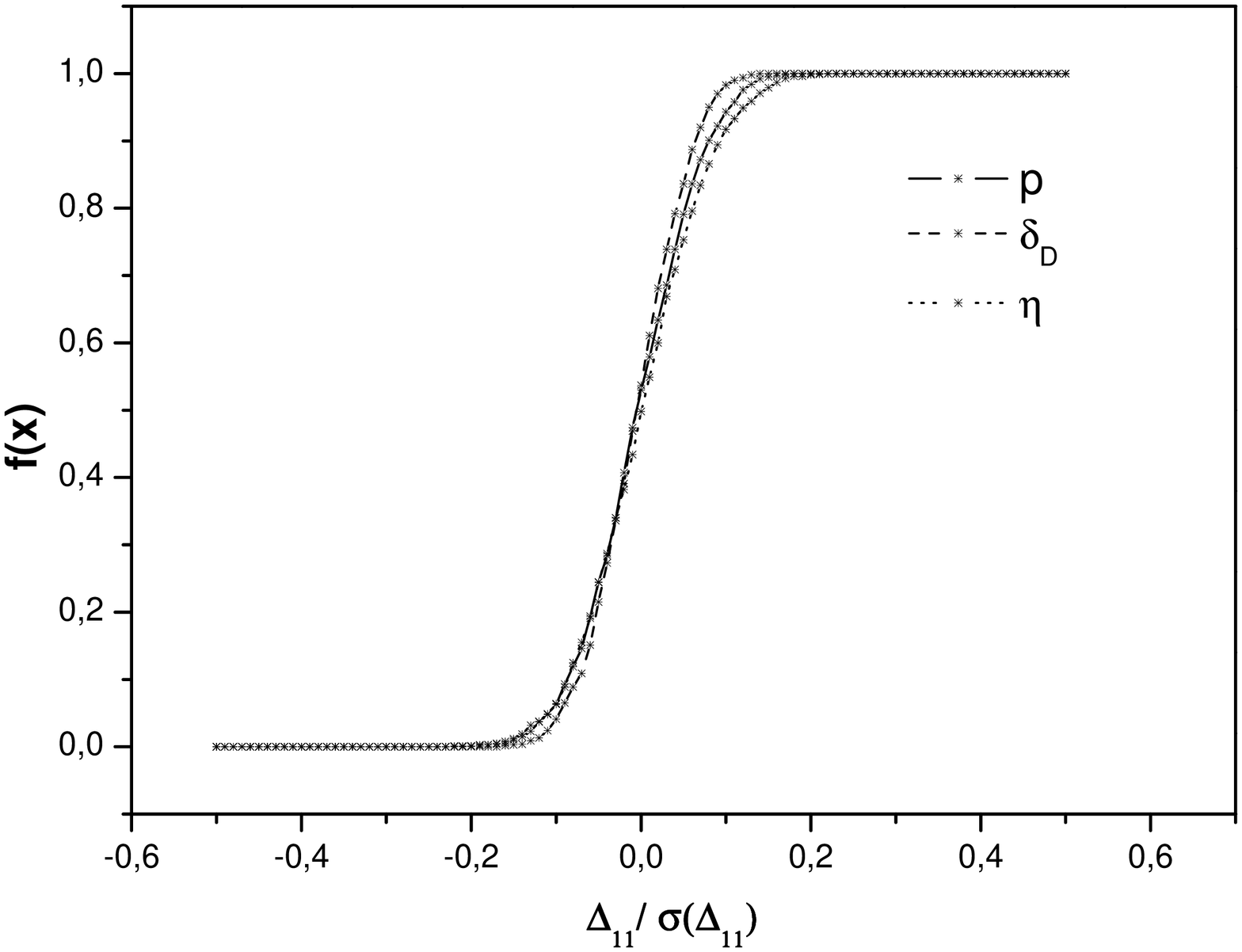}\\
\includegraphics[angle=270,scale=0.30]{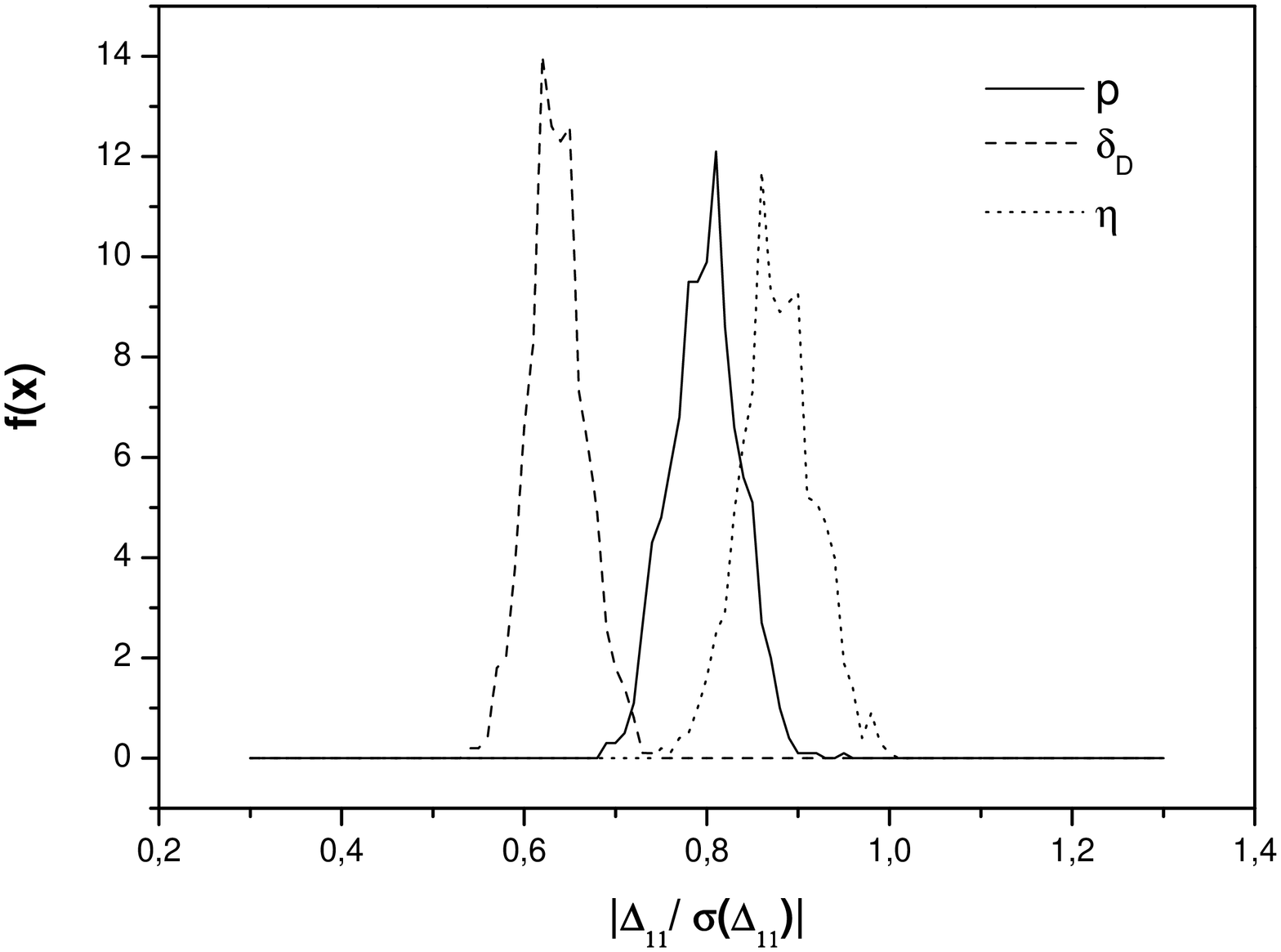}
\includegraphics[angle=270,scale=0.30]{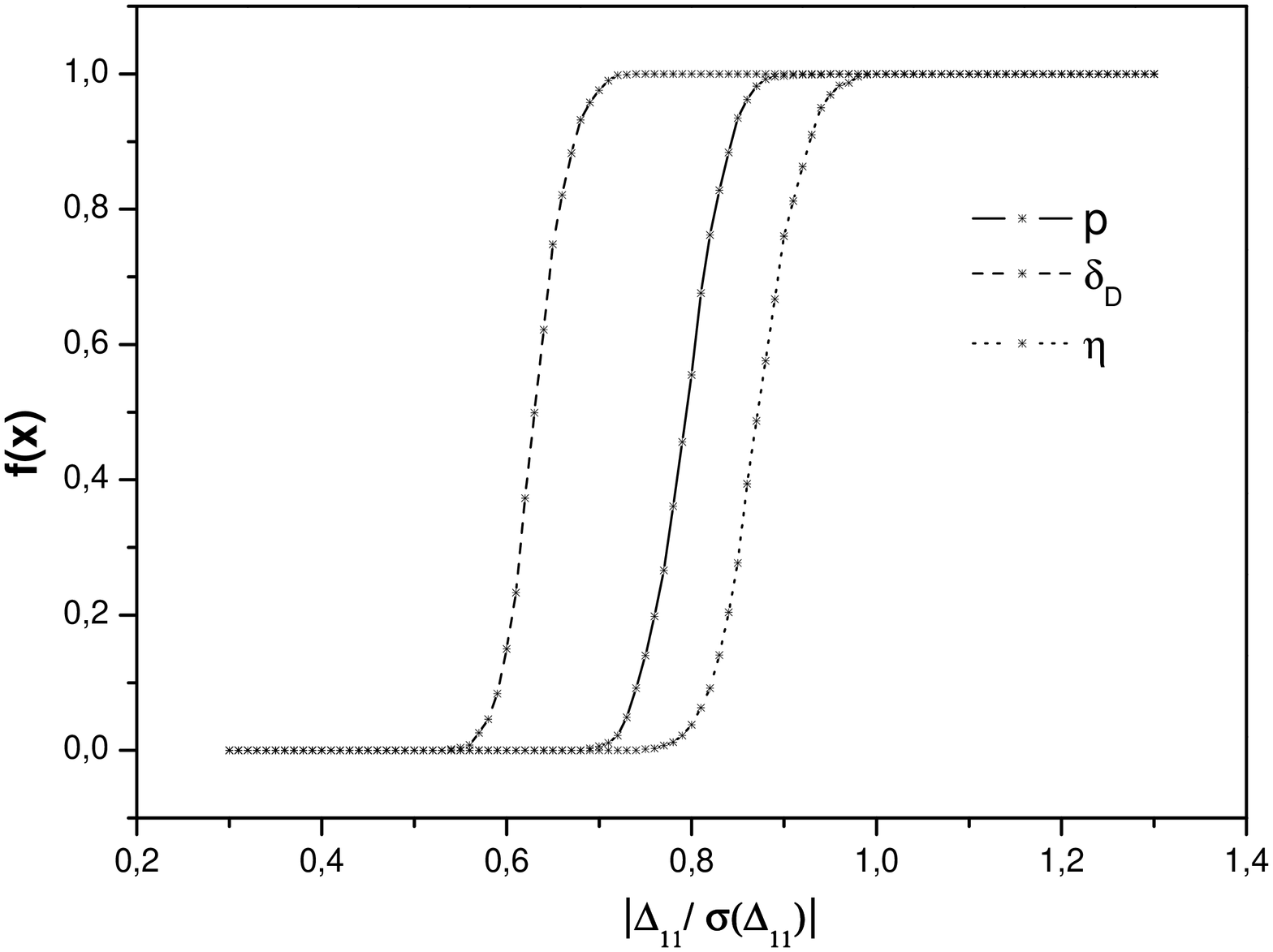}\\
\caption{The Probability Density Function (PDF) (left panel) and
Cumulative Distribution Function (CDF) (right panel) for analyzed statistics.
The figure was obtained from 1000 simulations of samples of 247 clusters
each with number of members galaxies the same as in the real clusters.
From up to down we present statistics:
$\Delta_{11}/\sigma(\Delta_{11})$ and $|\Delta_{11}/\sigma(\Delta_{11})|$
\label{fig:f4}}
\end{figure}

In contrary to analysis of the position angles $p$ where for each galaxy we have one 
solution, during the analysis of spatial orientation of galaxies we have two solutions for each 
galaxy i.e we have two possible values of angles $\delta_D$ and $\eta$ for each galaxy.
It should be pointed out, that till now, nobody  assumed that there could be any difference 
between the values of analyzed statistics in the case of the  $p$, $\delta_D$ and $\eta$ 
angles. However the analysis of the Tables \ref{tab:t1a} - \ref{tab:t1c} (as well as 
Figures \ref{fig:f1} - \ref{fig:f4}) indicates the presence of such difference.

The significance of the above observation can be investigated using Kolmogorov - Smirnov 
test. We have chosen for testing the statistics $\bar{\chi^2}$ 
because we have a good theoretical predictions about both the mean values and variances of 
this statistics. The analysis of \citet{g2011b}, caried in the case of analysis of the position 
angles $p$,  showed that the statistics $\bar{\chi^2}$  were normal  distributed (even though 
the $\chi^2$ statistics was  not normal distributed) with the mean and standard deviation as 
expected from theoretical analysis i.e. $E(\bar{\chi^2})=35$ and $\sigma^2(\bar{\chi^2})=0.2834$. 
In order to reject the $H_0$ hypothesis that the distribution is Gaussian with the mean 
value and variance as assumed, the value of observed statistics $\lambda$ should be greater 
than $\lambda_{cr}$. At the significance level $\alpha = 0.05$ the value $\lambda_{cr}$ = 1.358. 
In the case of analysis of position angles the obtained values of $\lambda$ statistic
was less than critical one, what means that we can not exclude the $H_0$ hyphothesis 
\citep{g2011b}. Now, the similar analysis for $\delta_D$ and $\eta$ angles shows the opposite 
result because the obtained values of $\lambda$ are greater than critical ones. As a result we 
were able to exclude the hypothesis that distributions are Gaussian with theoretical parameter 
as noted above. 

The next step is to test the new hypothesis $H_0$ that the analyzed statistics are normally 
distributed with parameters as obtained from simulations. During such an investigation the 
problem that usually arise is, as shown by \citet{Massey51} and  \citet{Lilliefors67}, 
that the standard tables used for the Kolmogorov-Smirnov test are valid only in the case of 
analysis a completely specified continuous distribution. When we test if the distribution 
is normal, but parameters of the distribution are estimated from the sample, the modification
of the classical Kolmogorov-Smirnov test, known as Kolmogorov - Lilliefors test, should be used 
instead \citep{Lilliefors67}. The significance of this problem for investigation of 
the galaxy alignment was discused in detail in \citet{g2011b}. 

In the present analysis we conclude that in the case of all $11$ analyzed  statistics the 
values of  $\lambda$ are significantly less than critical values $D_{cr}$ (for our case 
i.e. $n=1000$  and the significance level $\alpha = 0.05$, $D_{cr}=0.028$, \citet{g2011b}) which
means that we can not  exclude our $H_0$ hypothesis. Summarizing, we conclude that the obtained 
results are not in conflict with our prediction that the statistics is normally distributed 
with parameters as obtained from simulations.

Because of relatively small numbers of galaxies in some clusters, we repeated our 
analysis with 1000 simulations of 247 fictitious clusters, each cluster with the 
number of member galaxies the same as in the real clusters (Figures \ref{fig:f1} 
- \ref{fig:f4}). It is easy to see the differences between distributions of analyzed 
statistics for $p$, $\delta_D$ and $\eta$ angles. One could observe that usually the 
analysis of the $\eta$ angles gives the higher values of observed statistics as in other 
cases. The exception is the analysis of $\Delta_{11}/\sigma(\Delta_{11})$ statistics, 
where for all  analyzed  angles PDF and CDF are very similar.

One should note that we have performed this procedure twice, first with galaxies in the 
clusters with coordinates distributed as in the real clusters and second independently 
for galaxies randomly distributed around the whole celestial sphere. 
Firstly, we have compared the distribution of the position angles $p$ and the results are
presented in the Table \ref{tab:t2a}. When we analyzed the sample of 2360 galaxies the
difference between the case of galaxies in clusters distributed as in the real clusters 
and the case of galaxies randomly distributed around the whole celestial sphere is in all
cases less than $2\,\sigma$ and usually is on $1\,\sigma$ level. For the sample of clusters 
with real number of galaxies the situation is similar, but the differences are
a little bit higher, up to $3\,\sigma$. The exception is only for $\Delta/\sigma(\Delta)$ and
for $\lambda_c$ statistics. The reason is (as it is noted above) that the variance 
of $\Delta/\sigma(\Delta)$ is obtained from linear approximation hence the mean value 
of $\Delta/\sigma(\Delta)$ is only approximated, while the simulated value of $\lambda_c$ 
depends on the binning process. When we compared the results of the statistics obtained for  
cluster with the real number of galaxies with that obtained for fictitious cluster 2360 
galaxies each, the differences between statistics are typically on the $2\,\sigma$ level,
but in any case are less than $3\,\sigma$. 
The above results have clearly showed that during 
analysis of the alignment of galaxies in clusters, we had to compare the observational distribution 
of the analyzed angles with the results of numerical simulations based on catalogues that contain 
the clusters populated the same as real ones, but could not base only on pure theoretical 
predictions. Of course, a good simulated catalog should also have the same possible systematic 
effect as the real one.

\begin{figure}
\includegraphics[angle=270,scale=0.30]{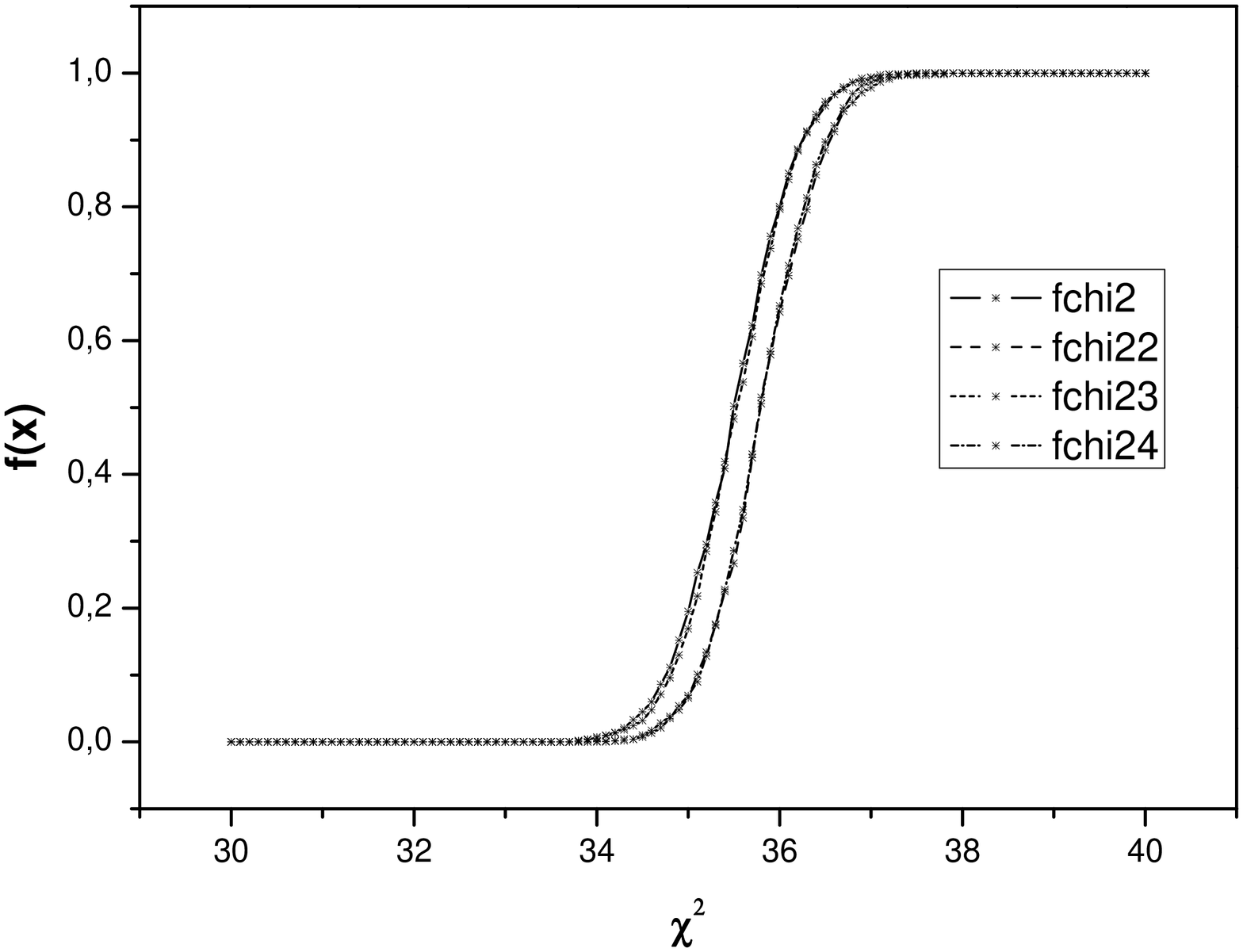}
\includegraphics[angle=270,scale=0.30]{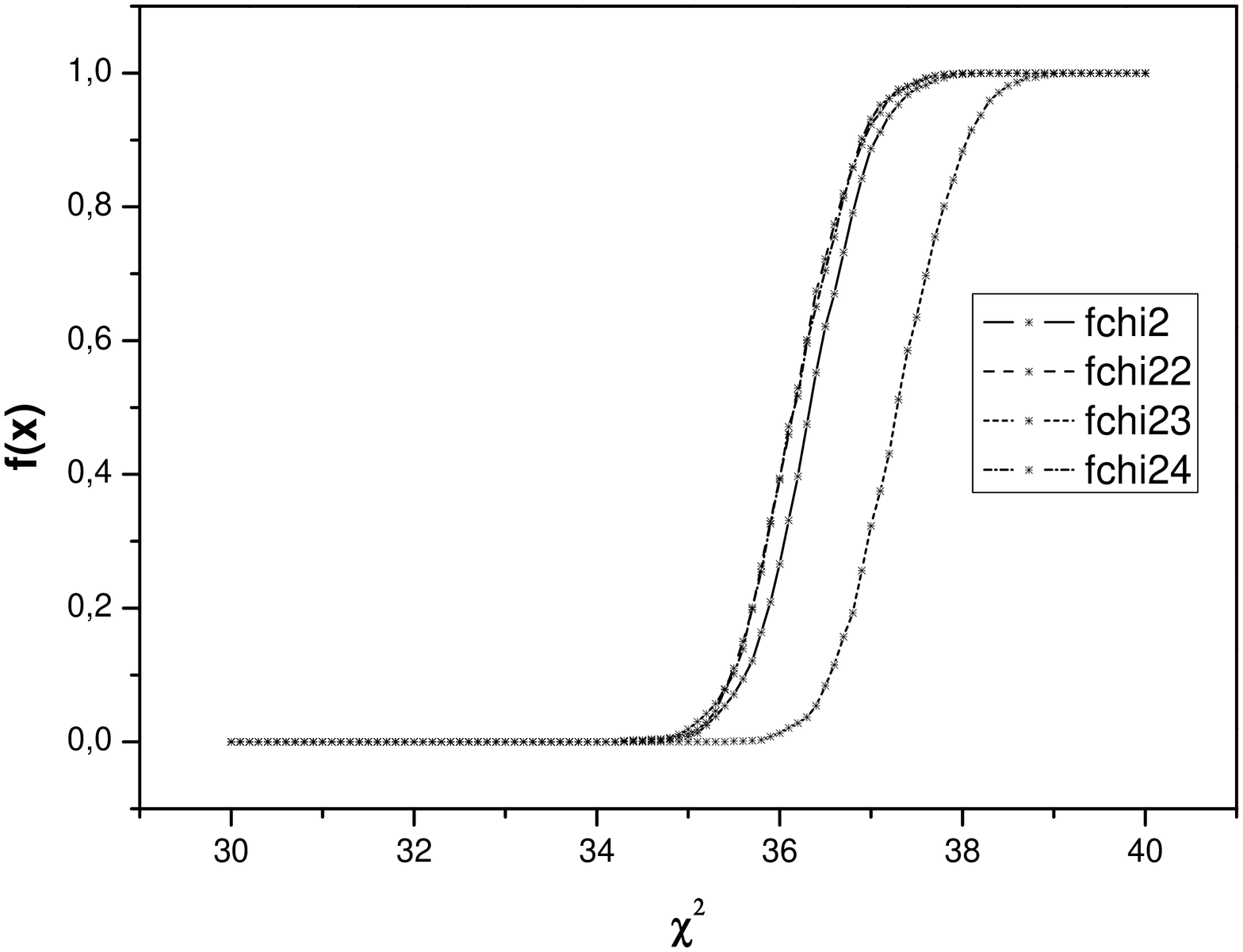}\\
\includegraphics[angle=270,scale=0.30]{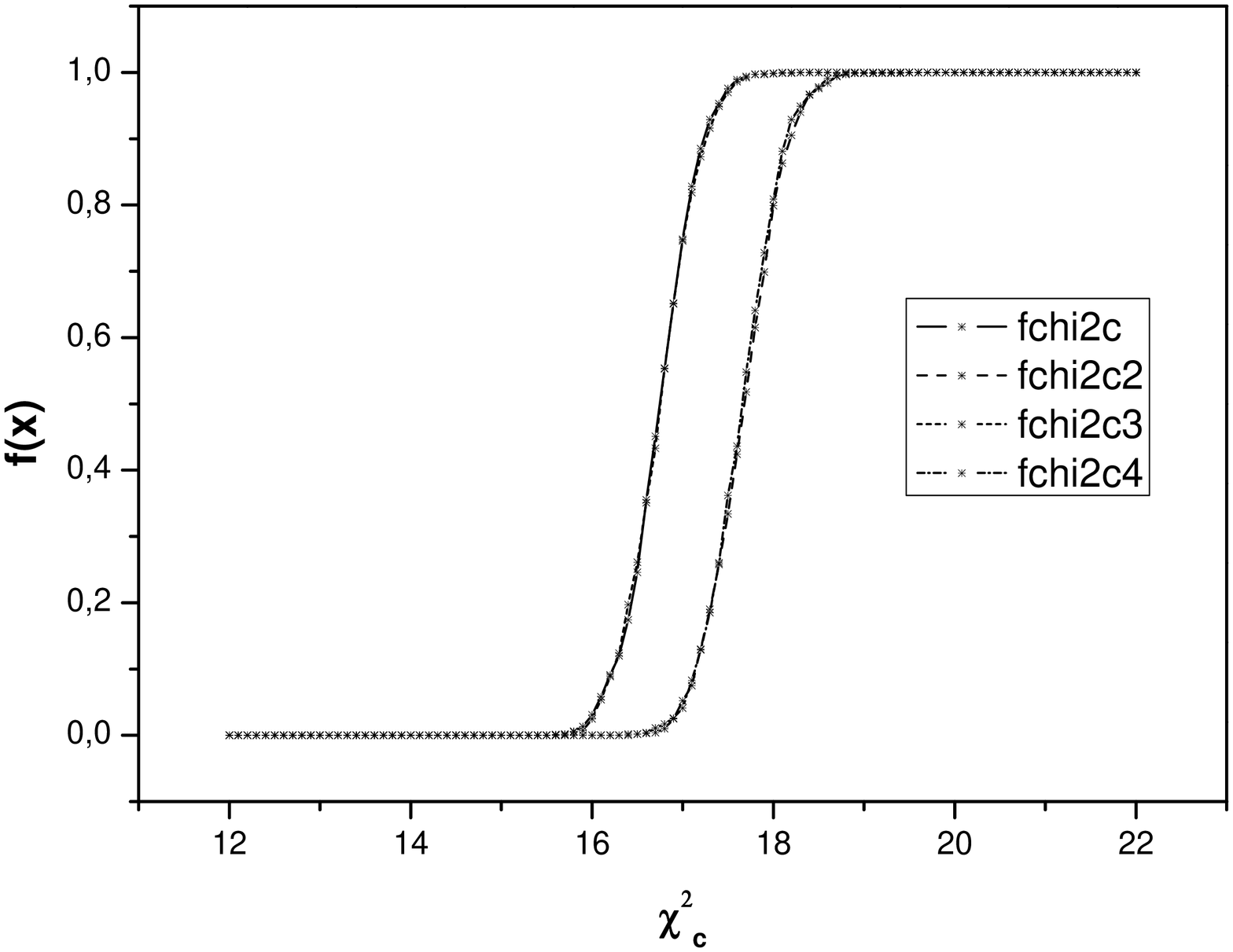}
\includegraphics[angle=270,scale=0.30]{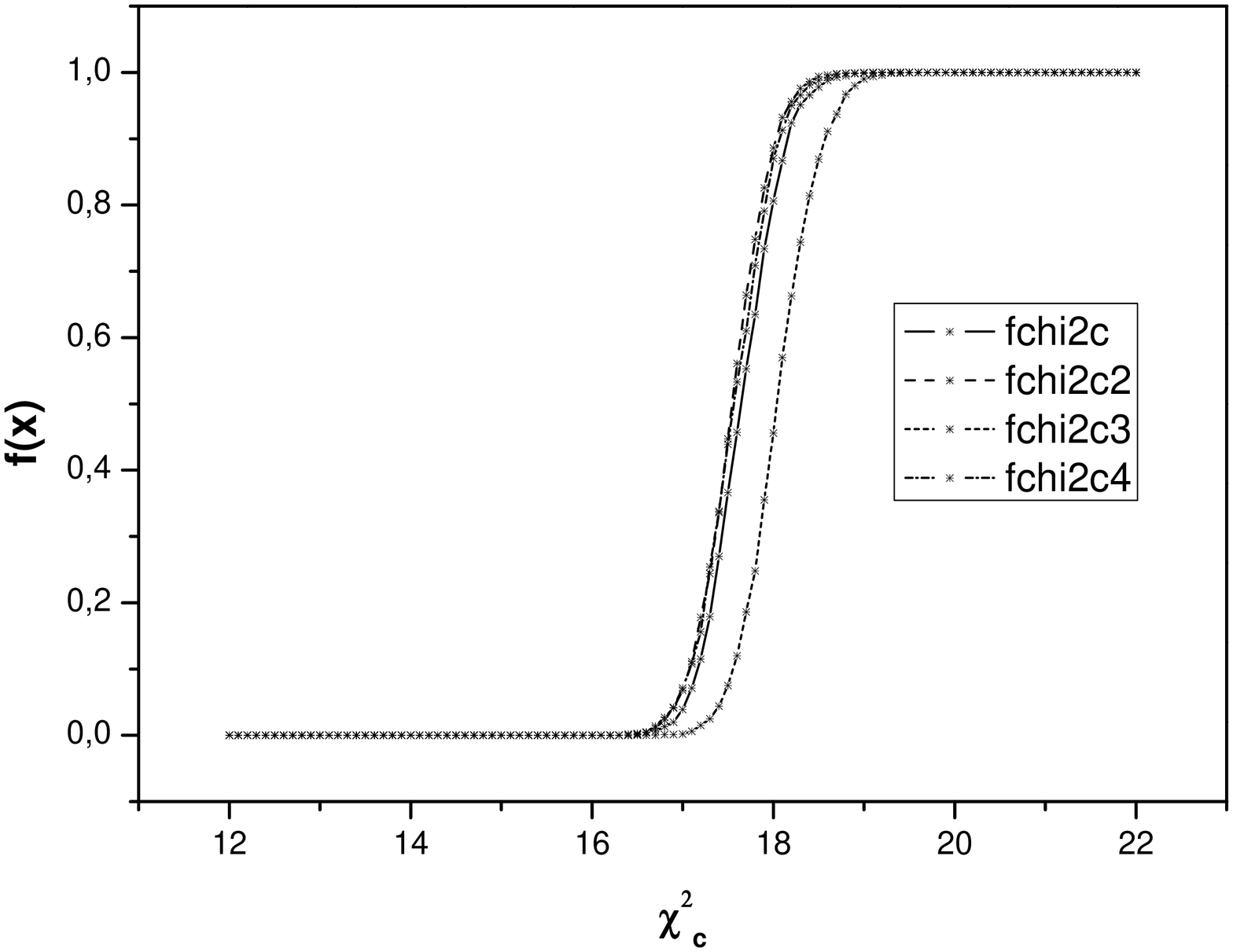}\\
\includegraphics[angle=270,scale=0.30]{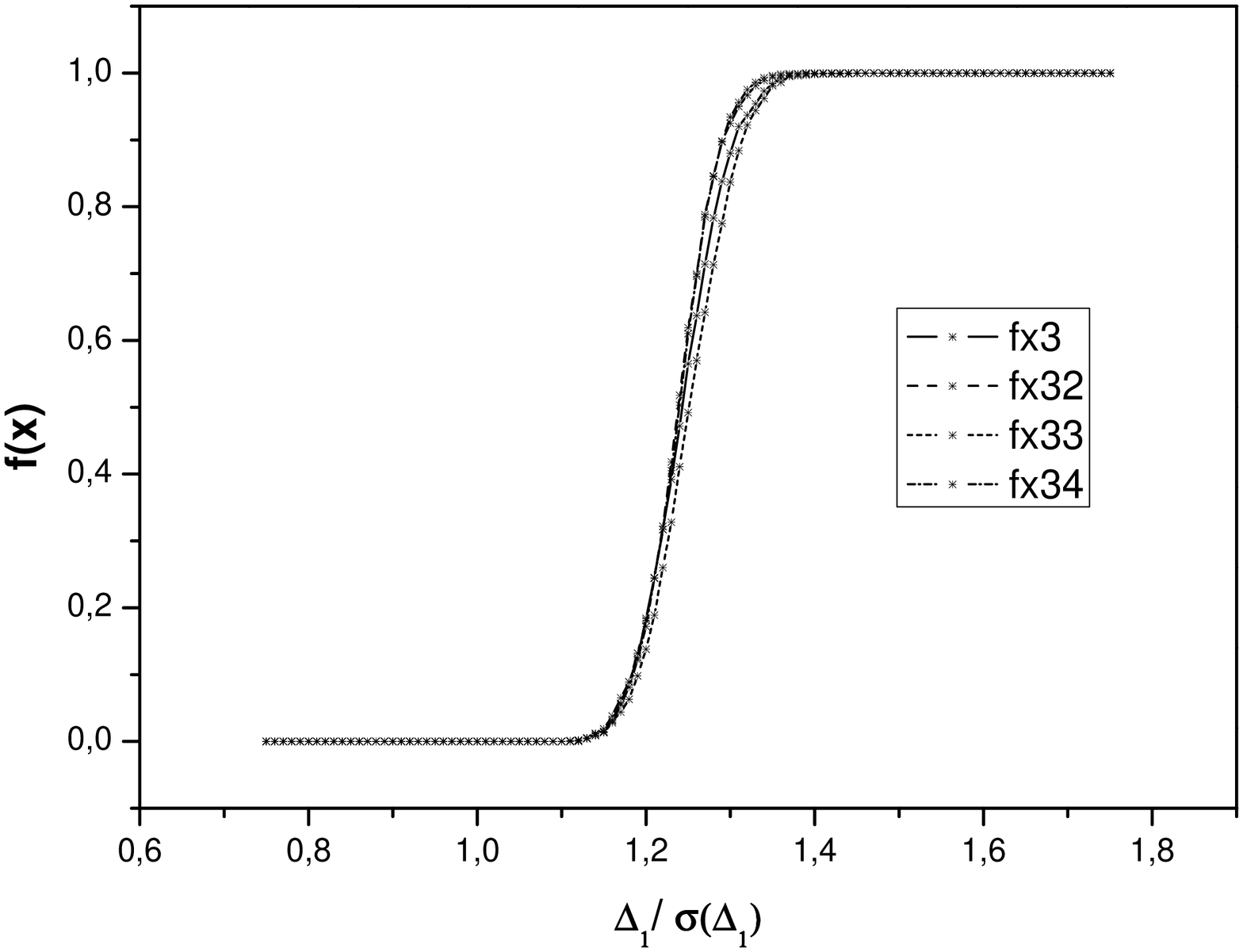}
\includegraphics[angle=270,scale=0.30]{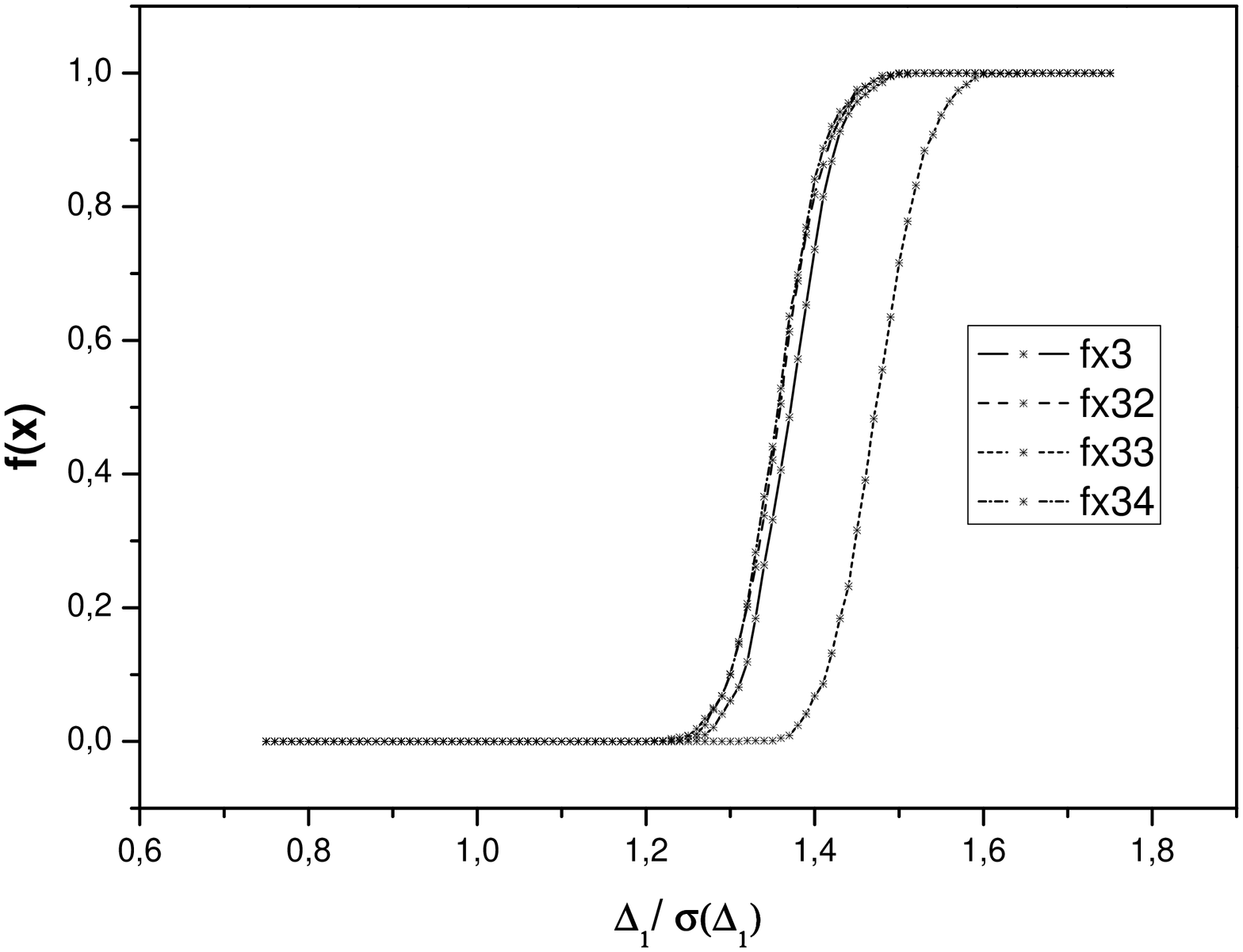}\\
\caption{The Cumulative Distribution Function (CDF) for
$\delta_D$ (left panel) and $\eta$ (right panel) for analyzed statistics.
The figure was obtained from 1000 simulations of samples of 247 clusters.
Each simulation was done 4 times, with the number of members galaxies the 
same as in the real cluster, and with 2360 Galaxies. In both cases we used 
coordinates distributed as in the real clusters and  independently coordinates 
of galaxies randomly distributed around the whole celestial sphere.
From up to down we present statistics:
$\chi^2$,$\chi_c^2$, $\Delta_{1}/\sigma(\Delta_{1})$.
\label{fig:f5}}
\end{figure}

\begin{figure}
\includegraphics[angle=270,scale=0.30]{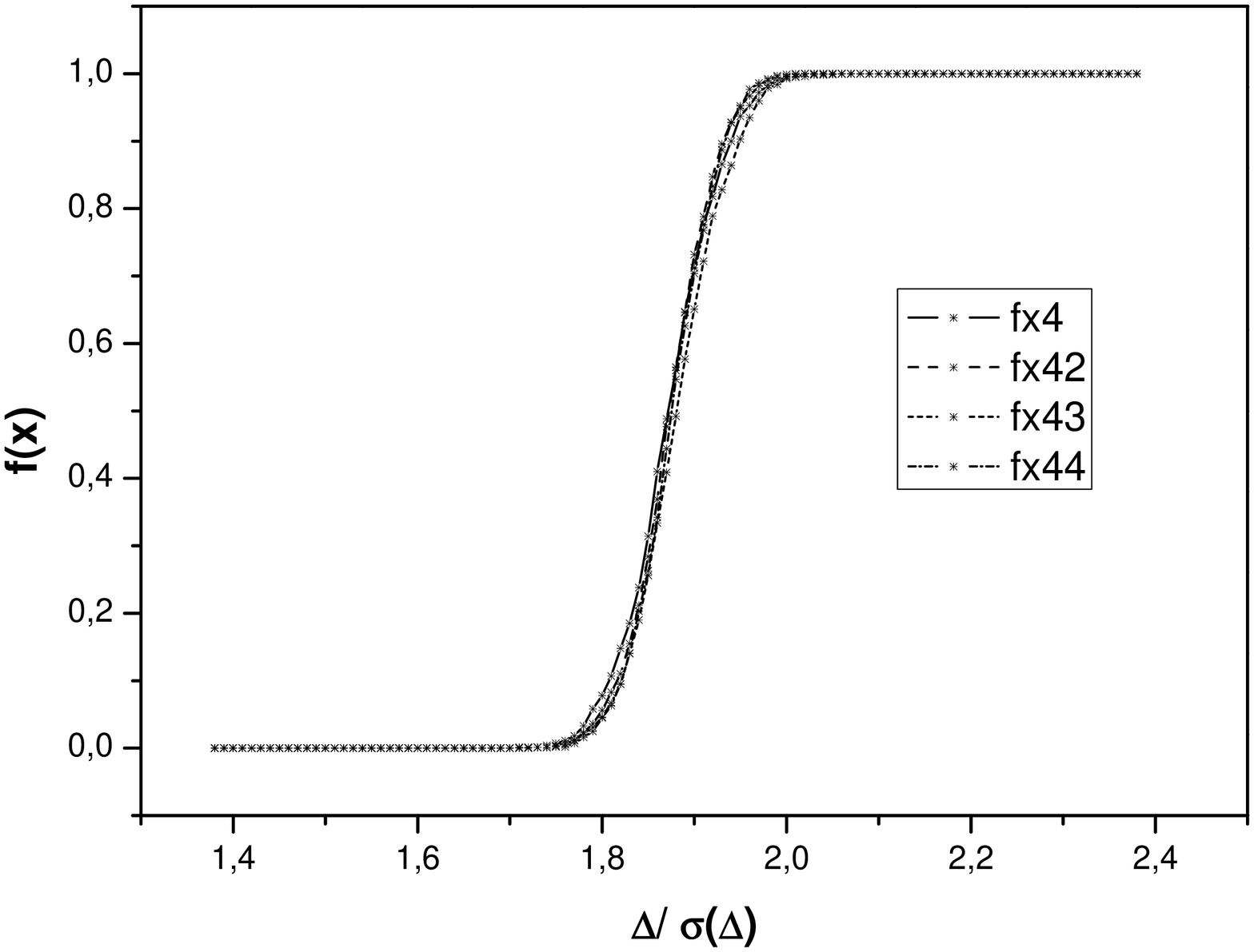}
\includegraphics[angle=270,scale=0.30]{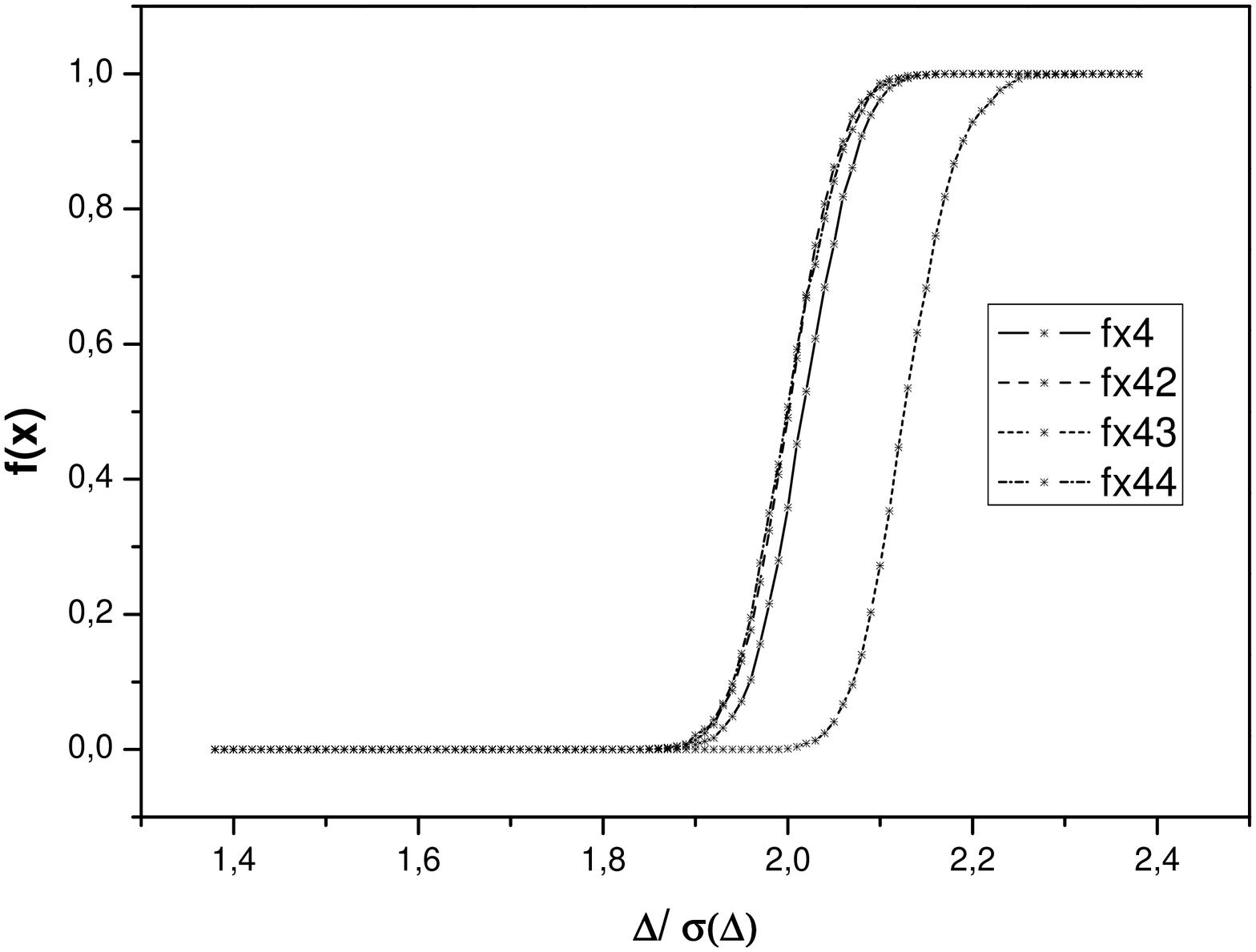}\\
\includegraphics[angle=270,scale=0.30]{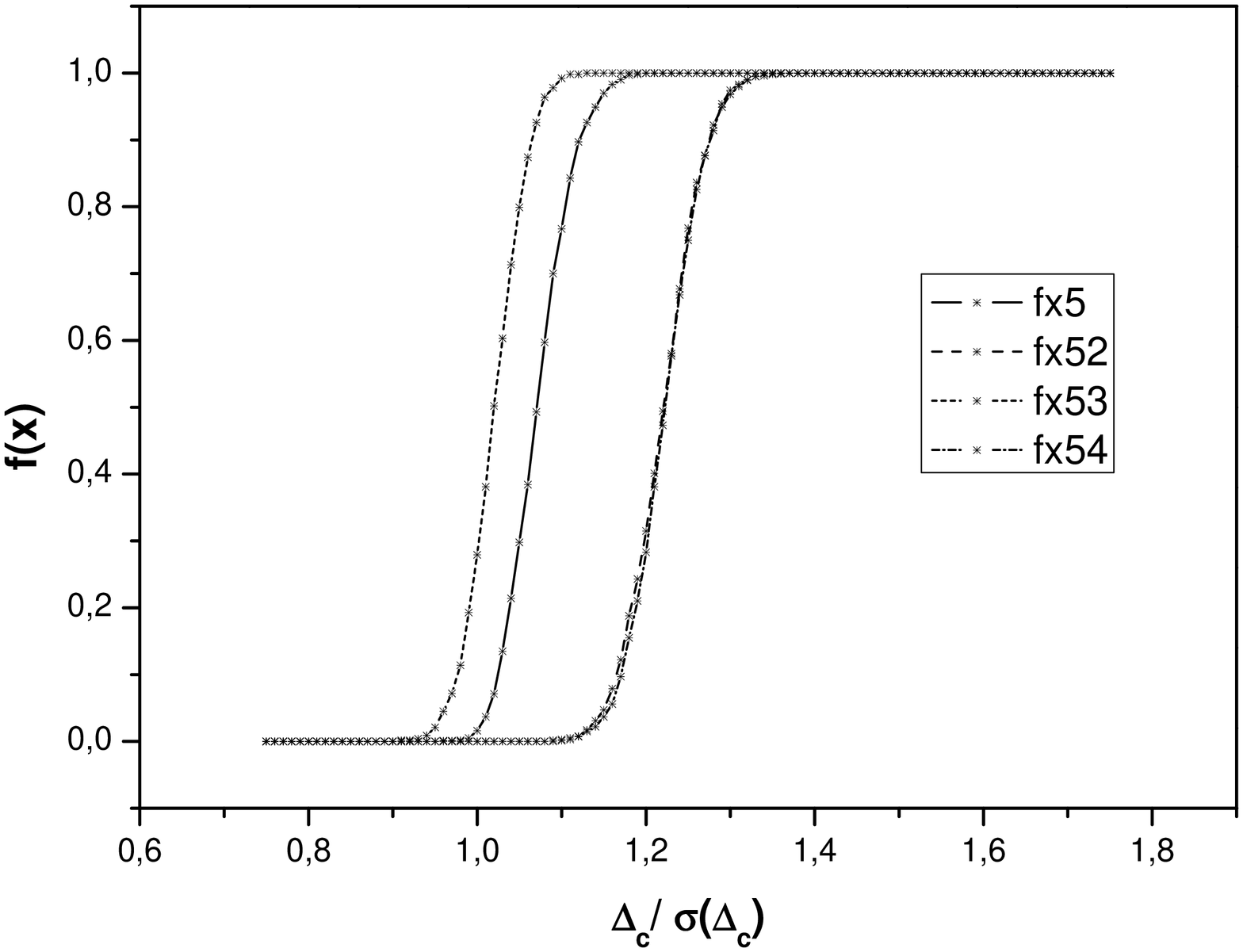}
\includegraphics[angle=270,scale=0.30]{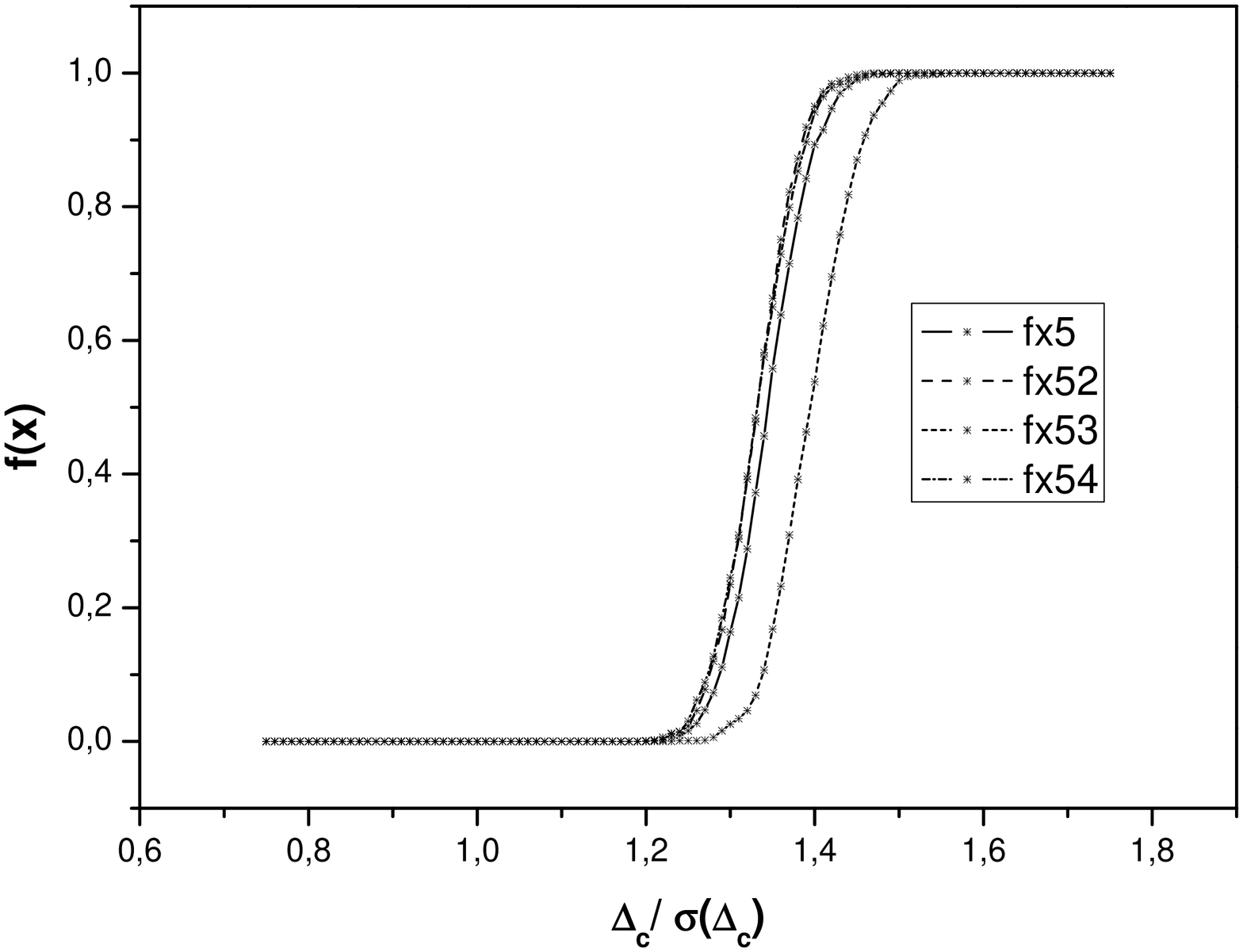}\\
\includegraphics[angle=270,scale=0.30]{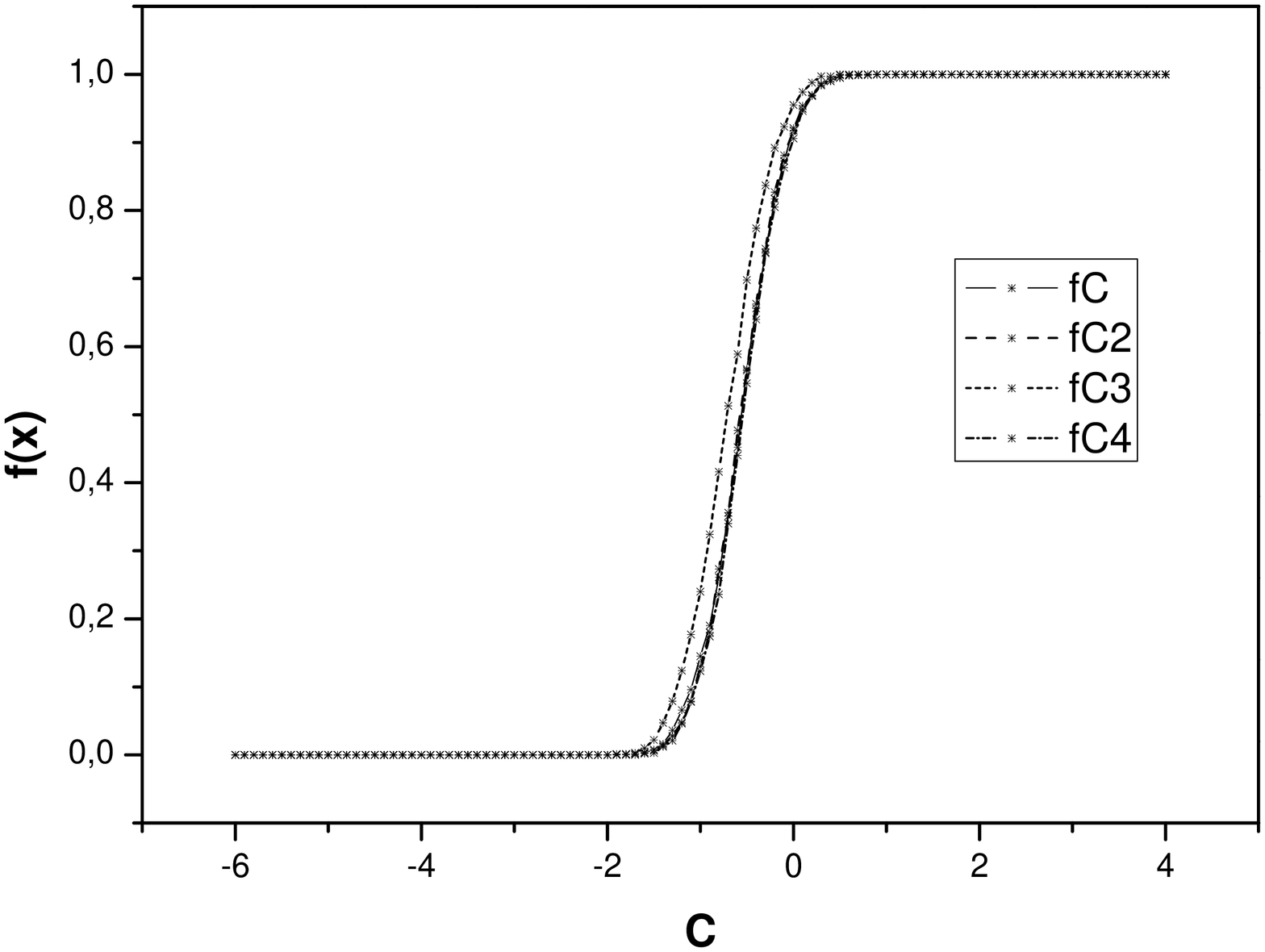}
\includegraphics[angle=270,scale=0.30]{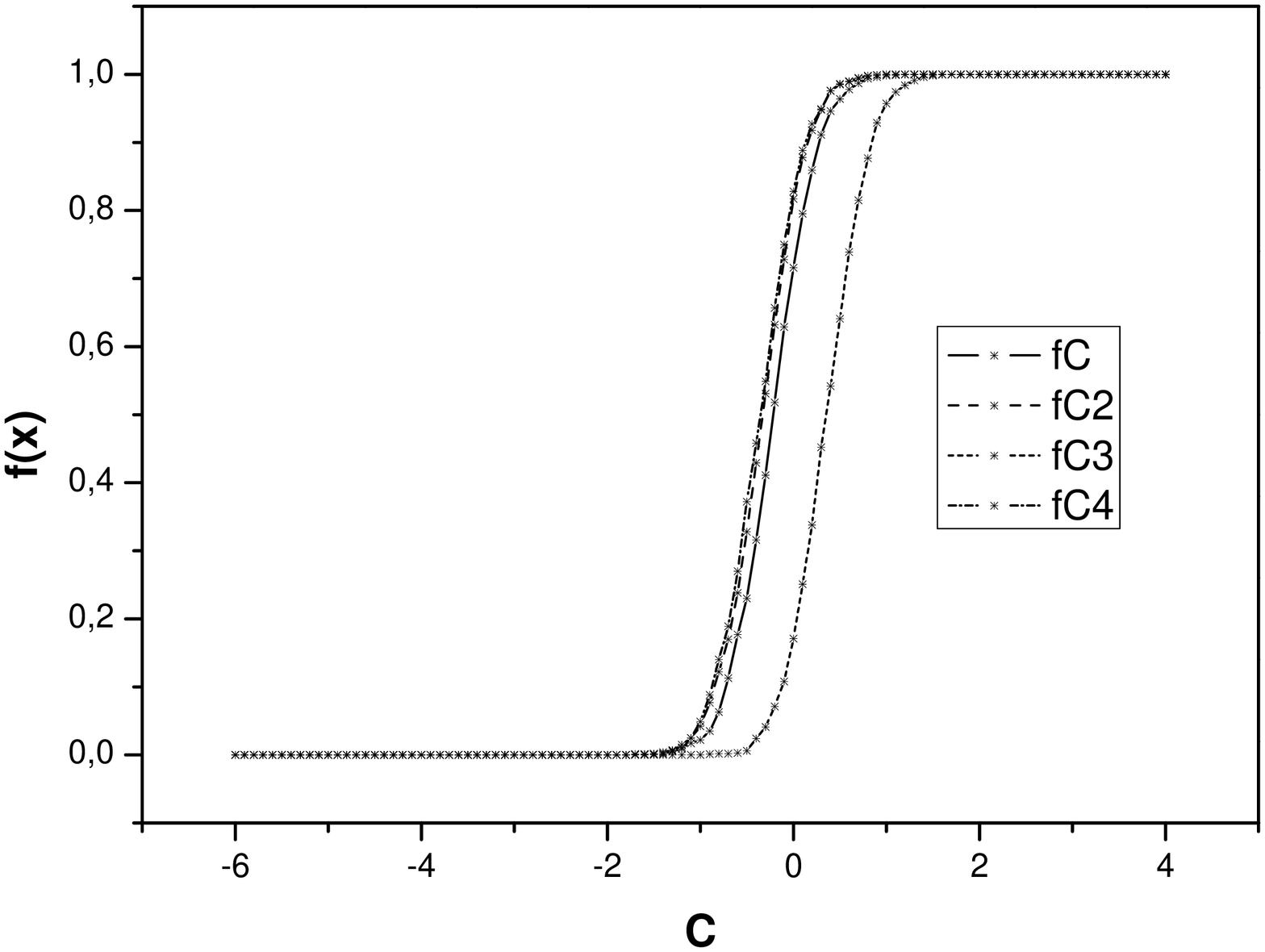}\\
\caption{The Cumulative Distribution Function (CDF) for
$\delta_D$ (left panel) and $\eta$ (right panel) for analyzed statistics.
The figure was obtained from 1000 simulations of samples of 247 clusters.
Each simulation was done 4 times, with the number of members galaxies the 
same as in the real cluster, and with 2360 Galaxies. In both cases we used 
coordinates distributed as in the real clusters and  independently coordinates 
of galaxies randomly distributed around the whole celestial sphere.
From up to down we present statistics:
$\Delta/\sigma(\Delta)$,$\Delta_c/\sigma(\Delta_c)$, $C$.
\label{fig:f6}}
\end{figure}

\begin{figure}
\includegraphics[angle=270,scale=0.30]{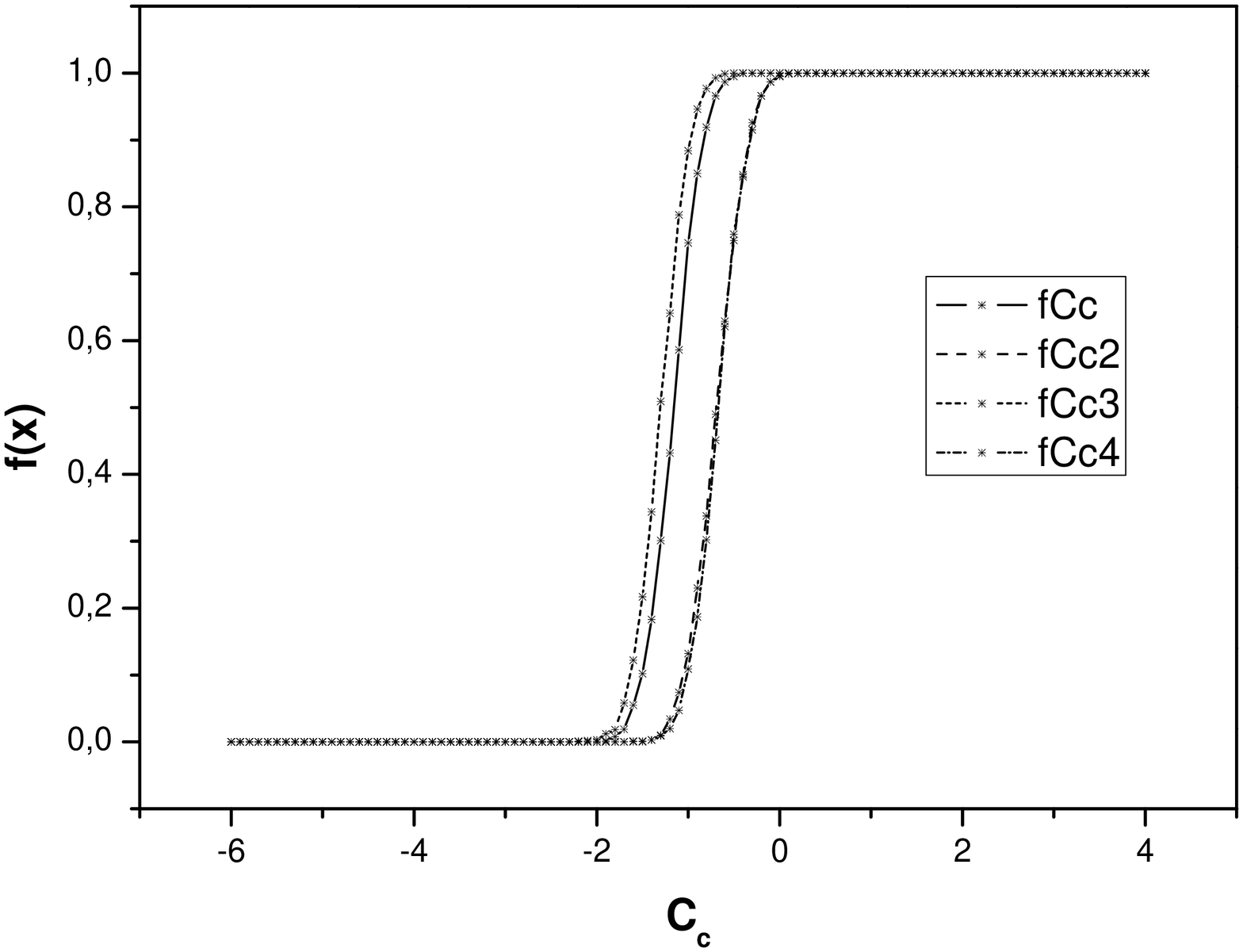}
\includegraphics[angle=270,scale=0.30]{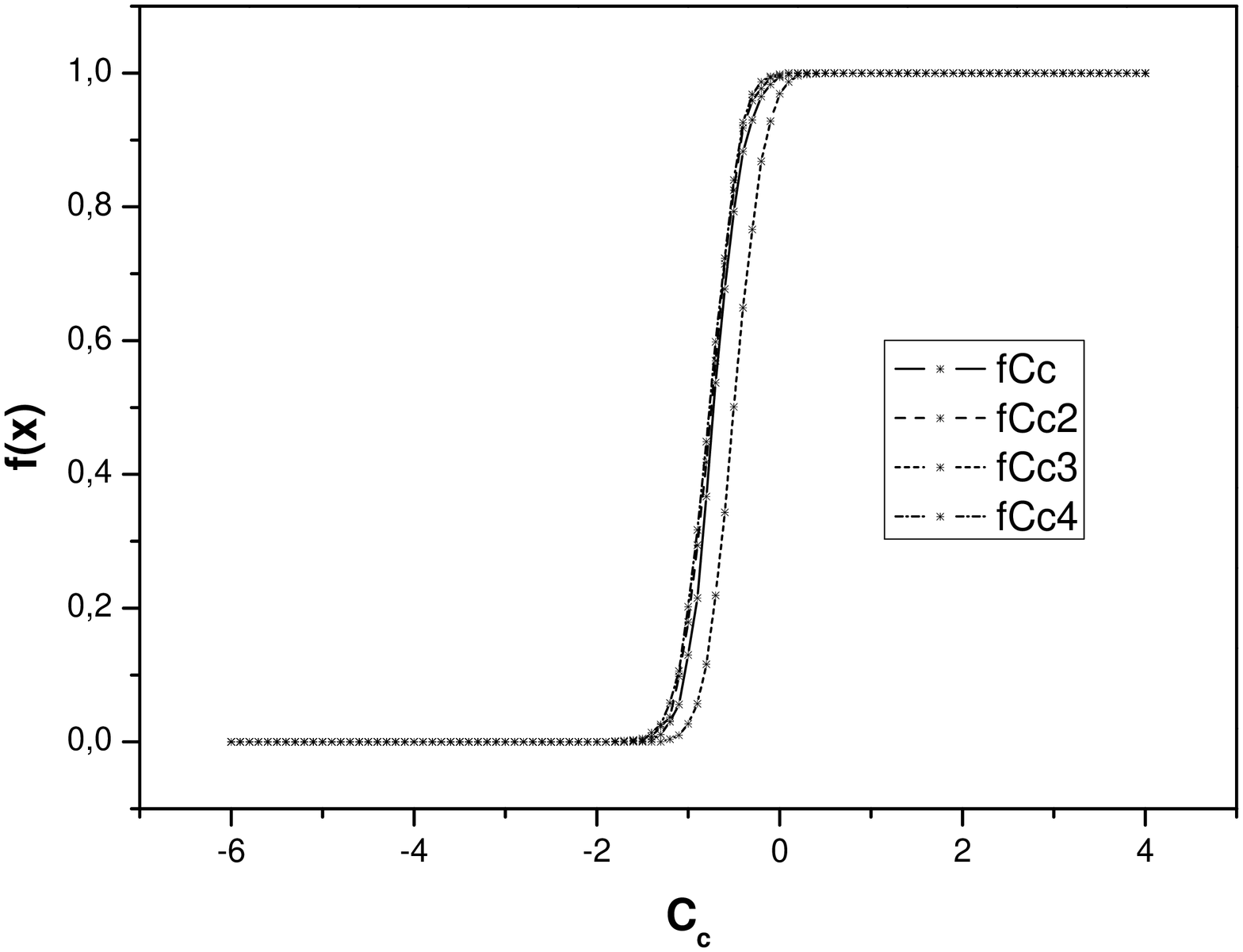}\\
\includegraphics[angle=270,scale=0.30]{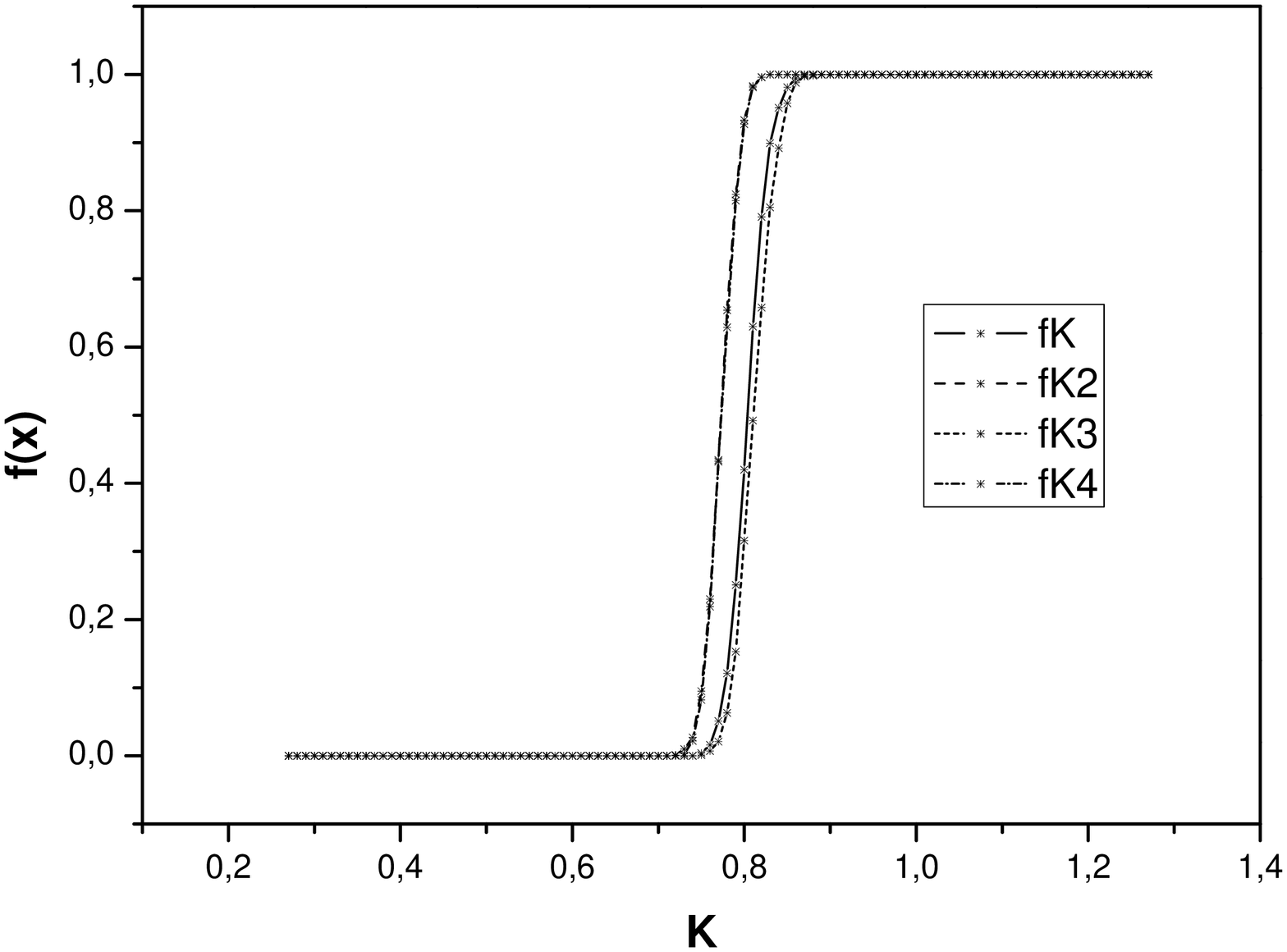}
\includegraphics[angle=270,scale=0.30]{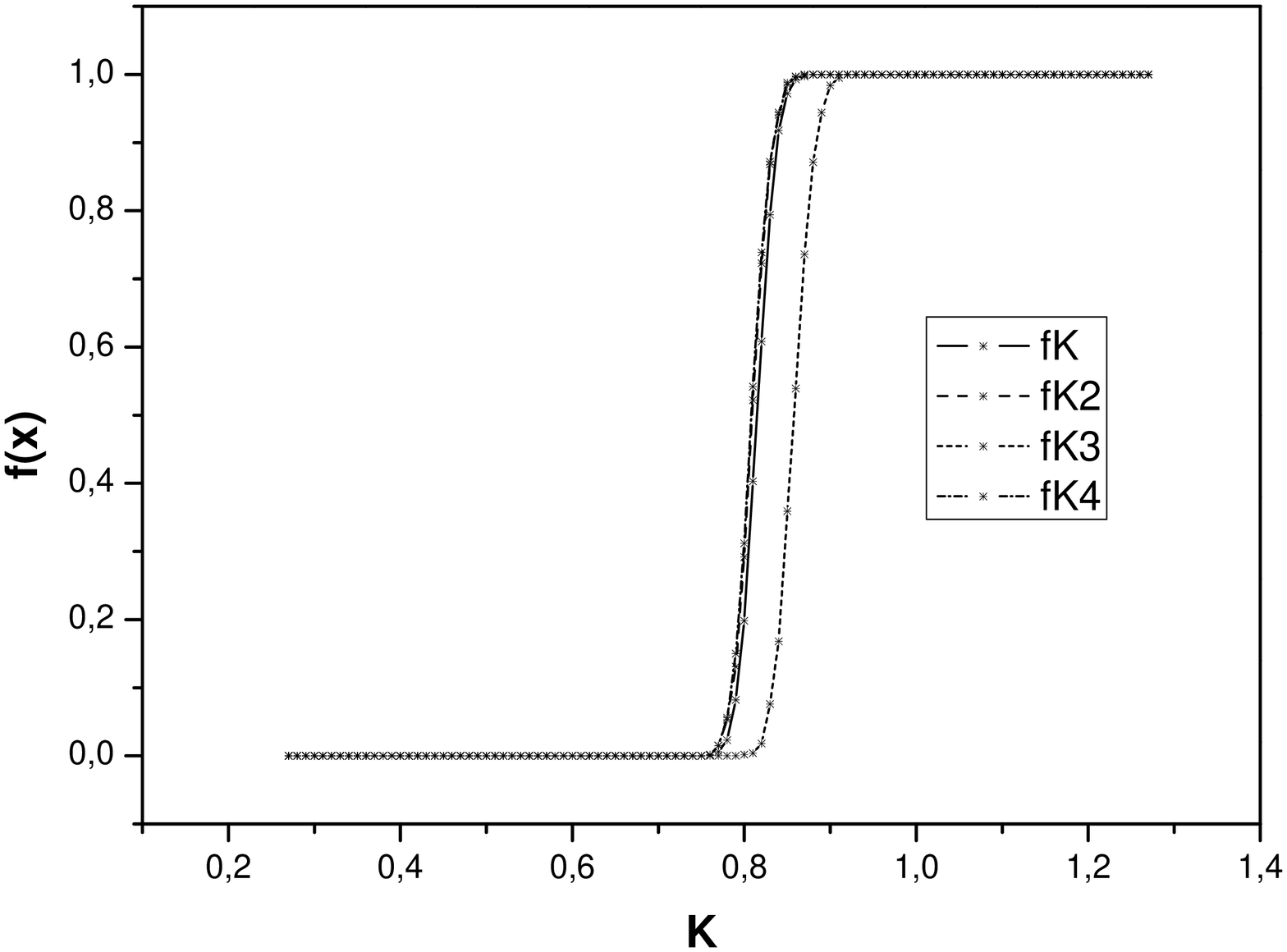}\\
\includegraphics[angle=270,scale=0.30]{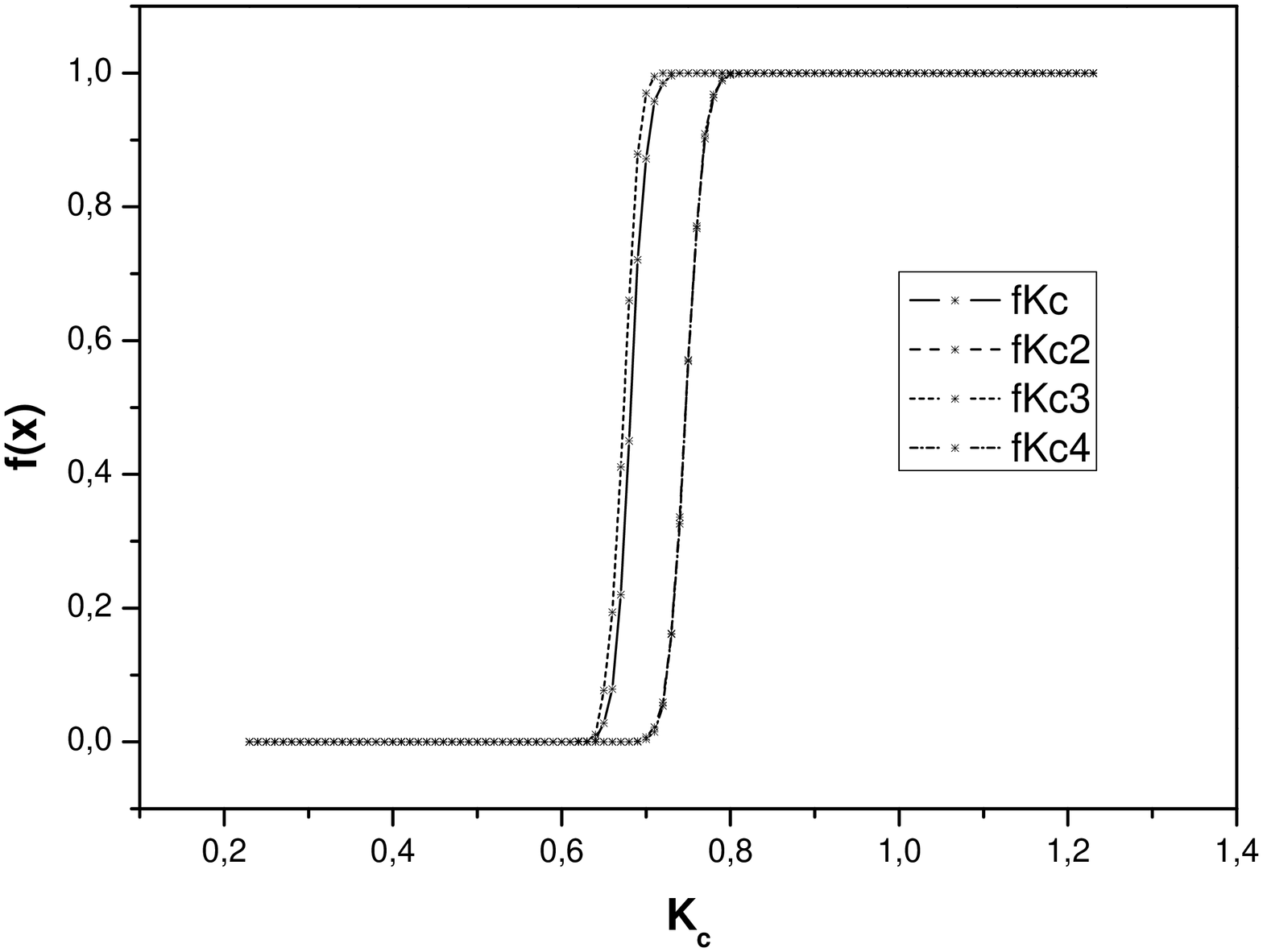}
\includegraphics[angle=270,scale=0.30]{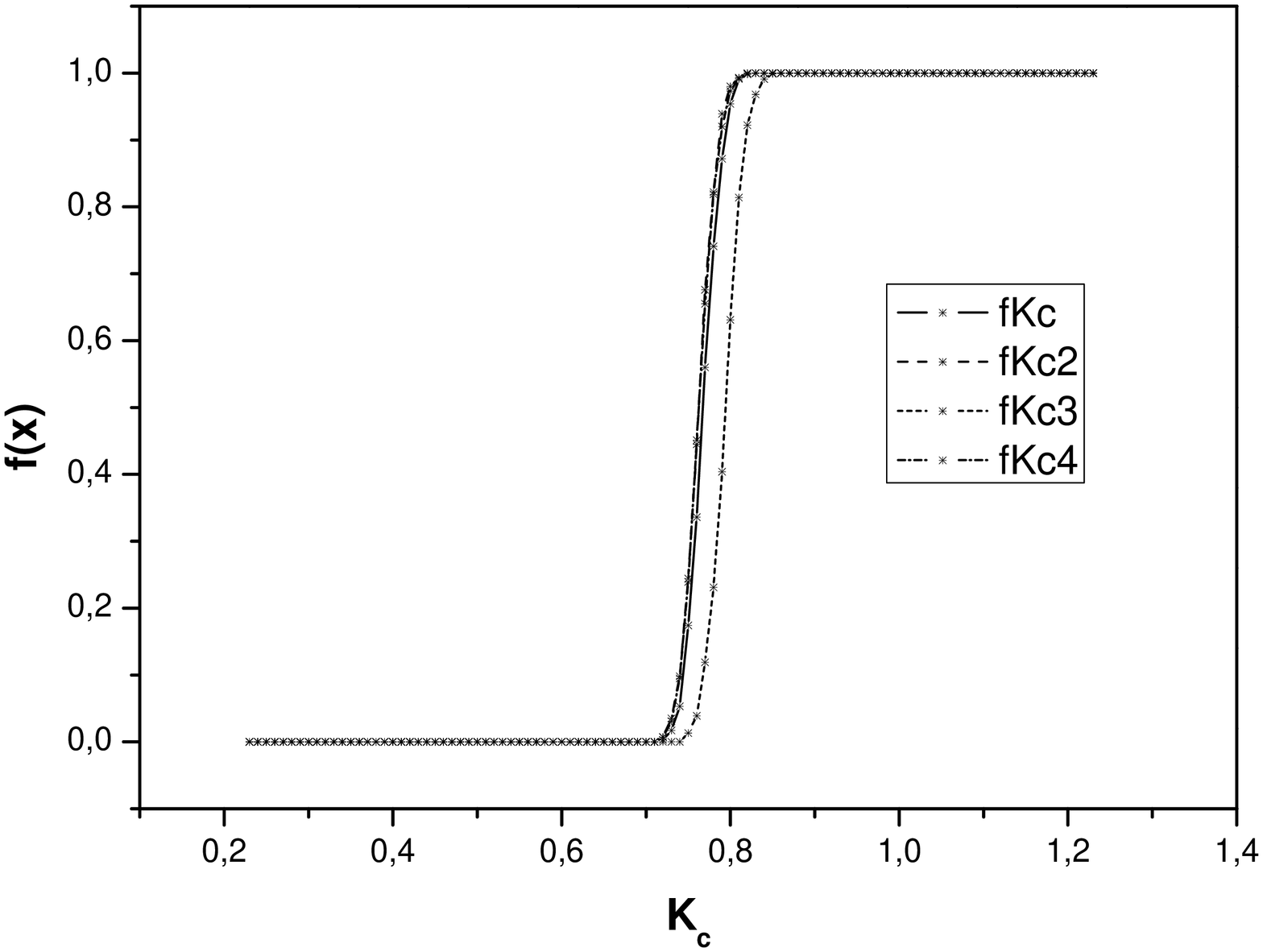}\\
\caption{The Cumulative Distribution Function (CDF) for
$\delta_D$ (left panel) and $\eta$ (right panel) for analyzed statistics.
The figure was obtained from 1000 simulations of samples of 247 clusters.
Each simulation was done 4 times, with the number of members galaxies the 
same as in the real cluster, and with 2360 Galaxies. In both cases we used 
coordinates distributed as in the real clusters and  independently coordinates 
of galaxies randomly distributed around the whole celestial sphere.
From up to down we present statistics:
$C_c$, $\lambda$, $\lambda_c$.
\label{fig:f7}}
\end{figure}

\begin{figure}
\includegraphics[angle=270,scale=0.30]{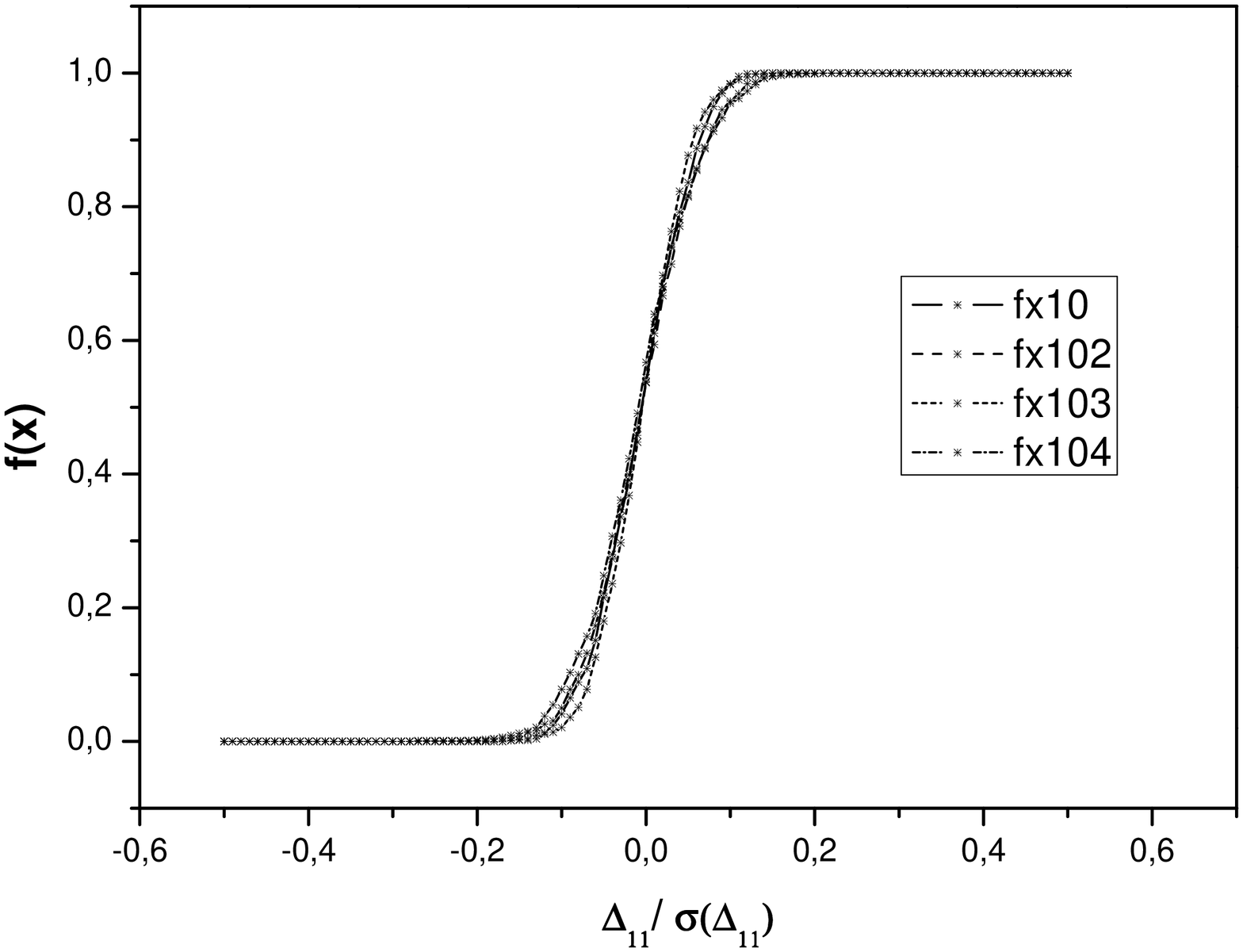}
\includegraphics[angle=270,scale=0.30]{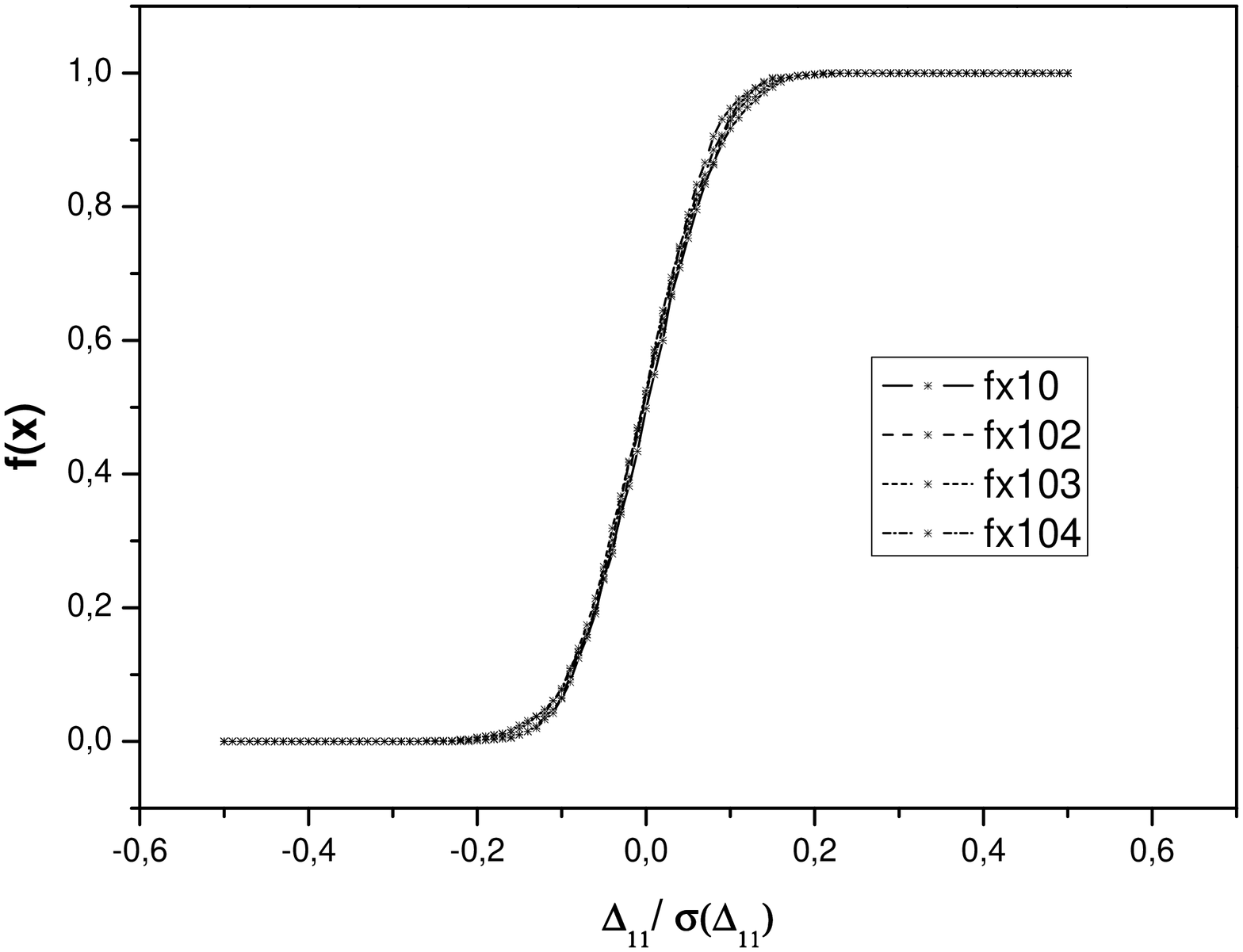}\\
\includegraphics[angle=270,scale=0.30]{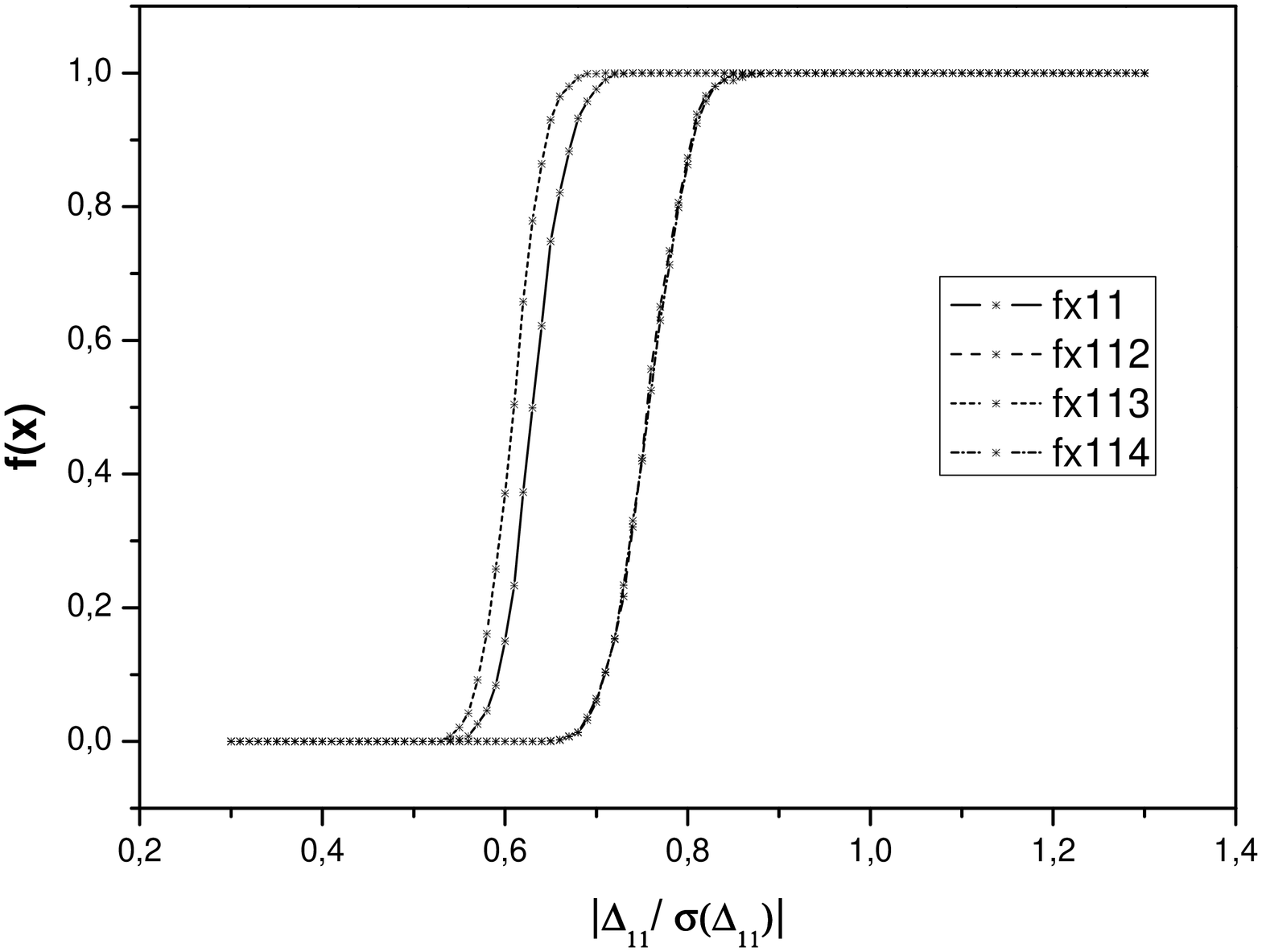}
\includegraphics[angle=270,scale=0.30]{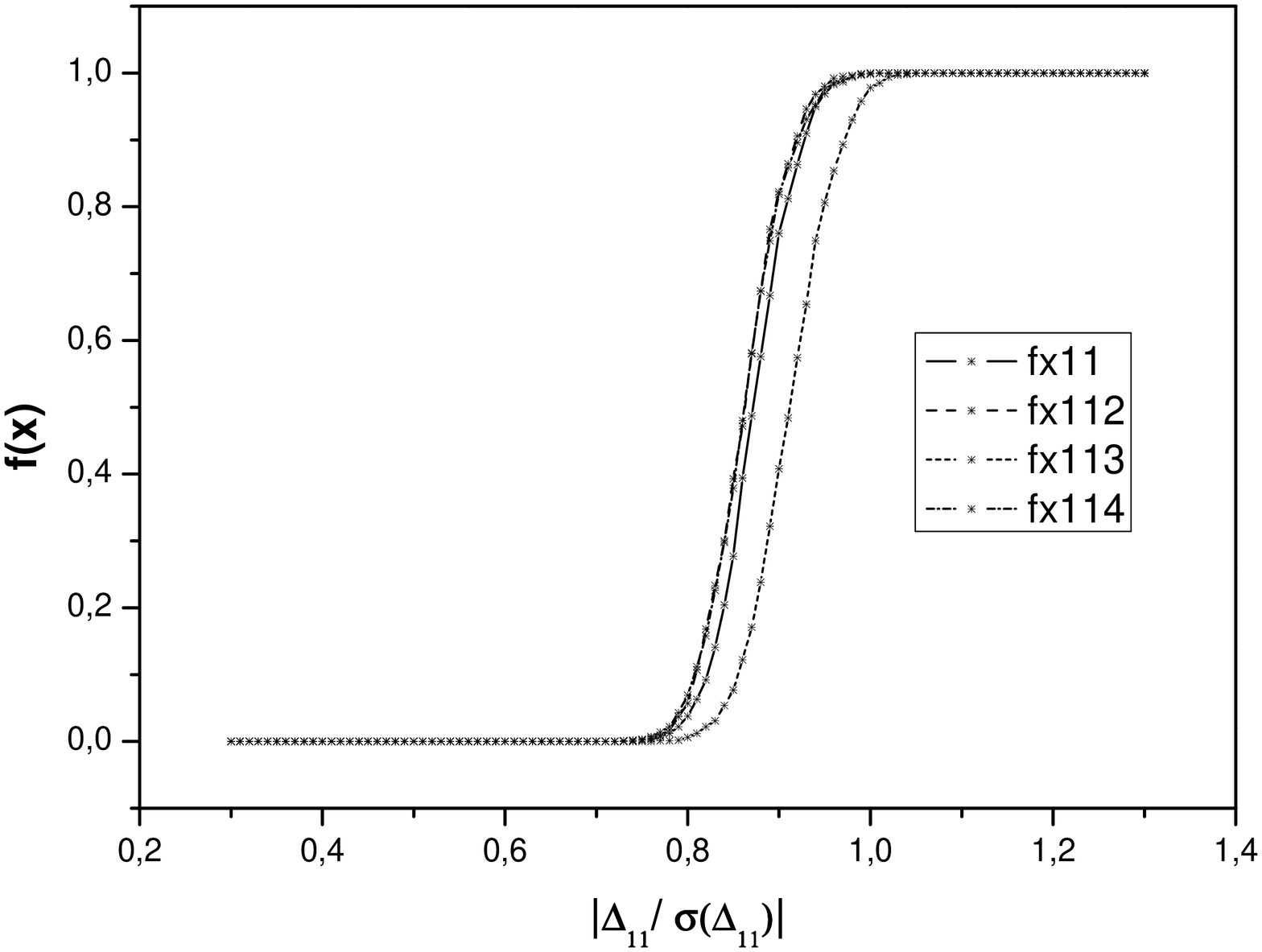}\\
\caption{The Cumulative Distribution Function (CDF) for
$\delta_D$ (left panel) and $\eta$ (right panel) for analyzed statistics.
The figure was obtained from 1000 simulations of samples of 247 clusters.
Each simulation was done 4 times, with the number of members galaxies the 
same as in the real cluster, and with 2360 Galaxies. In both cases we used 
coordinates distributed as in the real clusters and  independently coordinates 
of galaxies randomly distributed around the whole celestial sphere.
From up to down we present statistics:
$\Delta_{11}/\sigma(\Delta_{11})$ and $|\Delta_{11}/\sigma(\Delta_{11})|$
\label{fig:f8}}
\end{figure}

Result of analogous analysis for angles giving the spatial orientation of galaxies i.e 
angles $\delta_D$ and $\eta$ are presented in Tables \ref{tab:t2b} - \ref{tab:t2c} 
and in Figures \ref{fig:f5} - \ref{fig:f8}. 
Again we have observed the differences between the cases when we analyzed a huge populated 
cluster and the cluster with the real number of galaxies in clusters as well as in the case 
of cluster with galaxies distributed around the whole celestial sphere and the case 
when the galaxies are distributed as in the real clusters only with the exception of 
$\Delta_{11}/\sigma(\Delta_{11})$ statistic.  The crucial observation is that the differences 
in the latter cases are much higher than in the former one. The presence  of the above differences 
shows that the results of analysis of alignment in real clusters should be rather compared 
with the numerical simulations instead of pure theoretical predictions. In our opinion 
the reason for such differences is  mostly caused by the fact that during the process 
of deprojection of the spatial orientation of galaxies from its optical images we 
obtain two possible orientations - see Equations \ref{eq:a1} - \ref{eq:a3}. From 
analysis of these equations it is easy to see that solutions are not independent 
and as a  result the distribution of analyzed statistics is modified and must be 
obtained from numerical simulations.

For the  investigation of alignment we are able to analyze the distribution of the $p$,
$\delta_D$ and $\eta$ angles. Unfortunately, if we want to analyze the distribution of 
the real values, the following problem of $\delta_D$ and $\eta$ angles arises. If we do 
not know the morphological types of galaxies, we have to assume the real axial ratio. 
This is usually done by assuming, during calculation the inclination angle, the average 
value $q_0=0.2$, but then the effect of deprojection masks any possible alignment as it 
is shown in \citet{g5,g2011a,g11c,Paj12}. 
In the above papers it was shown that this problem can be solved when we know the
morphological type of individual galaxies and use true values $ q_0$ depending 
on the morphological type (according to  \citep{Hei72} with the help of  \citep{Fou85}  
corrections of $q$ to standard photometrical axial ratios) of axial ratio instead of the
average value $q_0=0.2$. Unfortunately MRSS \citep{MRSS03} does not provide information 
about morphological type of individual galaxies. 
Therefore  in the present paper we have analyzed, like in \citet{g2011b}, only the distribution of the 
position angles $p$ in the sample ($A$) of 247 rich Abell clusters both in Equatorial 
and Supergalactic coordinate systems. 
Moreover we have analyzed the restricted sample ($B$) in which only galaxies brighter than 
$m_3+3^m$ are taken into account. The results are presented in Table \ref{tab:t3}.

Our null hypothesis $H_0$ is that the mean value of the analyzed statistics is as expected 
in the cases of a random distribution of the position angles, against $H_1$ hypothesis 
that the analyzed values are different than predicted in the case of random distribution. 
For nearly all performed  tests the result are significant on at least $3\,\sigma$ level. 
In all cases there are no significant differences when we analyzed the distribution of 
Equatorial position angles $p$ and Supergalactic position angles $P$. One can see from 
PDF and CDF presented in the Figures \ref{fig:f1} -  \ref{fig:f4} that the probability that  
such results are coming from random distributions is less than $0.1\%$. The exception is 
only $\chi_c^2$ where the effect is on $2\,\sigma$ level and for $\Delta_{11}/\sigma(\Delta_{11})$ 
statistic where we have seen no effect.

Moreover, we have checked our result using Watson test.
The teoretical simulations show that for the sample of 247 clusters each with 
numbers of galaxies as in real clusters the expected mean value of $U^2$ statistic is $0.0280$
with standard deviation equal $0.0011$. For the real sample of 247 cluster we have obtained 
the value $\bar{U}^2=0.0642$ with $\sigma(\bar{U}^2)=0.0035$, hence this test rejects $H_0$ 
hypothesis on the $10\,\sigma$ level.

The statistics of $\Delta_{11}/\sigma(\Delta_{11})$ shows the direction of deviation from
isotropy with respect to the assumed coordinate system main plane. Our results show that in 
the case of $\Delta_{11}/\sigma(\Delta_{11})$ test we can not exclude our null hypothesis 
$H_0$ that the mean  value of  statistic is the same as predicted for the case of random
distribution. We have obtained the results for both Equatorial and Supergalactic coordinate 
systems. Of course, there should be no physical reason that the detected alignment could be 
connected with equatorial plane. Moreover, since we have analyzed the sample of clusters with 
redshift up to $z=0.12$, which is much more distant than the Local Supercluster, there is 
also no reason to expect the special meaning of the Local Supercluster equator. Because in 
both cases the obtained values of $\Delta_{11}/\sigma(\Delta_{11})$ statistic are close to 
zero, it increases the probability that the observed alignment is related  not to a particular 
global plane, but with the alignment with respect to galaxy cluster or cluser's parent  
supercluser planes. Therefore, our result confirms  the prediction that the detected 
alignment is not connected with equatorial plane nor with Supergalactic plane.  The final 
interpretation of this phenomenon, especially in the context of the evolution 
of galaxies and their structures   needs a detailed future study. 

We separately analyzed  the sample $B$ (only with galaxies brighter than $m_3+3^m$). Our 
investigation confirms the conclusion obtained by \citep{g2011b} that the observational 
alignment is weaker than in the case of sample A where all galaxy cluster members were 
analyzed,  but still significant. As above, $\Delta_{11}/\sigma(\Delta_{11})$ is close to 
zero which is in agreement with the predictions of our null hypothesis $H_0$. For $\chi^2$ 
test the result is significant at $2\,\sigma$ level while for remaining tests the 
results are very significant i.e. on  $3\,\sigma$ level. This result is important 
because, for avoiding possible role of background object, the restricted sample $B$ 
took into account only galaxies brighter than $m_3+3^{m}$.
This leads to the conclusion that the presence of background objects has no significant 
effect for all our results.

It is also necessary to investigate possible influence of errors in 
measurement of the position angles. For that, we  repeat our analysis presented
above in Table \ref{tab:t3} adding uncertainties in measurements of the 
position angles. We assumed standard error $\sigma(P)=2^{\circ}$. It means that even
on the $3\sigma$ level, the deviation will be $6^{\circ}$, what is more than in the 
worst case of the uncertainties ($5^{\circ}$). One notes that for real data, the 
uncertainties quickly increase for rounded images. However, this is not important 
in our analysis because face-on galaxies with axial ratio $q>0.75$ are excluded from 
the analysis. From Table \ref{tab:t3a} it is easy to conclude that uncertainty in 
determining  position angles does not significantly affect the results we have received.

The  alternative method to investigate the impact of errors and  possible  
influence of background object is to use the jackknife method  \citep{Efr79,Hawkins02}. 
The jackknife technique is based on drawing all possible samples of $N - 1$ values from 
the $N$ data points and repeating the test of $x$ statistic calculations on them, which 
allows us to calculate the standard deviations in the analyzed values of $x$, $\sigma_j(x)$. 
The best estimator for the standard errors in the value of $x$ is then just $\sqrt{N-1}\sigma_j$ 
\citep{Hawkins02}.   

Now, we can use the jackknife technique for the analyzed galaxy clusters and see if the 
errors are as predicted by the theory. Please note that now contrary to the analysis presented 
in Tables \ref{tab:t3} and \ref{tab:t3a}, statistics errors for individual clusters are not 
the same. In such a case average value of statistic should be obtained by weighted arithmetic mean
(see for example \citet{Brandt97} i.e 
$\bar{x}={\sum_{i = 1}^n {x_i \over \sigma^2_i} \over \sum_{i = 1}^n {1 \over \sigma^2_i}}$.
Respectively, the weighted average uncertainty  is given by formulae: 
$\sigma(\bar{x})=\left(\sqrt{\sum_{i = 1}^n {1 \over \sigma^2_i}}\right)^{-1}$.
We present obtained results in Table \ref{tab:t3b} where additionaly  we present 
diferences between obtained weighted arithmetic mean and values expected from simulations 
(presented in the Tables \ref{tab:t2a}, column $Real Number$,  values $\bar{x})$ divided by 
the weighted average uncertainty. From Table \ref{tab:t3b} it is easy to conclude that 
again $\Delta_{11}/\sigma(\Delta_{11})$ is close to zero which is in agreement with the 
predictions of our null hypothesis $H_0$. Generally, the recieved weighted average uncertainties 
are similar, but a little greater than errors presented in Table \ref{tab:t3}.  
Only the $\chi^2$ test does not survive jacknife procedure while for other tests the results are still 
deviating from prediction of $H_0$ hypothesis. One could observe that again effects for
sample $B$ are weaker than in the case of sample A, but is still significant.
These results confirm above conclusion that that uncertainty in determining  position angles does
not significantly affect our results and moreover lead to conclusion that the influence of 
background objects is not significant for our results.

Finally we have analyzed the differences between clusters with different BM types 
(Table \ref{tab:t4}). For this, we have used the means and standard deviations for all 
subsamples and compare the mean values of the statistic using the following well know 
statistical  test. When comparing  the mean values from  subsamples with standard deviations 
known, we use the following statistics:   
\begin{equation}
\label{eq:rr1}
U={\bar{X_1}-\bar{X_2} \over  \sqrt{{\sigma_1^2 \over n_1} + {\sigma_2^2 \over n_2}}}
\end{equation}
where $\bar{X_1}$ and $\bar{X_2}$ are the mean values of samples (subsamples) and $n_1$ 
and $n_2$ are the samples size. Under the assumption of the new null hypothesis $H_0$  
that the real mean values of $\bar{X_1}$ and of $\bar{X_2}$ are equal, the statistic $U$ 
has standard normal distribution. One should note however, that in the real case the 
standard deviation is not a'priori known and is estimated from the samples. Hence the 
crucial check is if both standard deviations are equal to each other. We do it by using 
a very well known $F$ Fisher test. In this test we make use of the 
statistic $F$:
\begin{equation}
\label{eq:rr2}
F={S_1^{*2} \over S_2^{*2}}
\end{equation}
where the variance estimators are: 
$S_i^{*2}(x)={n_1 \over n_1-1}S_i^2(x)= {1 \over n_i-1} \sum_{j=1}^{n_i} (X_{ij}-\bar{X_i})^2$.
Under $H_0$ hypothesis that the analyzed variances are equal to each other, 
the $F$ statistics has $F$ Snedecor distribution with $(n_1-1,n_2-1)$
degrees of freedom. 
In Table \ref{tab:t4} we presented the estimator of the standard deviation for the mean 
values, i.e. $S_{Mi}(\bar{x})=\sqrt{\sum_{j=1}^{n_i} (X_{ij}-\bar{X_i})^2/[n_i\,(n_i-1)]}$. Our 
analysis shows that in majority cases  we can not exclude $H_0$ that $\sigma_1^2=\sigma_2^2$. 

Then, for testing the significance 
of the differences of the mean values in subsamples we should use the well know Student  test. 
The statistics $t$
\begin{equation}
\label{eq:rr3}
t={\bar{X_1}-\bar{X_2} \over  \sqrt{{n_1 S_1^2+n_2 S_2^2 \over n_1+n_2-2}{n_1+n_2 \over n_1 n_2}}}
\end{equation}
has Student distribution with $n_1+n_2-2$ degrees of 
freedom under the assumption of $H_0$ hypothesis that the real mean values of $\bar{X_1}$ 
and of $\bar{X_2}$ are equal. Please note that this test is valid only in the case when standard 
deviations are equal.

In the few cases, when standard deviations in subsamples do not fulfill this condition, as for the
comparison BM II-III subsample with B II and BI-II subsamples in the case of control tests (with 
exception of $\chi_c^2$ test) for the comparison of the mean values  of the statistics we 
have to use the full Cochrane-Cox test \citep{satt46,coh57,Tou95,Krys98}. The statistics is 
given by the formula:
\begin{equation}
\label{eq:rr4}
CC= {\bar{X_1}-\bar{X_2} \over  \sqrt{{S_1^2 \over n_1-1} + {S_2^2 \over n_2-1}}} 
\end{equation}
with approximated critical value
\begin{equation}
\label{eq:rr5}
c(p,n1,n2) \cong {{S_1^2 \,t(p,n_1-1) \over n_1-1} + {S_2^2 \,t(p,n_2-1) \over n_2-1} \over {S_1^2 \over n_2-1} + {S_2^2 \over n_2-1}}
\end{equation}
where $t(p,n_i-1)$ are critical values (quantiles) of Student test. 

Our analysis shows that in majority cases we cannot exclude the $H_0$ hypothesis that mean 
values of analyzed statistics for different BM types are equal.  One should note that only for 
cluster type BM II-III the mean 
deviates from the case when cluster have other morphological type. The value of statistics
for this type is higher than for other ones. In some cases these differences are significant,
but only when we compare BM II and BM II-III statistics more then a half (i.e. $6$) tests show
that the differences is significant. 

Finally, we have been able to check if the value of statistics for subsamples of cluster 
belonging to  particular  BM types deviates from the mean values obtained for the whole 
population. The possible difference is observed only in the case BM II-III type, while for 
other subsamples at most one test shows possible differences.

Similarly as in the case of the whole sample, also in the case of clusters with 
different BM types, we have investigated possible influence of errors in measurement of the position 
angles. The result are presented in Table \ref{tab:t4a}. One should note that also in 
this case, the uncertainty in determining position angles does not significantly affect the results 
we have received.

\section{Discussion}

Any statistical study involving orientations must take into account that 
directional variables are cyclic \citep{Feigelson12}; i.e. $359^0$ is close to $1^0$ 
or if the analyzed range is  $180^0$ (as in the case of positional 
angles) that $179^0$ is close to $1^0$. It is one of the reason to include the procedure 
considered with the sign of the expression for $S$ (equations \ref{eq:a1}, \ref{eq:a2}, \ref{eq:a3}
 and the paragraph below them).  Fourier transform (see equation 
 \ref{eq:f9}) and Peeble’s first auto-correlation test (equation \ref{eq:c2}) 
take it into account (the exception is the auto-correlation control test but the correct 
values and Cumulative Distribution Functions are obtained from simulations. 

Another aspect of the problem is that when we study the distribution of cyclic variables, 
then it becomes important to determine the analogy of the Gaussian distribution. Usually, 
such a role is fulfilled by the von Mises circular distribution because it is a close approximation 
to the wrapped normal distribution 
(i.e. a wrapped probability distribution that results from the "wrapping" of the normal 
distribution around the unit circle)  which is the circular analogue of the normal 
distribution \citep{Fisher93}. In the context of research of the  galaxy orientation, 
von Mises circular distribution  was used for example during testing the effects of theoretical 
model of galactic formations \citep{g92,g93,g94}.

The most important claim raised against the use of the Fourier test  \citep{h4}
is that from theoretical point of view the Fourier transform   does not lead to 
reliable statistics tests because the conditions under which the exponential 
formulae (eq \ref{eq:f7}) are rarely valid in practice (see Chpt 5 of \citet{Percival93}).
This problem is well known and has been discussed many times in the literature 
in context of usability of power spectrum analysis (PSA) \citep{Yu69,Webster76,Guthrie90, Lake72,Newman94,Hawkins02,Godlowski06}, 
together with the Rayleigh test \citep{Mardia72,Batschelet81}.
For a given frequency, the Rayleigh  power 
spectrum corresponds to the Fourier power spectrum.
\citet{Newman94} pointed out that \citet{Yu69} 
version of Power Spectrum Analysis can be applied correctly only when 
a uniform distribution function is tested. In other cases,  test statistics is significantly 
modified \citep{Percival93, Newman94} and must be obtained from simulations (see also \citet{Godlowski06}).
Fortunately, in case of position angles 
the theoretical distribution is uniform, hence the Fourier transform \citep{h4} 
should work well in this case and therefore the  exponential formulae (Equation \ref{eq:f7}) 
are valid in practice.

In the papers   \citep{g10a,g2011a,Flin11} it was shown the alignment of galaxies in clusters 
increases with their  richness. It allowed to conclude that  the angular momentum of the cluster 
increases with the the numbers of galaxies in clusters i.e. with the mass of the structure. 
With such dependency it is expected that in the sample of  rich clusters we should observe the  
significant alignment. The result of the present paper confirmed such predictions.
One should note that the increase of alignment of bright galaxy members (central galaxies)
with the clusters richness has recently been found by \citet{Huang16}. They state that 
alignment of central galaxies  may originate from the filamentary accretion processes, but 
also possibly affected by the tidal field. 

If the alignment is increasing with the richness of the cluster the question which arises is what 
is its shape and reason. The relation between the angular momentum and the mass of the structure 
has been discussed since a long time 
and usually is presented as  $J\sim M^{5/3}$ \citep{Wesson79,Wesson83,Carrasco82,Brosche86}. The 
explanation of this phenomenon in the light of the structure formation scenarios is not clear but 
it could be explained by the Li model \citep{Li98,Godlowski03,Godlowski05} in which galaxies form in 
the rotating universe or by tidal torque scenario in hierarchical clustering model as suggested by 
\citet{HP88} and \citet{Catelan96} (see also \citet{Noh06a,Noh06b}). For that, extending the idea 
of \citet{HP88} and \citet{Catelan96} we use a novel theoretical approach \citet{sg15,sg17} in which 
the distribution function of dynamic characteristics of galaxies ensembles is calculated via tidal 
(shape-distorting) quadrupolar (and also higher multipolar) interaction between the galaxies. 
This function, among other things, may be used to a  better statistical treatment 
of observational data, which permits to discriminate observationally the relevance of available 
theories of galaxies formation. The calculation of the average galaxies angular momenta with the 
help of the above distribution function permits to study theoretically their orientations. 
In the papers \citet{sg15,sg17}
it was shown that with a reasonable assumption given, the angular momentum of galaxy  structures 
increases with their richness, however the final form of the dependence (not necessary $J\sim M^{5/3}$) 
depends on the assumption about cluster morphology. The preliminary results given in \citet{sg17}
shows that the present data does not allow to discriminate between different dependence between 
angular momentum and mass. The results of this paper are in agreement with such theoretical 
predictions.  In particular, this shows that the 
relation between alignment of galaxies in clusters and their mass is more complicated than simple 
increasing according to formulae $J\sim M^{5/3}$. 

We also  should note that during the studies of the angular momentum of galaxy cluster we 
have some complications that make analysis not so easy. Recall that generally clusters do not 
rotate \citep{Hwang07}, so the angular
momentum of such structures is connected with galaxy members alignment, but there 
are small number of clusters with intrinsic rotation. A sample of six \citet {Hwang07} rotating
clusters was analyzed by \citet{Aryal13} who did not found any alignment for that cluster, so the
angular momentum of such structures is coming from orbital movement, not from alignment. Finally we 
should note that  recently \citet{Aryal17} analyzed the sample of dynamically unstable Abell clusters 
founding a random orientation of galaxies inside these clusters.
One should note that recently was found that the alignment of galaxies evolves in time 
\citep{sg15,sg17,Schmitz18}. In the papers \citet{sg15,sg17} it was shown that angular momentum of 
galaxies in cluster increases with time. This could  explain the result of \citet{Hao11} who found 
that alignment of Brightest Cluster Galaxies decreases with redshift. This could also potentially 
explain \citet{Aryal17} result because it seems to be reasonable to assume that the dynamically 
unstable \citet{Aryal17} clusters are young, hence their angular momenta are still small. Such 
predictions could also explain the  results of \citep{song12} paper,  who found that the alignment 
profile of cluster galaxies drops faster at higher redshifts. Our results shows that although the exceptions 
exist, they do not significantly influence the statistics of the whole sample analyzed in this paper.

During the  investigation of alignment in clusters, the important problem is the influence of environmental 
effects to the origin of galaxy angular momenta. \citet{gpf11} shows  possible impact of the membership
clusters on the superclusters. Also \citet{Huang16,Huang18,Wang18} indicates possible role of environmental 
effects on central galaxy and radial alignments. Moreover, \citet{Huang16,Huang18} pointed out the role of 
central dominating galaxies in cluster and merger process  events which tend to destroy alignment.

In particular, the discussion shows that even though the tidal torque theory is at the moment 
supported by observations, it is still a significant simplification. There exist rich clusters that undergone 
many mergers and accretion events in history and cannot be well modeled by this theory. In this way, numerical 
simulations of structure formation which capture
some of the complexity may be more compelling.

One should note however, that even now, the investigation of the spatial 
orientation of galaxies in clusters would be possible with the use of information about 
frequency of galaxy occurrence in clusters with particular morphological types, since 
galaxy proportions with different spectral types can be estimated on the basis of density 
profiles in cluster, even if we do not have the information about morphological types for 
each particular galaxy \citep{Dre80,cal12,coe12,Hoy12}. In clusters, we may be able to estimate 
the fraction of galaxies having the particular morphological type. In numerical simulations, 
it will be taken into consideration the information about the frequency of occurrence of galaxies 
with particular morphological types in each cluster. Galaxy proportions with different 
spectral types in various cluster areas would be estimated on the basis of density profiles 
in cluster \citet{Dre80,cal12,coe12,Hoy12}.  Subsequently, from the formula: 
$cos^2 i=(q^2 -q^2_0 )/(1-q^2_0)$ the observed value of $q=a/b$ could be calculated for each 
galaxy. Using this value generated on the assumption of isotropy and $q_0=0.2$, the new values 
$cos^2 i$, as well as $\delta_D$ and $\eta$ angles would be enumerated. In such way we will 
obtain a new theoretical isotropic distributions for $\delta_D$ and $\eta$  angles in which 
the information about frequency of appearing galaxies with individual morphological types 
in clusters will be already included. Only with these corrected theoretical isotropic distributions 
we will be able to apply for testing the isotropy of galaxy orientations hypothesis when 
morphological types of particular galaxies are unknown. In this way we will compare obtained 
"theoretical isotropic distribution", which will take into consideration both the information 
about galaxy proportions of occurrence of different morphological types in cluster and the 
value of average galactic axial ratio $q_0$ with the observational distributions obtained 
as well with the assumption $q_0=0.2.$

Although such alternative solution exists, 
it requires precise and complicated numerical simulations and has never been used in practice 
before. For the above reason we decide to postpone it to future investigation and in the 
present paper we have analyzed, like in \citet{g2011b}, only the distribution of the 
position angles $p$ in the sample ($A$) of 247 rich Abell clusters both in Equatorial 
and Supergalactic coordinate systems.

\section{Conclusions}

The motivating theoretical goal of the project has been to give an improvement 
in the discrimination among different models of galaxy formation. A general idea has been to 
anylyze the angular momentum of galaxies in clusters 
and check if the results agree with scenarios predictions.
That is why, in this paper we have focused on how we perform the analysis of the 
alignment of galaxies in clusters.
In the original method presented in \citet{g2011b} the distributions 
of the position angles for galaxies in each cluster were analyzed using statistical tests: $\chi^2$ 
test, Fourier tests, Autocorrelation test and Kolmogorov test. The mean value of the analyzed 
statistics was compared with theoretical predictions as well as with results obtained from 
numerical simulations. The method allows to check if the mean value of analyzed statistics is the 
same as expected in the case of random distribution of the position angles of galaxies. 

In the present paper we have analyzed this method in detail, giving proposal of some significant 
improvements and introducing new statistical tests into the method. 
We have considered how the tests changes if we assume various expected values of galaxies in bins. 
In particular, in the autocorrelation test, the values of statistics slightly changes.
However, in the Fourier test, not only the formulas for coefficients changes but also the 
coefficients need not be independent and we consider this in the analysis. 
We have also analyzed the properties of the Kolmogorov-Smirnov test 
applied to the analysis of the alignment of galaxies in clusters and 
finally we introduced control tests to all considered tests.
In all cases the theoretical predictions have been compared with the numerical simulations.

The second major advantage of the present paper in comparison to the previous investigation is 
that our analysis allowed us to expand the investigation of alignment of galaxies in cluster 
from the analysis of position angles only to the angles giving spatial orientation of galaxy planes, 
that has never been done before. The main difference is that the analysis 
of the position angles gives the information about orientation of galaxies only for edge-on galaxies. 
The analysis of the spatial orientation has allowed us to include all galaxies especially 
face-on galaxies.
  The difficulty that arises is that  during the process of deprojection of the spatial orientation 
of galaxies from its optical images we obtain two  possible orientations, and because we are able to
find which solution is correct only in the small number of galaxies, both solutions must be taken 
into account during further analysis.

Another crucial problem during analysis of the angles giving spatial orientation of galaxies is 
that if for any reason we exclude from analysis any type of galaxies (for example face-on galaxies), 
then the theoretical distribution of analyzed angles will be modified, even in the case when the 
distribution of galaxy planes is random and isotropic. In this case, a random  distribution of 
analyzed angles which is the base of comparison with the real one must be, in practice, obtained 
from numerical simulations. This problem was analyzed for example by \citet{g2,Ar00} (for modern 
analysis see for example \citet{Flin11}). However, we have noticed that nobody took care of the 
fact that both obtained solutions for orientations are not independent of each other (only 
\citet{Panko13} made a remark that such problem could arise). Consequently, nobody analyzed if 
statistical test gives, even for "random" distributions, the same values of statistics as it is 
predicted in the case of the position angles. Our results have clearly showed that the expected 
mean values of the statistics for $\delta_D$ and $\eta$ angles varied from that obtained during 
analysis of the position angles. It means that "theoretical random distribution" must be modified 
this time. This is very easy to be observed in the example of nearly face-on galaxies, when
both possible orientations are similar, which leads to the situation that both obtained values 
for $\delta_D$ and $\eta$ angles are similar. As a result in this case the first solution strongly
affects the second. This  phenomenon is also responsible for the results that during analysis
of the spatial orientation of galaxies, we found the significant difference between the case when 
we assumed the real coordinates for galaxies in clusters and the case  of the analysis of the 
fictitious clusters with coordinates distributed around the whole celestial sphere. 
In the paper we have analyzed this problem theoretically as well as show using numerical 
simulations how it affects the real data.

In this paper we have analyzed the sample of 247 rich Abell clusters containing at least 100
members using a significantly improved method of the investigation of the  orientation of galaxies 
in clusters.  We found that  the mean values of tested statistics, obtained on the base of the 
analyzed sample, significantly deviated from the  expected in the case of the random distributions. 
As a result, we could conclude that the orientations of galaxies in analyzed clusters are not 
random. It means that we genuinely confirmed an existence of the alignment of galaxies in rich 
Abells' galaxy clusters, suggested by  \citep{g2011b}, especially by results of Cr\'amer-von Mises 
and Watson tests. Moreover, we have shown that the above results are not due to errors in measurement 
of position angles nor influence of background objects (also done by jackknife method).

The results that the aligmnent is increasing with richness and is observed in rich 
clusters supports the scenarios that predict such a thing (Li model, tidal torque scenario in 
the hierarchical clustering model). The other scenarios like Zeldovich pancakes \citep{Zeldovich70} and primordial 
turbulence \citep{Silk83} cannot explain such alignment and hence are not supported by our results.

It was natural to expect that observed alignment could not be connected with
equatorial plane. Indeed, the obtained values of $\Delta_{11}/\sigma(\Delta_{11})$ statistics  
does not show any deviation from zero, as predicted  in such cases. This result was obtained
both in the case of the analysis in Equatorial and Supergalactic coordinate systems, what means that
observed alignment is also not connected with Local Supercluster plane.

This result is generally independent from the clusters Bautz-Morgan types. Only cluster 
type BM II-III shows possible deviation from results obtained for other morphological 
types especially if we compare BM II-III with BM II type clusters.  Our result clearly  
confirmed  \citet{g10a} opinion that, contrary to the suggestions of
\citet{Aryal04,Aryal05b,Aryal05c,Aryal06,Aryal07}, the alignment of the orientation
of galaxies is only weakly correlated with their morphological types according to the 
classification of Bautz-Morgan (BM). One should note that in the paper \citet{Bier15}, 
during the analysis of the Binggeli effect for sample of 6188 galaxy clusters also selected 
from \citet{Panko06} catalogue,  the differences was found with the Binggeli effect for BM type 
II clusters. It sugests that both of this observations  could be connected with different
 morphological populations of the clusters 
i.e. the late type clusters (BM II-III and BM III) are spiral-rich clusters.

It is important to note that the present observational results obtained for the sample of rich Abell
 clusters is based only on the analysis of the positions angles. It is due to the fact that 
we have no information connected with morphological type of members galaxies, hence the process of
 deprojection of the spatial orientation of galaxies from its 
optical images is a source of errors that are difficult to be controlled. 
Therefore, during the comparison of the real data with theoretical 
predictions and numerical simulations, we have concentrated on the analysis of the position 
angles only and postponed the spatial analysis of the real clusters to future studies. 
We should point out that our method of analysing the spatial angles is now well developed 
theoretically. 

In the future studies, we will investigate the real samples also with the analysis of the 
distribution of the angles $\delta_D$ and $\eta$ giving spatial orientation of galaxies. 
The future investigation will be possible with 
the use of information about frequency of galaxy occurrence in clusters with particular 
morphological types, since galaxy proportions with different spectral types can be 
estimated on the basis of density profiles in cluster \citet{Dre80,cal12,coe12,Hoy12}.
We are also planning to extend our research to fewer galaxy clusters.

Finally, we would like to conclude  that now we have a well-tested method of studying the 
orientation of galaxies in clusters that can be used for research on other data sets, 
such as these from the new Kilo-Degree Survey.

\section*{Acknowledgments}
 
 The authors thanks anonymous referee for detailed 
remarks which helped to improve the original manuscript.

\section{Appendix}

If we denote:
\begin{eqnarray}
\label{eq:f25}
&&A= \sum_{k=1}^n N_{0,k} \cos^2{2 \theta_k}, \quad
  B= \sum_{k=1}^n N_{0,k} \sin^2{2 \theta_k} \nonumber \\
&&C= \sum_{k=1}^n N_{0,k} \cos^2{4 \theta_k}, \quad
  D= \sum_{k=1}^n N_{0,k} \sin^2{4 \theta_k}, \nonumber \\
&&U= \sum_{k=1}^n N_{0,k} \cos{2 \theta_k}\cos{4 \theta_k}, \quad
  W= \sum_{k=1}^n N_{0,k} \sin{2 \theta_k}\sin{4 \theta_k} \nonumber \\
&&K= \sum_{k=1}^n (N_k -N_{0,k})\cos{2 \theta_k}, \quad
  L= \sum_{k=1}^n (N_k -N_{0,k})\sin{2 \theta_k}  \nonumber \\
&&M= \sum_{k=1}^n (N_k -N_{0,k})\cos{4 \theta_k}, \quad
  N= \sum_{k=1}^n (N_k -N_{0,k})\sin{4 \theta_k}
\end{eqnarray}
and moreover
\begin{eqnarray}
\label{eq:f32}
&&Y= \sum_{k=1}^n N_{0,k} \cos{2 \theta_k}\sin{2 \theta_k} \quad
Z= \sum_{k=1}^n N_{0,k} \cos{4 \theta_k}\sin{4 \theta_k} \nonumber \\
&&V= \sum_{k=1}^n N_{0,k} \cos{2 \theta_k}\sin{4 \theta_k}  \quad
X= \sum_{k=1}^n N_{0,k} \sin{2 \theta_k}\cos{4 \theta_k}  
\end{eqnarray}
then, in the most general general case, we obtain the folowing solution of Equation \ref{eq:f20}.
The inverse matrix  to the covariance matrix  of ${\bf x}$ (i.e. cooeficients  $\Delta_{ij}$) has a form:
\begin{equation}
\label{eq:f34}
G =
\left(
\begin{array}{cccc}
A&Y&U&V\\
Y&B&X&W\\
U&X&C&Z\\
V&W&Z&D
\end{array}
\right)
\end{equation}

While we introduce auxiliary vector $H$:
\begin{equation}
\label{eq:f33}
H =
\left(
\begin{array}{c}
K\\
L\\
M\\
N
\end{array}
\right)
\end{equation}
then the resulting vector $I$ is equal:
\begin{equation}
\label{eq:f35}
I =
\left(
\begin{array}{c}
\Delta_{11} \\
\Delta_{21} \\
\Delta_{12} \\
\Delta_{22}
\end{array}
\right)= G^{-1}\cdot H
\end{equation}
The amplitude $\Delta \equiv J=\sum_i \sum_j I_{i}^T {G_{ij} I_{j}}$
is described by $4D$ Gaussian distribution and
expression for the required probability is the folowing:
\begin{equation}
\label{eq:f36}
P(>\Delta)=(1+J/2)\exp{(-J/2)}.
\end{equation}

Even if we, like  \citep{h4} take into account only first Fourier mode 
(i.e. only coefficients affiliated with
$cos 2 \theta$ and $sin 2 \theta$) situation is not simple.
In such a case the resulting vector $I$ (Equation \ref{eq:f19}) is reduced to the form:
\begin{equation}
\label{eq:f19a}
I=
\left(
\begin{array}{c}
\Delta_{11} \\
\Delta_{21} 
\end{array}
\right)
\end{equation}
while auxiliary vector $H$ for:
\begin{equation}
\label{eq:f33a}
H =
\left(
\begin{array}{c}
K\\
L
\end{array}
\right)
\end{equation}

In this case $G$ matrix has form:
\begin{equation}
\label{eq:f37}
G =
\left(
\begin{array}{cc}
A&Y\\
Y&B\\
\end{array}
\right)
\end{equation}
while responsible covariance matrix $C=G^1$ is equal:
\begin{equation}
\label{eq:f38}
C=G^{-1} =
\left(
\begin{array}{cc}
{B\over AB-Y^2}&{-Y\over AB-Y^2}\\
{-Y\over AB-Y^2}&{A\over AB-Y^2}\\
\end{array}
\right)
\end{equation}
and we obtain the following expression for the
$\Delta_{i1}$ coefficients:
\begin{equation}
\label{eq:f39}
\Delta_{11} = {BK-YL \over AB-Y^2}
\end{equation}
\begin{equation}
\label{eq:f40}
\Delta_{21} = {AL-YK \over AB-Y^2}
\end{equation}





One should remember that then we take into account only first Fourier mode,
then during computation of the probability, we have again $2$D not $4$D Gaussian
distribution. So now the expression for the required probability is:
\begin{equation}
\label{eq:f44}
P(>\Delta)=\exp{(-J/2)}.
\end{equation}
Such general situation as discussed above is rather unusual in practical applications
and corresponds to the theoretical situation when theoretical distribution
of $N_{0,k}$ is not symmetric i.e $N_{0,k} \ne N_{0,n-k}$. In the case
of the analysis of the $\delta_D$ angle ($\theta= \delta_D +\pi/2$) it answers the
situation when theoretical distribution is not symmetric according to the value of
$\delta_D=0$. Such theoretical models seem a bit strange, but an example could be
a model with angular momentum pointed out directly to the Local Supercluster
Center (Virgo Cluster center) i.e. hedgehog model. During analysis of such a
model, we could take into account the fact that our Galaxy is not directly lying in the
Local Supercluster plane and as a result the coordinates of the Virgo Cluster
center in this supergalactic coordinate system are not $L=0, B=0$ but
 $L = 0, B = -3.19^o$ \citep{f4}.

The case that $N_{0,k}$ are symmetric i.e $N_{0,k} = N_{0,n-k}$
(what means that for  $\delta_D$ angle theoretical distribution is symmetric according 
value of $\delta_D=0$) was analysed in details by \citet{g3} 
\footnote{However please note that there are printed errors in God{\l}owski (1994). 
Most important is that Eq. 18 should have form $P(\Delta)=(1+J/2)\exp{(-J/2)}$}
In such a case the most important simplifications is that all auxiliary values
given in formulae \ref{eq:f32} (i.e. Y, V, X, Z) are equal zero.
As a result, if we analyze first and second Fourier mods together,
solutions for  coefficiences $\Delta_{ij}$ have a form:
\begin{eqnarray}
\label{eq:f24}
&&\Delta_{11} = {CK-UM \over AC-U^2 },\nonumber \\
&&\Delta_{21} = {DL-WN \over BD-W^2 },\nonumber \\
&&\Delta_{12} = {-UK+AM \over AC-U^2},\nonumber \\
&&\Delta_{22} = {-WL+BN \over BD-W^2}
\end{eqnarray}

with covariance matrix $ C_{cov}({\bf x})$:
\begin{equation}
\label{eq:f27}
C_{cov}({\bf x})=
\left(
\begin{array}{cccc}
{C \over (AC-U^2)} & 0 & {-U \over(AC-U^2)} & 0  \\
0 & {D \over (BD-W^2)} & 0 & {-W \over (BD-W^2)} \\
{-U \over (AC-U^2)} & 0 & {A \over (AC-U^2)} & 0 \\
 0 & {-W \over (BD-W^2)} & 0 & {B \over (BD-W^2)}.
\end{array}
\right)
\end{equation}

Please note when we analyzed first and second Fourier modes seperately solutions are 
reduced to the explicit form  \citet{g3}:
\begin{equation}
\label{eq:f10a}
\Delta_{1j} = {\sum_{k = 1}^n (N_k-N_{0,k})\cos{2J \theta_k} \over
\sum_{k = 1}^n N_{0,k} \cos^2{2J \theta_k}},
\end{equation}
and
\begin{equation}
\label{eq:f11a}
\Delta_{2j} = { \sum_{k = 1}^n (N_k-N_{0,k})\sin{2J \theta_k} \over
\sum_{k = 1}^n N_{0,k} \sin^2{2J \theta_k}},
\end{equation}
with the standard deviation
\begin{equation}
\label{eq:f12a}
\sigma(\Delta_{1j}) =
\left( {\sum_{k = 1}^n N_{0,k} \cos^2{2J \theta_k} } \right)^{-1/2} \approx
 \left( {2 \over n N_{0,k}} \right)^{1/2},
\end{equation}
and
\begin{equation}
\label{eq:f13a}
\sigma(\Delta_{2j}) =
 \left( {\sum_{k = 1}^n N_{0,k} \sin^2{2J \theta_k} } \right)^{-1/2} \approx
 \left( {2 \over n N_{0,k}} \right)^{1/2}.
\end{equation}
while the  probability that the amplitude
\begin{equation}
\label{eq:f6b}
\Delta_J = \left( \Delta_{1j}^2 + \Delta_{2j}^2 \right)^{1/2}
\end{equation}
is greater than a fixed value is now given by the formula:
\begin{equation}
\label{eq:f7b}
P(>\Delta_j ) =
\exp{\left( -{1 \over 2} \left({\Delta_{1j}^2\over
\sigma(\Delta_{1j}^2)}+{\Delta_{2j}^2 \over \sigma(\Delta_{2j}^2)}\right)
\right)}
\approx \exp{\left( -{n \over 4} N_0 \Delta_1^2 \right)}
\end{equation}
with standard deviation being approximately:
\begin{equation}
\label{eq:f8b}
\sigma(\Delta_1) \approx \left( {2 \over n N_0} \right)^{1/2}
\end{equation}.

Finally the case then all $N_{0,k}=N_0$ are equal was analyzed in details by 
\citet{g2011b}. Please note that in such a case formulae for $\Delta_{ij}$
coefficients are reduced to the explicit form 
\begin{equation}
\label{eq:f10}
\Delta_{1j} = {\sum_{k = 1}^n N_k\cos{2J \theta_k} \over
\sum_{k = 1}^n N_{0} \cos^2{2J \theta_k}},
\end{equation}
and
\begin{equation}
\label{eq:f11}
\Delta_{2j} = { \sum_{k = 1}^n N_k\sin{2J \theta_k} \over
\sum_{k = 1}^n N_{0} \sin^2{2J \theta_k}},
\end{equation}
while formulae for probability (\ref{eq:f36})
are reduced to the explicit form 
\begin{equation}
\label{eq:f17}
P(>\Delta ) =
\left(1+{n \over 4} N_0 \Delta_j^2 \right) \exp{\left( -{n \over 4} N_0 \Delta_j^2 \right)}.
\end{equation}
where amplitude $\Delta$:
\begin{equation}
\label{eq:f16}
\Delta =
 \left( \Delta_{11}^2 + \Delta_{21}^2+\Delta_{12}^2 + \Delta_{22}^2 \right)^{1/2}
\end{equation}

One should note that in the case when we analysed only first Fourier mode the 
formulae  (\ref{eq:f10} and \ref{eq:f11}) for $\Delta_{i1}$   are exactly the same
as originaly obtained by \citet{h4}. 
Also the formulae for probability that the amplitude
\begin{equation}
\label{eq:f6}
\Delta_1 = \left( \Delta_{11}^2 + \Delta_{21}^2 \right)^{1/2}
\end{equation}
is greater than a fixed value:
\begin{equation}
\label{eq:f7}
P(>\Delta_1 ) = \exp{\left( -{n \over 4} N_0 \Delta_1^2 \right)}
\end{equation}
are exactly the same as obtained by \citet{h4}.

During analysis of the distribution of position angles ($\theta \equiv p$)
$\Delta_{11}<0$ means an excess of galaxies with position angles near $90^o$
- parallel to main plane of the coordinate system (equatorial or supergalactic
in our case). It indicates the rotation axis tends to be perpendicular to the main
plane. If $\Delta_{11}>0$, then the  excess of objects with position angles
perpendicular to the main plane of the coordinate system is observed. Therefore,
for $\Delta_{11}>0$ the rotation axis tends to be parallel to the main plane.

We could do similar analysis for the angles giving information on the spatial
orientation of galaxies. For $\theta \equiv \eta$, the positive sign $\Delta_{11}$
($\Delta_{11} > 0$) means that projection of rotation axis for that plane tends
to be directed toward $\eta = 0$. Therefore, for $\Delta_{11} < 0$ the projection
of rotation axes tends to be perpendicular to zero point, respectively.

One can also deduce  the direction of the departure from isotropy from the sign  of
$\Delta_{11}$  for distribution of $\delta_D$ angle ($\theta \equiv \delta_D+\pi/2$).
 If $\Delta_{11} < 0$, then  an excess of galaxies with rotation axes parallel
to the coordinate system main plane is observed, while for $\Delta_{11} > 0$
 rotation axes tend to be perpendicular to the coordinate system main plane.

One should note that alternatively the direction of deviation from isotropy 
could be also obtained by computaion of Directional Mean and Rayleigh's Z 
statistics  \citep{Mardia72,Batschelet81}. It was proposed to use it for 
analysis of the orientation by \citet{Kindl87}  who  proposed to use 
phase angle for any preferred orientation given by formulae: 
$\Theta_J=(2J)^{-1}\cdot \arctan(\Delta_{2J}/\Delta_{1J})$. Although the parameter
is statistically interesting, it is not used in practice in the investigation of alignment of galaxies.

\clearpage

\clearpage

\clearpage

\clearpage

\clearpage

\clearpage

\clearpage

\clearpage

\clearpage

\begin{table}
\begin{center}
\caption{
The result of Kolmogorov- Smirnov test for analysis of $\delta_D$, 
$\eta$  and $P$ angles. The distribution of tested 
statistics for 1000 simulations of the sample of 2227 galaxies.
}
\label{tab:t0}
\begin{tabular}{cccc}
\hline
\hline
Test&$P$&$\delta_D$&$\eta$\\
\hline
$\chi^2$                        &$0.885$&$1.170$&$1.202$\\
$C$                             &$1.123$&$1.628$&$0.776$\\
$\Delta_{11}$                   &$0.442$&$1.150$&$0.556$\\
\hline
\end{tabular}
\end{center}
\end{table}


\begin{table}
\begin{center}
\caption{The result of numerical simulation - sample of 247 cluster
each with 2360 galaxies}
\label{tab:t1a}
\begin{tabular}{ccccc}
\hline
\hline
Test&$\bar{x}$&$\sigma(x)$&$\sigma(\bar{x})$&$\sigma(\sigma(x))$\\
\hline
angle $p$&&&&\\
\hline
$\chi^2$                           &$34.9978$&$0.5442$&$0.0172$&$0.0128$\\
$\chi_c^2$                         &$16.9984$&$0.3550$&$0.0112$&$0.0079$\\
$\Delta_{1}/\sigma(\Delta_{1})$    &$ 1.2524$&$0.0424$&$0.0013$&$0.0009$\\
$\Delta/\sigma(\Delta)$            &$ 1.8794$&$0.0460$&$0.0014$&$0.0010$\\
$\Delta_c/\sigma(\Delta_c)$        &$ 1.2549$&$0.0419$&$0.0013$&$0.0009$\\
$C$                                &$-0.9917$&$0.3899$&$0.0123$&$0.0087$\\
$C_c$                              &$-0.9916$&$0.2509$&$0.0079$&$0.0056$\\
$\lambda$                          &$ 0.7708$&$0.0166$&$0.0005$&$0.0004$\\
$\lambda_c$                        &$ 0.7314$&$0.0166$&$0.0005$&$0.0004$\\
$\Delta_{11}/\sigma(\Delta_{11})$  &$-0.0010$&$0.0643$&$0.0020$&$0.0014$\\
$|\Delta_{11}/\sigma(\Delta_{11})|$&$ 0.7984$&$0.0393$&$0.0012$&$0.0009$\\
\hline
\end{tabular}
\end{center}
\end{table}


\begin{table}
\begin{center}
\caption{Results of numerical simulations - sample of 247 clusters
each with 2360 galaxies}
\label{tab:t1b}
\begin{tabular}{ccccc}
\hline
\hline
Test&$\bar{x}$&$\sigma(x)$&$\sigma(\bar{x})$&$\sigma(\sigma(x))$\\
\hline
angle $\delta_D$&&&&\\
\hline
$\chi^2$                           &$35.5837$&$0.5739$&$0.0181$&$0.0128$\\
$\chi_c^2$                         &$16.8006$&$0.3882$&$0.0123$&$0.0087$\\
$\Delta_{1}/\sigma(\Delta_{1})$    &$ 1.2570$&$0.0484$&$0.0015$&$0.0011$\\
$\Delta/\sigma(\Delta)$            &$ 1.8870$&$0.0494$&$0.0016$&$0.0011$\\
$\Delta_c/\sigma(\Delta_c)$        &$ 1.0254$&$0.0343$&$0.0012$&$0.0008$\\
$C$                                &$-0.6594$&$0.4100$&$0.0130$&$0.0092$\\
$C_c$                              &$-1.2522$&$0.2529$&$0.0080$&$0.0057$\\
$\lambda$                          &$ 0.8164$&$0.0216$&$0.0009$&$0.0005$\\
$\lambda_c$                        &$ 0.6779$&$0.0151$&$0.0005$&$0.0003$\\
$\Delta_{11}/\sigma(\Delta_{11})$  &$ 0.0001$&$0.0475$&$0.0016$&$0.0011$\\
$|\Delta_{11}/\sigma(\Delta_{11})|$&$ 0.6138$&$0.0284$&$0.0009$&$0.0006$\\
\hline
\end{tabular}
\end{center}
\end{table}


\begin{table}
\begin{center}
\caption{Results of numerical simulations - sample of 247 clusters
each with 2360 galaxies}
\label{tab:t1c}
\begin{tabular}{ccccc}
\hline
\hline
Test&$\bar{x}$&$\sigma(x)$&$\sigma(\bar{x})$&$\sigma(\sigma(x))$\\
\hline
angle $\eta$&&&&\\
\hline
$\chi^2$                           &$37.3439$&$0.5806$&$0.0184$&$0.0130$\\
$\chi_c^2$                         &$18.1021$&$0.3955$&$0.0125$&$0.0088$\\
$\Delta_{1}/\sigma(\Delta_{1})$    &$ 1.4786$&$0.0479$&$0.0015$&$0.0011$\\
$\Delta/\sigma(\Delta)$            &$ 2.1339$&$0.0473$&$0.0015$&$0.0011$\\
$\Delta_c/\sigma(\Delta_c)$        &$ 1.4009$&$0.0475$&$0.0015$&$0.0011$\\
$C$                                &$ 0.4096$&$0.3793$&$0.0120$&$0.0085$\\
$C_c$                              &$-0.4442$&$0.2626$&$0.0083$&$0.0059$\\
$\lambda$                          &$ 0.8630$&$0.0195$&$0.0006$&$0.0004$\\
$\lambda_c$                        &$ 0.7985$&$0.0192$&$0.0006$&$0.0004$\\
$\Delta_{11}/\sigma(\Delta_{11})$  &$ 0.0021$&$0.0718$&$0.0023$&$0.0016$\\
$|\Delta_{11}/\sigma(\Delta_{11})|$&$ 0.9167$&$0.0448$&$0.0014$&$0.0010$\\
\hline
\end{tabular}
\end{center}
\end{table}


\begin{table}
\begin{center}
\caption{Results of numerical simulations - sample of 247 clusters
each with 2360 galaxies, with numbers of galaxies as in real clusters
simulated both with coordinates distributed as in real clusters and
for galaxies randomly distributed around the whole celestial sphere.}
\label{tab:t2a}
\begin{tabular}{c|cc|cc|cc|cc}
\hline
\hline
\multicolumn{1}{c}{Test}&
\multicolumn{2}{c}{$2360$ Galaxies}&
\multicolumn{2}{c}{$2360_{coordinates}$}&
\multicolumn{2}{c}{$Real Number$}&
\multicolumn{2}{c}{$Real_{coordinates}$}\\
\hline
Test&$\bar{x}$&$\sigma(\bar{x})$&$\bar{x}$&$\sigma(\bar{x})$&$\bar{x}$&$\sigma(\bar{x})$&$\bar{x}$&$\sigma(\bar{x})$\\
\hline
angle $p$&&&&&&&&\\
\hline
$\chi^2$                           &$34.9978$&$0.0172$&$35.0281$&$0.0166$&$34.9798$&$0.0170$&$35.0080$&$0.0168$\\
$\chi_c^2$                         &$16.9984$&$0.0112$&$17.0129$&$0.0115$&$16.9741$&$0.0115$&$17.0052$&$0.0116$\\
$\Delta_{1}/\sigma(\Delta_{1})$    &$ 1.2524$&$0.0013$&$ 1.2538$&$0.0013$&$ 1.2550$&$0.0013$&$ 1.2577$&$0.0013$\\
$\Delta/\sigma(\Delta)$            &$ 1.8794$&$0.0014$&$ 1.8804$&$0.0014$&$ 1.8788$&$0.0014$&$ 1.8846$&$0.0013$\\
$\Delta_c/\sigma(\Delta_c)$        &$ 1.2549$&$0.0013$&$ 1.2536$&$0.0012$&$ 1.2529$&$0.0013$&$ 1.2580$&$0.0013$\\
$C$                                &$-0.9917$&$0.0123$&$-1.0062$&$0.0118$&$-1.0195$&$0.0119$&$-0.9918$&$0.0117$\\
$C_c$                              &$-0.9916$&$0.0079$&$-0.9963$&$0.0081$&$-1.0100$&$0.0077$&$-0.9983$&$0.0080$\\
$\lambda$                          &$ 0.7708$&$0.0005$&$ 0.7713$&$0.0005$&$ 0.7720$&$0.0005$&$ 0.7725$&$0.0005$\\
$\lambda_c$                        &$ 0.7314$&$0.0005$&$ 0.7320$&$0.0005$&$ 0.7301$&$0.0005$&$ 0.7327$&$0.0005$\\
$\Delta_{11}/\sigma(\Delta_{11})$  &$-0.0010$&$0.0020$&$ 0.0013$&$0.0019$&$ 0.0014$&$0.0020$&$-0.0006$&$0.0020$\\
$|\Delta_{11}/\sigma(\Delta_{11})|$&$ 0.7984$&$0.0012$&$ 0.7992$&$0.0012$&$ 0.7983$&$0.0012$&$ 0.8025$&$0.0012$\\
\hline
\end{tabular}
\end{center}
\end{table}


\begin{table}
\begin{center}
\caption{Results of numerical simulations - sample of 247 clusters
each with 2360 galaxies, with numbers of galaxies as in real clusters
simulated both with coordinates distributed as in real clusters and
for galaxies randomly distributed around the whole celestial sphere.}
\label{tab:t2b}
\begin{tabular}{c|cc|cc|cc|cc}
\hline
\hline
\multicolumn{1}{c}{Test}&
\multicolumn{2}{c}{$2360$ Galaxies}&
\multicolumn{2}{c}{$2360_{coordinates}$}&
\multicolumn{2}{c}{$Real Number$}&
\multicolumn{2}{c}{$Real_{coordinates}$}\\
\hline
Test&$\bar{x}$&$\sigma(\bar{x})$&$\bar{x}$&$\sigma(\bar{x})$&$\bar{x}$&$\sigma(\bar{x})$&$\bar{x}$&$\sigma(\bar{x})$\\
\hline
angle $\delta_D$&&&&&&&&\\
\hline
$\chi^2$                           &$35.5837$&$0.0181$&$35.8522$&$0.0172$&$35.5568$&$0.0186$&$35.8683$&$0.0179$\\
$\chi_c^2$                         &$16.8006$&$0.0123$&$17.7032$&$0.0127$&$16.7960$&$0.0122$&$17.7239$&$0.0128$\\
$\Delta_{1}/\sigma(\Delta_{1})$    &$ 1.2570$&$0.0015$&$ 1.2440$&$0.0013$&$ 1.2490$&$0.0015$&$ 1.2427$&$0.0013$\\
$\Delta/\sigma(\Delta)$            &$ 1.8870$&$0.0016$&$ 1.8818$&$0.0014$&$ 1.8779$&$0.0016$&$ 1.8788$&$0.0014$\\
$\Delta_c/\sigma(\Delta_c)$        &$ 1.0254$&$0.0012$&$ 1.2275$&$0.0013$&$ 1.0772$&$0.0011$&$ 1.2250$&$0.0013$\\
$C$                                &$-0.6594$&$0.0130$&$-0.4865$&$0.0125$&$-0.5077$&$0.0126$&$-0.5057$&$0.0123$\\
$C_c$                              &$-1.2522$&$0.0080$&$-0.6211$&$0.0083$&$-1.1155$&$0.0081$&$-0.6405$&$0.0086$\\
$\lambda$                          &$ 0.8164$&$0.0009$&$ 0.7790$&$0.0005$&$ 0.8089$&$0.0006$&$ 0.7781$&$0.0006$\\
$\lambda_c$                        &$ 0.6779$&$0.0005$&$ 0.7524$&$0.0006$&$ 0.6870$&$0.0005$&$ 0.7521$&$0.0006$\\
$\Delta_{11}/\sigma(\Delta_{11})$  &$ 0.0001$&$0.0016$&$-0.0031$&$0.0020$&$-0.0009$&$0.0017$&$ 0.0006$&$0.0019$\\
$|\Delta_{11}/\sigma(\Delta_{11})|$&$ 0.6138$&$0.0009$&$ 0.7626$&$0.0012$&$ 0.6366$&$0.0010$&$ 0.7618$&$0.0011$\\
\hline
\end{tabular}
\end{center}
\end{table}


\begin{table}
\begin{center}
\caption{Results of numerical simulations - sample of 247 clusters
each with 2360 galaxies, with numbers of galaxies as in real clusters
simulated both with coordinates distributed as in real clusters and
for galaxies randomly distributed around the whole celestial sphere.}
\label{tab:t2c}
\begin{tabular}{c|cc|cc|cc|cc}
\hline
\hline
\multicolumn{1}{c}{Test}&
\multicolumn{2}{c}{$2360$ Galaxies}&
\multicolumn{2}{c}{$2360_{coordinates}$}&
\multicolumn{2}{c}{$Real Number$}&
\multicolumn{2}{c}{$Real_{coordinates}$}\\
\hline
Test&$\bar{x}$&$\sigma(\bar{x})$&$\bar{x}$&$\sigma(\bar{x})$&$\bar{x}$&$\sigma(\bar{x})$&$\bar{x}$&$\sigma(\bar{x})$\\
\hline
angle $\eta$&&&&&&&&\\
\hline
$\chi^2$                           &$37.3439$&$0.0184$&$36.2309$&$0.0181$&$36.3915$&$0.0179$&$36.2255$&$0.0176$\\
$\chi_c^2$                         &$18.1021$&$0.0125$&$17.6242$&$0.0124$&$17.7058$&$0.0124$&$17.6016$&$0.0119$\\
$\Delta_{1}/\sigma(\Delta_{1})$    &$ 1.4786$&$0.0015$&$ 1.3619$&$0.0015$&$ 1.3761$&$0.0014$&$ 1.3640$&$0.0015$\\
$\Delta/\sigma(\Delta)$            &$ 2.1339$&$0.0015$&$ 2.0053$&$0.0015$&$ 2.0218$&$0.0015$&$ 2.0053$&$0.0015$\\
$\Delta_c/\sigma(\Delta_c)$        &$ 1.4009$&$0.0015$&$ 1.3369$&$0.0014$&$ 1.3498$&$0.0014$&$ 1.3363$&$0.0013$\\
$C$                                &$ 0.4096$&$0.0120$&$-0.3124$&$0.0124$&$-0.1704$&$0.0125$&$-0.2839$&$0.0123$\\
$C_c$                              &$-0.4442$&$0.0083$&$-0.7211$&$0.0084$&$-0.6574$&$0.0084$&$-0.7053$&$0.0085$\\
$\lambda$                          &$ 0.8630$&$0.0006$&$ 0.8140$&$0.0006$&$ 0.8201$&$0.0006$&$ 0.8147$&$0.0006$\\
$\lambda_c$                        &$ 0.7985$&$0.0006$&$ 0.7684$&$0.0006$&$ 0.7731$&$0.0006$&$ 0.7677$&$0.0006$\\
$\Delta_{11}/\sigma(\Delta_{11})$  &$ 0.0021$&$0.0023$&$ 0.0005$&$0.0022$&$ 0.0069$&$0.0022$&$ 0.0017$&$0.0021$\\
$|\Delta_{11}/\sigma(\Delta_{11})|$&$ 0.9167$&$0.0014$&$ 0.8679$&$0.0014$&$ 0.8778$&$0.0013$&$ 0.8672$&$0.0013$\\
\hline
\end{tabular}
\end{center}
\end{table}


\begin{table}
\begin{center}
\caption {The value of analyzed statistics for position angles $p$, the real sample of 247 Abell clusters.}
\label{tab:t3}
\begin{tabular}{c|c|cc|cc}
\hline
\hline
\multicolumn{1}{c}{}&
\multicolumn{1}{c}{}&
\multicolumn{2}{c}{Equatorial coordinates}&
\multicolumn{2}{c}{Supergalactic coordinates}\\
\hline
Sample&Test&$\bar{x}$&$\sigma(\bar{x})$&$\bar{x}$&$\sigma(\bar{x})$\\
\hline
A&$\chi^2$                           &$36.8591$&$0.5924$&$36.7899$&$0.6315$\\
 &$\chi_c^2$                         &$17.7579$&$0.4030$&$18.0619$&$0.4355$\\
 &$\Delta_{1}/\sigma(\Delta_{1})$    &$ 1.7046$&$0.0622$&$ 1.7021$&$0.0626$\\
 &$\Delta/\sigma(\Delta)$            &$ 2.2663$&$0.0594$&$ 2.2746$&$0.0591$\\
 &$\Delta_c/\sigma(\Delta_c)$        &$ 1.4619$&$0.0540$&$ 1.5682$&$0.0562$\\
 &$C$                                &$ 1.1940$&$0.4530$&$ 1.1220$&$0.4237$\\
 &$C_c$                              &$-0.1030$&$0.3003$&$ 0.1904$&$0.2990$\\
 &$\lambda$                          &$ 0.9177$&$0.0240$&$ 0.9138$&$0.0220$\\
 &$\lambda_c$                        &$ 0.8365$&$0.0242$&$ 0.8561$&$0.0248$\\
 &$\Delta_{11}/\sigma(\Delta_{11})$  &$-0.0005$&$0.0855$&$ 0.0940$&$0.0924$\\
 &|$\Delta_{11}/\sigma(\Delta_{11})|$&$ 1.0347$&$0.0543$&$ 1.1206$&$0.0588$\\
\hline
B&$\chi^2$                           &$36.4000$&$0.6072$&$36.2919$&$0.6124$\\
 &$\chi_c^2$                         &$17.5943$&$0.3963$&$17.8530$&$0.4216$\\
 &$\Delta_{1}/\sigma(\Delta_{1})$    &$ 1.6283$&$0.0577$&$ 1.6316$&$0.0578$\\
 &$\Delta/\sigma(\Delta)$            &$ 2.2055$&$0.0565$&$ 2.2199$&$0.0554$\\
 &$\Delta_c/\sigma(\Delta_c)$        &$ 1.4522$&$0.0521$&$ 1.5070$&$0.0525$\\
 &$C$                                &$ 0.8843$&$0.4355$&$ 0.7863$&$0.4212$\\
 &$C_c$                              &$-0.1070$&$0.3012$&$-0.0671$&$0.3063$\\
 &$\lambda$                          &$ 0.8928$&$0.0224$&$ 0.8934$&$0.0210$\\
 &$\lambda_c$                        &$ 0.8313$&$0.0228$&$ 0.8360$&$0.0228$\\
 &$\Delta_{11}/\sigma(\Delta_{11})$  &$ 0.0023$&$0.0826$&$ 0.0810$&$0.0866$\\
 &|$\Delta_{11}/\sigma(\Delta_{11})|$&$ 1.0079$&$0.0519$&$ 1.0700$&$0.0565$\\
\hline
\end{tabular}
\end{center}
\end{table}


\begin{table}
\begin{center}
\caption {
The value of analyzed statistics for position angles $p$, the real sample of 247 Abell clusters.
Random erorrs in  position angles $p$ are included.
}
\label{tab:t3a}
\begin{tabular}{c|c|cc|cc}
\hline
\hline
\multicolumn{1}{c}{}&
\multicolumn{1}{c}{}&
\multicolumn{2}{c}{Equatorial coordinates}&
\multicolumn{2}{c}{Supergalactic coordinates}\\
\hline
Sample&Test&$\bar{x}$&$\sigma(\bar{x})$&$\bar{x}$&$\sigma(\bar{x})$\\
\hline
A&$\chi^2$                              &$36.6984$&$0.5855$&$36.1891$&$0.5704$\\
 &$\chi_c^2$                            &$17.8344$&$0.3979$&$17.6854$&$0.4036$\\
 &$\Delta_{1}/\sigma(\Delta_{1})$       &$ 1.7012$&$0.0623$&$ 1.7045$&$0.0622$\\
 &$\Delta/\sigma(\Delta)$               &$ 2.2698$&$0.0590$&$ 2.2651$&$0.0583$\\
 &$\Delta_c/\sigma(\Delta_c)$           &$ 1.4560$&$0.0544$&$ 1.5608$&$0.0553$\\
 &$C$                                   &$ 0.7062$&$0.4719$&$ 1.5293$&$0.4213$\\
 &$C_c$                                 &$-0.2488$&$0.3175$&$ 0.5138$&$0.3012$\\
 &$\lambda$                             &$ 0.9216$&$0.0234$&$ 0.9164$&$0.0219$\\
 &$\lambda_c$                           &$ 0.8286$&$0.0242$&$ 0.8603$&$0.0244$\\
 &$\Delta_{11}/\sigma(\Delta_{11})$     &$-0.0016$&$0.0857$&$ 0.0930$&$0.0922$\\
 &|$\Delta_{11}/\sigma(\Delta_{11})|$   &$ 1.0336$&$0.0549$&$ 1.1214$&$0.0585$\\
\hline
B&$\chi^2$                              &$36.7267$&$0.6101$&$36.3227$&$0.6332$\\
 &$\chi_c^2$                            &$17.8267$&$0.4066$&$17.7057$&$0.4175$\\
 &$\Delta_{1}/\sigma(\Delta_{1})$       &$ 1.6246$&$0.0573$&$ 1.6326$&$0.0574$\\
 &$\Delta/\sigma(\Delta)$               &$ 2.2140$&$0.0551$&$ 2.2045$&$0.0549$\\
 &$\Delta_c/\sigma(\Delta_c)$           &$ 1.4670$&$0.0512$&$ 1.5030$&$0.0518$\\
 &$C$                                   &$ 0.1138$&$0.4225$&$ 0.4789$&$0.4146$\\
 &$C_c$                                 &$-0.6032$&$0.3047$&$-0.2049$&$0.2890$\\ 
 &$\lambda$                             &$ 0.8905$&$0.0215$&$ 0.9011$&$0.0206$\\
 &$\lambda_c$                           &$ 0.8290$&$0.0222$&$ 0.8254$&$0.0229$\\
 &$\Delta_{11}/\sigma(\Delta_{11})$     &$ 0.0070$&$0.0827$&$ 0.0793$&$0.0870$\\
 &|$\Delta_{11}/\sigma(\Delta_{11})|$   &$ 1.0116$&$0.0517$&$ 1.0760$&$0.0536$\\
\hline
\end{tabular}
\end{center}
\end{table}


\begin{table}
\begin{center}
\caption {
Jacknife analysis for position angles $p$, the real sample of 247 Abell clusters.
}
\label{tab:t3b}
\begin{tabular}{c|ccc|ccc}
\hline
\hline
\multicolumn{1}{c}{}&
\multicolumn{3}{c}{Sample A}&
\multicolumn{3}{c}{Sample B}\\
\hline
Test&$\bar{x}$&$\sigma_j(\bar{x})$&$(\bar{x}-E(\bar{x}))/\sigma_j(\bar{x})$&$\bar{x}$&$\sigma_j(\bar{x})$&$(\bar{x}-E(\bar{x}))/\sigma_j(\bar{x})$ \\
\hline
$\chi^2$                              &$34.1996$&$0.7052$&$<0. $&$33.7071$&$0.6906$&$<0. $\\
$\chi_c^2$                            &$15.8378$&$0.4903$&$<0. $&$15.8050$&$0.4785$&$<0. $\\
$\Delta_{1}/\sigma(\Delta_{1})$       &$ 1.7176$&$0.0621$&$7.44$&$ 1.6304$&$0.0617$&$6.09$\\
$\Delta/\sigma(\Delta)$               &$ 2.2761$&$0.0621$&$6.40$&$ 2.2044$&$0.0610$&$5.34$\\
$\Delta_c/\sigma(\Delta_c)$           &$ 1.4583$&$0.0517$&$3.30$&$ 1.4175$&$0.0595$&$2.76$\\
$C$                                   &$ 0.2654$&$0.5168$&$2.49$&$ 0.0262$&$0.5080$&$2.05$\\
$C_c$                                 &$-0.4880$&$0.3447$&$1.51$&$-0.4784$&$0.3406$&$1.56$\\
$\lambda$                             &$ 0.8670$&$0.0257$&$3.69$&$ 0.8390$&$0.0252$&$2.66$\\
$\lambda_c$                           &$ 0.7823$&$0.0263$&$1.99$&$ 0.7690$&$0.0254$&$1.53$\\
$\Delta_{11}/\sigma(\Delta_{11})$     &$ 0.0024$&$0.0642$&$0.02$&$ 0.0060$&$0.0642$&$0.07$\\
$|\Delta_{11}/\sigma(\Delta_{11})|$   &$ 1.0335$&$0.0642$&$3.67$&$ 1.0081$&$0.0642$&$3.26$\\
\hline
\end{tabular}
\end{center}
\end{table}


\begin{table}
\begin{center}
\caption {The value of analyzed statistics position angles $p$, the real sample of 247 Abell clusters, BM types,
Equatorial coordinates, sample A.}
\label{tab:t4}
\begin{tabular}{c|cc|cc|cc|cc|cc}
\hline
\hline
\multicolumn{1}{c}{}&
\multicolumn{2}{c}{BM I}&
\multicolumn{2}{c}{BM I-II}&
\multicolumn{2}{c}{BM II}&
\multicolumn{2}{c}{BM II-III}&
\multicolumn{2}{c}{BM III}\\
\hline
\multicolumn{1}{c}{Members}&
\multicolumn{2}{c}{35}&
\multicolumn{2}{c}{53}&
\multicolumn{2}{c}{59}&
\multicolumn{2}{c}{53}&
\multicolumn{2}{c}{47}\\
\hline
Test&$\bar{x}$&$\sigma(\bar{x})$&$\bar{x}$&$\sigma(\bar{x})$&$\bar{x}$&$\sigma(\bar{x})$&$\bar{x}$&$\sigma(\bar{x})$&$\bar{x}$&$\sigma(\bar{x})$\\
\hline
$\chi^2$                           &$36.603$&$1.546$&$  36.975$&$1.286$&$  36.232$&$1.134$&$  37.888$&$1.177$&$  36.544$&$1.609$\\
$\chi_c^2$                         &$17.837$&$1.024$&$  16.992$&$0.930$&$  18.035$&$0.696$&$  18.596$&$0.915$&$  17.268$&$1.003$\\
$\Delta_{1}/\sigma(\Delta_{1})$    &$ 1.658$&$0.154$&$   1.593$&$0.114$&$   1.539$&$0.113$&$   1.920$&$0.154$&$   1.827$&$0.160$\\
$\Delta/\sigma(\Delta)$            &$ 2.328$&$0.150$&$   2.252$&$0.109$&$   2.060$&$0.111$&$   2.492$&$0.143$&$   2.239$&$0.152$\\
$\Delta_c/\sigma(\Delta_c)$        &$ 1.393$&$0.150$&$   1.446$&$0.091$&$   1.413$&$0.102$&$   1.689$&$0.138$&$   1.333$&$0.128$\\
$C$                                &$ 1.486$&$1.162$&$   0.778$&$0.974$&$   0.279$&$0.888$&$   2.875$&$1.011$&$   0.696$&$1.075$\\
$C_c$                              &$-0.969$&$0.815$&$   0.093$&$0.483$&$  -0.215$&$0.525$&$   1.170$&$0.792$&$  -0.974$&$0.750$\\
$\lambda$                          &$ 0.938$&$0.063$&$   0.877$&$0.045$&$   0.850$&$0.044$&$   0.985$&$0.055$&$   0.955$&$0.063$\\
$\lambda_c$                        &$ 0.784$&$0.059$&$   0.814$&$0.044$&$   0.819$&$0.044$&$   0.934$&$0.063$&$   0.809$&$0.058$\\
$\Delta_{11}/\sigma(\Delta_{11})$  &$-0.287$&$0.205$&$   0.120$&$0.163$&$   0.125$&$0.164$&$   0.203$&$0.219$&$  -0.311$&$0.196$\\
$|\Delta_{11}/\sigma(\Delta_{11})|$&$ 0.932$&$0.137$&$   0.968$&$0.093$&$   0.996$&$0.100$&$   1.222$&$0.142$&$   1.021$&$0.134$\\
\hline
\end{tabular}
\end{center}
\end{table}


\begin{table}
\begin{center}
\caption {
The value of analyzed statistics position angles $p$, the real sample of 247 Abell clusters, BM types,
Equatorial coordinates, sample A. Random erorrs in  position angles $p$ are included.
}
\label{tab:t4a}
\begin{tabular}{c|cc|cc|cc|cc|cc}
\hline
\hline
\multicolumn{1}{c}{}&
\multicolumn{2}{c}{BM I}&
\multicolumn{2}{c}{BM I-II}&
\multicolumn{2}{c}{BM II}&
\multicolumn{2}{c}{BM II-III}&
\multicolumn{2}{c}{BM III}\\
\hline
\multicolumn{1}{c}{Members}&
\multicolumn{2}{c}{35}&
\multicolumn{2}{c}{53}&
\multicolumn{2}{c}{59}&
\multicolumn{2}{c}{53}&
\multicolumn{2}{c}{47}\\
\hline
Test&$\bar{x}$&$\sigma(\bar{x})$&$\bar{x}$&$\sigma(\bar{x})$&$\bar{x}$&$\sigma(\bar{x})$&$\bar{x}$&$\sigma(\bar{x})$&$\bar{x}$&$\sigma(\bar{x})$\\
\hline
$\chi^2$                              &$36.909$&$1.676$&$36.488$&$1.038$&$35.523$&$1.129$&$38.434$&$1.275$&$36.279$&$1.584$\\
$\chi_c^2$                            &$16.594$&$1.034$&$18.130$&$0.790$&$17.878$&$0.668$&$18.709$&$0.944$&$17.383$&$1.083$\\
$\Delta_{1}/\sigma(\Delta_{1})$       &$ 1.621$&$ 0.155$&$ 1.592$&$0.116$&$  1.545$&$0.113$&$  1.928$&$0.154$&$ 1.825$&$0.160$\\
$\Delta/\sigma(\Delta)$               &$ 2.325$&$ 0.150$&$ 2.233$&$0.110$&$  2.054$&$0.109$&$  2.507$&$0.143$&$ 2.275$&$0.150$\\
$\Delta_c/\sigma(\Delta_c)$           &$ 1.367$&$0.145$&$ 1.400$&$0.094$&$  1.430$&$0.102$&$  1.703$&$0.143$&$ 1.341$&$0.126$\\
$C$                                   &$ 0.767$&$1.188$&$-0.057$&$1.001$&$  0.425$&$0.890$&$  2.028$&$1.143$&$ 0.383$&$1.005$\\
$C_c$                                 &$-0.523$&$0.622$&$-0.826$&$0.648$&$ -0.300$&$0.589$&$  1.220$&$0.883$&$-0.985$&$0.683$\\ 
$\lambda$                             &$ 0.930$&$0.061$&$  0.859$&$0.043$&$  0.872$&$0.042$&$  1.004$&$0.054$&$ 0.957$&$0.064$\\
$\lambda_c$                           &$ 0.763$&$0.058$&$ 0.808$&$0.045$&$  0.820$&$0.041$&$  0.943$&$0.064$&$ 0.781$&$0.060$\\
$\Delta_{11}/\sigma(\Delta_{11})$     &$-0.255$&$0.199$&$ 0.128$&$0.164$&$  0.125$&$0.167$&$  0.192$&$0.225$&$-0.336$&$0.192$\\
$|\Delta_{11}/\sigma(\Delta_{11})|$   &$ 0.909$&$0.131$&$ 0.958$&$0.098$&$  1.005$&$0.103$&$  1.249$&$0.145$&$ 1.004$&$0.132$\\
\hline
\end{tabular}
\end{center}
\end{table}

\end{document}